\documentclass[12pt,a4paper]{article}            
\usepackage[skins,theorems]{tcolorbox}
\tcbset{highlight math style={enhanced, colframe=red,colback=white,arc=0pt,boxrule=1pt}}
\usepackage[bookmarksopen, bookmarksnumbered, bookmarksopenlevel=2,linktocpage]{hyperref}
\usepackage{tikz}
\usepackage[svgnames]{xcolor}
\usepackage{tikz-3dplot}
\usepackage{multirow}
\usetikzlibrary{calc}
\usetikzlibrary{decorations} %
\usepackage[UKenglish]{babel}
\usepackage[toc,page]{appendix}
\usepackage{amsmath}
\usepackage{amssymb}
\usepackage{graphicx}
\usepackage{hhline}
\usepackage[bf]{caption}
\usepackage{cite}
\usepackage[vcentermath]{youngtab}
\usepackage{geometry}
\usepackage{slashed}
\usepackage{color}
\usepackage{stackrel}
\usepackage{tikz-cd} 
\usepackage{cancel} 
\usepackage{multirow}
\usepackage{multicol}
\usepackage[normalem]{ulem}
\usepackage{empheq}
\usepackage{arydshln}
\usepackage{tablefootnote}

\newenvironment{eqn*}{\begin{equation*}\begin{aligned}}{\end{aligned}\end{equation*}\noindent}
\newcommand{\bqa}{\begin{eqnarray}}
\newcommand{\eqa}{\end{eqnarray}}

\usepackage{footnote}

\hypersetup{
    pdftitle={},
    pdfauthor={},
    pdfsubject={}
}
\numberwithin{equation}{section}
\numberwithin{table}{section}

\makeatletter

\newcommand{\be}{\begin{equation}}
\newcommand{\ee}{\end{equation}}
\newcommand{\beq}{\begin{equation}}
\newcommand{\eeq}{\end{equation}}
\newcommand{\ba}{\begin{aligned}}
\newcommand{\ea}{\end{aligned}}

\newcommand{\bea}{\begin{eqnarray}}
\newcommand{\eea}{\end{eqnarray}}

\newcommand{\cO}{\mathcal{O}}

\newcommand{\cL}{\mathcal{L}}

\newcommand{\cN}{\mathcal{N}}

\newcommand{\cF}{\mathcal{F}}

\newcommand{\cV}{\mathcal{V}}

\newcommand{\cM}{\mathcal M}

\newcommand\bi{\begin{itemize}}
\newcommand\ei{\end{itemize}}

\usepackage[dvipsnames]{xcolor}

\renewcommand{\a}{{\alpha}}
\renewcommand{\b}{{\beta}}

\renewcommand{\k}{{\kappa}}

\def\ad{\mathop{\mathrm{ad}}\nolimits}

\def\diag{\mathop{\mathrm{diag}}\nolimits}
\def\Im{\mathop{\mathrm{Im}}\nolimits}

\def\Re{\mathop{\mathrm{Re}}\nolimits}

\def\tr{\mathop{\mathrm{tr}}\nolimits}

\def\unit{{1\kern-.65ex {\rm l}}}
\def\1{{1\kern-.65ex {\rm l}}}

\def\bbP{\mathbb{P}}

\def\ch{\mathrm{ch}}
\def\rk{\mathrm{rk}}
\def\td{\mathrm{Td}}

\def\CF{{\cal F}}

\def\CM{{\cal M}}

\def\bbC{{\mathbb{C}}}

\def\bbH{{\mathbb{H}}}

\def\bbP{{\mathbb{P}}}
\def\bbQ{{\mathbb{Q}}}

\def\bbZ{{\mathbb{Z}}}

\newcount\hour \newcount\minute
\hour=\time \divide \hour by 60
\minute=\time
\def\now{%
\ifnum \hour<13
  \ifnum \hour=0 \advance \hour by 12 \number\hour:\else \number\hour:\fi%
     \ifnum \minute<10 0\fi%
     \number\minute%
\ A.M.%
\else \advance \hour by -12 \number\hour:%
  \ifnum \minute<10 0\fi%
  \number\minute%
  \ P.M.%
\fi%
}

\makeatother

\begin{document}

\rightline{ \small IFT-UAM/CSIC-26-156}
\rightline{ \small ZMP-HH/26-35}

\vskip 30pt
\begin{center}

\vspace{0cm}

{\Large \bf Non-Geometric Heterotic Vacua from Type II$_0$ Loci} 

 \vskip 11 mm


Christian Aoufia${}^1$, Giacomo Maggiorotti,${}^{2,3}$ and Max Wiesner${}^{2,3}$

\vskip 11 mm
${}^1${\it Instituto de Fisica Teórica UAM-CSIC, c/ Nicolás Cabrera 13-15, 28049 Madrid, Spain}  \\[3 mm]

${}^2$\textit{II. Institut f\"ur Theoretische Physik, Universit\"at Hamburg, Notkestrasse 9,\\ 22607 Hamburg, Germany} \\ [3mm]

${}^3$\textit{Zentrum f\"ur Mathematische Physik, Universit\"at Hamburg, \\ 
Bundesstrasse 55, 20146 Hamburg, Germany}

\end{center}

\vskip 7mm

\begin{abstract}
We determine 4d $\cN=2$ heterotic string backgrounds dual to Type IIB string theory compactified on Calabi--Yau threefolds containing a type II$_0$ locus in their complex structure moduli space. While previous work only established the existence of such a duality frame, we concretely identify the heterotic theories dual to one- and two-parameter models exhibiting such loci, yielding novel instances of Type IIB--heterotic duality. To establish the correspondence, we first match the massless $\mathcal{N}=2$ spectrum, which uniquely fixes the heterotic data. We then recover the light spectrum of BPS D3-branes wrapping cycles on the Type IIB side from the heterotic perspective, and finally match the duality groups governing the global structure of the vector multiplet moduli space.
All resulting heterotic theories are genuinely lower-dimensional, as they do not admit a decompactification limit to ten dimensions in their hypermultiplet moduli space. We identify two classes of duals. In the first, the heterotic string is compactified on a geometric K3 with a standard gauge bundle, demonstrating explicitly that type II$_0$ loci admit a consistent description as weakly coupled heterotic strings.  In the second, the heterotic string background is given by a string-scale K3 and is hence non-geometric and genuinely four-dimensional. In this case, directly establishing supersymmetry would require control over all perturbative and non-perturbative $\alpha'$ corrections. Here, the duality to Type IIB is most powerful, as it establishes the existence of new consistent, supersymmetric, and anomaly-free heterotic compactifications beyond the reach of conventional geometric methods. Finally, the Type IIB realization of the two-parameter models reveals that four-dimensional heterotic compactifications admit smooth transitions involving the nucleation of NS5-branes.

\end{abstract}

\vfill

\thispagestyle{empty}
\setcounter{page}{0}
 \newpage

\thispagestyle{empty}
{ \hypersetup{hidelinks} \tableofcontents }
\vspace{25pt} 
\setcounter{page}{1}
\section{Introduction}
Dualities play a key role in theoretical physics as they often-times enable explicit computations for strongly coupled systems and therefore provide a window into non-perturbative physics. In string theory, dualities are ubiquitous and underlie much of the success of describing theories of quantum gravity through string theory. Among the string dualities the most prominent and best understood examples are those relating the consistent ten-dimensional critical superstring theories~\cite{Polchinski:1995df,Schwarz:1995dk} and eleven-dimensional supergravity/M-theory~\cite{Witten:1995ex,Horava:1995qa,Horava:1996ma}. Still, once the ten-dimensional string theories are compactified to lower dimensions the web of possible string dualities becomes even more diverse
and allow to relate string theories that are not dual in ten dimensions, see~\cite{Aspinwall:1996mn} for a review. A classic example is T-duality which requires a compactification to at least nine dimensions~\cite{Dine:1988nrl}. The web of string dualities grows even further if the compactification manifold breaks a subset of the supercharges allowing, e.g., to relate  heterotic string theory and F-theory in eight dimensions~\cite{Vafa:1996xn} or the heterotic string and Type IIA string theory in six dimensions~\cite{Witten:1995ex, Sen:1995cj,Harvey:1995rn}. Also these instances of lower-dimensional dualities with less preserved supersymmetry are well-understood and part of the standard string-theory toolkit. 

The appeal of string dualities in the context of gravity is that they provide important insights into the nature of the gravitational theories with string theory as UV completion. Specifically, string dualities have experienced a renewed interest in the context of the Swampland program~\cite{Vafa:2005ui}, see~\cite{Palti:2019pca,vanBeest:2021lhn,Grana:2021zvf,Agmon:2022thq} for reviews, that aims to understand the consistency conditions that  quantum gravity imposes on effective theories of gravity. In this context, dualities are prominently featured in the Distance Conjecture~\cite{Ooguri:2006in} and its refinement, the Emergent String Conjecture (ESC)~\cite{Lee:2019oct}. These conjectures make statements about extreme regimes in the scalar field space of gravitational theories that have a consistent quantum gravity origin. The Distance Conjecture~\cite{Ooguri:2006in} states that, if we traverse large geodesic distances in this scalar field space, a tower of states becomes light exponentially, in Planck units, in the geodesic distance. The ESC refines this by specifying the nature of these light states. Concretely, the statement of the ESC is that the lightest tower of states becoming massless at infinite distance can either be a Kaluza--Klein (KK) tower, signalling a decompactification to a higher-dimensional theory of gravity, or the excitation tower of a critical string~\cite{Lee:2019oct}. Taken together, the distance conjecture and the ESC require that, for a gravitational theory to have a consistent UV completion, a weakly coupled dual description of the original theory has to emerge at infinite distance in its moduli space. The fundamental degrees of freedom of the emerging dual descriptions are then encoded in the tower of states that become light asymptotically. 

Both conjectures have been studied in great detail, in particular in string compactifications to flat space~\cite{Lee:2018urn,Lee:2019jan,  Lee:2019oct,Joshi:2019nzi, Lee:2019xtm, Marchesano:2019ifh, Baume:2019sry, Xu:2020nlh, Klaewer:2020lfg, Lee:2021qkx,Lee:2021usk, Alvarez-Garcia:2021pxo, Basile:2022zee, Rudelius:2023odg, Alvarez-Garcia:2023gdd, Blumenhagen:2023yws, Alvarez-Garcia:2023qqj, Chen:2024cvc, Aoufia:2024awo, Hassfeld:2025uoy, Monnee:2025ynn, Monnee:2025msf, Aoufia:2026bau,Raucci:2026fzp}, but also in AdS spacetimes~\cite{Baume:2020dqd,Perlmutter:2020buo,Baume:2023msm,Ooguri:2024ofs,Calderon-Infante:2024oed,Calderon-Infante:2026rkj,Mantegazza:2026spd}, and from the bottom-up perspective, without assuming a UV completion given by string theory~\cite{Basile:2023blg, Cribiori:2023ffn, Basile:2024dqq,Herraez:2024kux, Bedroya:2024ubj, Compratore:2024gqo,Grieco:2026yip,Compratore:2026rcn}. Most of the string-theoretic tests of the ESC make explicit use of the well-known superstring dualities. However, there exist setups for which the well-understood string dualities do not determine the duality frame that emerges in a given infinite distance limit in the moduli space. These are the more interesting cases as here the ESC provides non-trivial insights into the quantum gravitational UV completion of a theory. The reason is that for a given infinite distance limit in the moduli space without a standard dual interpretation, the ESC provides two logical consequences. First, the infinite distance limit may not persist at the quantum level since quantum corrections remove it from the actual moduli space. In this case, the classical infinite distance limit is said to be obstructed. Thus, if it can be shown that in a given classical infinite distance limit neither of the dual frames allowed by ESC can emerge, the ESC predicts the existence of quantum corrections that remove the infinite distance limit from the quantum moduli space. This interplay between the ESC and quantum corrections has been investigated in detail in~\cite{Baume:2019sry,Alvarez-Garcia:2021pxo,Alvarez-Garcia:2023gdd,Alvarez-Garcia:2023qqj,Compratore:2026fli,Compratore:2026mha,Compratore:2026tsy}, see also~\cite{Bonnefoy:2020uef}. A second logical possibility is that the classical infinite distance limit persists at the quantum level since quantum corrections are either suppressed or do not contribute at all. In this case, the ESC predicts the existence of a new string duality that does not derive from the well-established dualities. There is, of course, a third logical possibility namely that the Distance and Emergent String Conjectures in their original formulation are not valid and that they need to be suitably reformulated. For the setups that we discussed in this work, this possibility was put forward in~\cite{Hattab:2025aok,Hattab:2026lho}.

In this work, we take Calabi--Yau threefold compactifications of Type IIB string theory as our starting point. These compactifications lead to 4d $\cN=2$ effective theories of supergravity in the low-energy limit. The moduli space of the 4d $\cN=2$ theory of supergravity splits into a hypermultiplet and a vector multiplet moduli space. For Type IIB compactifications on Calabi--Yau threefolds, the vector multiplet sector is described by the complex structure deformations of the threefold and, crucially, does not receive any quantum corrections. Thus, according to the two logical consequences of the ESC outlined above, to any infinite distance limit in the complex structure sector of the threefold, there has to be a weakly coupled theory emerging that is dual to Type IIB on the Calabi--Yau threefold. Conversely, if this were not the case, the Distance and Emergent String Conjectures would be violated and would have to be refined. In this work, we provide further evidence that no such refinement is needed by explicitly constructing the required weakly-coupled dual description even in exotic corners of the 4d $\cN=2$ vector multiplet moduli space.

The infinite distance limits in the complex structure moduli space of Calabi--Yau threefolds have been investigated in detail in the context of the Distance Conjecture, starting with~\cite{Grimm:2018cpv,Grimm:2018ohb}. This led to a classification of infinite distance limits in terms of asymptotic Hodge theory, which describes the properties of the splitting of their middle cohomology in the asymptotic regime. Based on this classification, there are three classes of infinite distance degenerations which are referred as type II, III and IV.\footnote{Conversely, it has been shown recently in~\cite{Monnee:2026bsj} that all type II, III, and IV degenerations according to this classification are indeed at infinite distance in the K\"ahler metric on the moduli space.} As shown in~\cite{Compratore:2026rcn}, the classification of the infinite distance limits into type II, III, and IV based on asymptotic Hodge structure does not require a concrete UV completion and is a consequence of supergravity alone. However, the classification of infinite distance limits alone does not determine the duality frame that emerges in the respective limit. Instead, to determine the duality frame, the input from the UV completion is required. In the case of Calabi--Yau compactifications, this additional information is encoded in the geometry of the threefold. 

At infinite distance in its complex structure moduli space, a Calabi--Yau threefold undergoes certain types of degeneration. Using the geometry of such Calabi--Yau degenerations, it was shown in~\cite{Hassfeld:2025uoy,Monnee:2025ynn} that, indeed, all infinite distance limits in the vector multiplet moduli space of Type IIB Calabi--Yau threefold compactifications satisfy the emergent string conjecture, in the sense that there exists a consistent duality frame that emerges in these regimes. Concretely, it was shown in~\cite{Hassfeld:2025uoy} that type II degenerations correspond to emergent string limits in which a weakly coupled Type II or heterotic string emerges. In Type IIB string theory, this critical string can be constructed as a certain BPS string solution, dubbed EFT string in~\cite{Lanza:2020qmt,Lanza:2021udy}, of the 4d $\cN=2$ supergravity. Subsequently, it was then shown in~\cite{Monnee:2025ynn} that type III degenerations correspond to decompactifications to six dimensions and type IV degenerations to decompactifications to five dimensions. 

It is important to stress that the results of~\cite{Hassfeld:2025uoy,Monnee:2025ynn} show the existence of light states that can make up light degrees of freedom of a duality frame predicted by the ESC.
More precisely, in~\cite{Hassfeld:2025uoy} it was shown that there~\emph{exists} a weakly coupled critical string that becomes tensionless in type II limits. Similarly,~\cite{Monnee:2025ynn} demonstrated that there exist BPS states becoming light in type III or IV degenerations that have exactly the properties of a KK-tower associated with the compactification of a six- or five-dimensional theory, respectively. In neither case, however, do the results of~\cite{Hassfeld:2025uoy,Monnee:2025ynn} determine how the critical string theory dual to type II limits or the higher-dimensional theory emerging in type III or IV limits is concretely realized. Of course, in certain cases additional dualities can be used to give a concrete realization of the dual theory. This is particularly transparent for degenerations that are connected to the large complex structure point (LCS) in the complex structure moduli space. In the vicinity of this point, mirror symmetry relates Type IIB on a Calabi--Yau threefold $X$ to Type IIA on the mirror Calabi--Yau $Y$ in the large volume regime. For example, the five-dimensional theory arising in a type IV limit connected to LCS is then simply realized as M-theory on $Y$. However, in the complex structure moduli space of Calabi--Yau threefolds there are also infinite distance degenerations that are not connected to the LCS point. Hence, here mirror symmetry cannot be invoked to understand the realization of the emerging dual theory and the results of~\cite{Hassfeld:2025uoy,Monnee:2025ynn} only tell us that a consistent dual theory should exist. 

The goal of this work is to determine the exact heterotic realization of the theories dual to type II loci in the vector multiplet moduli space of Type IIB Calabi--Yau compactification. In this way, the ESC applied to Type IIB really establishes new string dualities beyond the well-studied ones. We are in particular interested in type II degenerations that are not connected to the LCS point. According to the results of~\cite{Hassfeld:2025uoy}, there is a string becoming tensionless in these limits, for which the massless spectrum on the worldsheet can be determined from its realization as a BPS solution to Type IIB supergravity and which reproduces the matter spectrum on the worldsheet of the heterotic string. A particular class of type II loci that are not connected to the LCS locus are the so-called type II$_0$ loci, which have previously been investigated in detail in~\cite{Joshi:2019nzi,Lee:2021qkx,Lee:2021usk,Alvarez-Garcia:2023gdd,Monnee:2025msf,Hattab:2025aok,Hattab:2026lho}. Unlike other type II loci, these already exist in one-parameter models where they are referred to as K-points. K-points for general one-parameter Calabi--Yau threefolds have been investigated in detail in~\cite{Bastian:2023shf} and have been prominently featured also in~\cite{Hattab:2025aok}. In this work, we focus on these loci and study how the dual heterotic theory is concretely realized. In view of the discussion above, this means not only to establish the existence of a string for which the worldsheet theory has the correct critical central charge~\cite{Hassfeld:2025uoy}, but to actually determine the worldsheet theory characterizing this dual string. In particular, this requires to work out the gauge data that specify the heterotic string theory, which cannot be directly inferred from the results of~\cite{Hassfeld:2025uoy}. The goal of this work is to provide such heterotic data and thereby determine the detailed realization of the duality frame that emerges at type II$_0$ loci in the complex structure moduli space of Calabi--Yau threefolds. In this way, we obtain, on the one hand, novel heterotic configurations and, on the other hand, new instances of Type II--heterotic dualities that cannot be obtained from one of the higher-dimensional Type II--heterotic dualities mentioned above. As a side product, we also show that, even though the heterotic dual theory may have some special features, from the heterotic perspective the type II$_0$ locus is indeed the weak-coupling locus in full accordance with the ESC.

The gauge theory data on the heterotic strings dual to type II$_0$ loci of Calabi--Yau threefolds are identified by matching the low-energy effective action of the Type IIB and heterotic configurations. In particular, this requires to go beyond the vector multiplet sector since the heterotic string theory has to also reproduce the correct number of hypermultiplets of the dual Type IIB compactification. Thus, a candidate for the worldsheet theory on the heterotic string can be identified already at the level of the effective supergravity theory, resulting in a concrete proposal for a duality between a specific heterotic string configurations and a Calabi--Yau compactifications of Type IIB string theory. 
This proposed duality can then be tested beyond the supergravity level. First of all, the perturbative heterotic BPS spectrum beyond the massless level can be matched with a certain subsector of the BPS spectrum on the Type IIB side. This match provides a test of the duality at the quantum level that is perturbative from the heterotic perspective. We can further test the duality at the non-perturbative heterotic level by also matching the duality group of the theory, i.e., by reproducing the action of the monodromy around the type II locus in the Type IIB complex structure moduli space on the heterotic side. By doing so, we demonstrate that the two theories do not only share the same Teichm\"uller space, but also that the actual moduli space, including the global structure, matches on both sides of the duality. This procedure thus establishes new dualities between heterotic string configurations and Type IIB compactifications on Calabi--Yau threefolds. 

In this paper, we explicitly work out this duality for four examples. First, we discuss the heterotic dual of Type IIB compactified on the three hypergeometric Calabi--Yau threefolds that have a K-point in their complex structure moduli space. These Calabi--Yau threefolds have a single complex structure parameter and are denoted by $X_{3,3}$, $X_{4,4}$ and $X_{6,6}$. As we show in this work, the heterotic dual of $X_{3,3}$ can simply be viewed as a compactification of the heterotic $SO(32)$ string on a K3$\times T^2$ for which the moduli of the $T^2$ are fixed to a specific value. In this case, the heterotic dual is simply a variant of the heterotic string configurations underlying other Type IIA--heterotic dualities~\cite{Kachru:1995wm}. The proposed duality also passes the perturbative and non-perturbative tests outlined above. Thus, the duality of the $X_{3,3}$ K-point to a weakly coupled heterotic string can be viewed as a proof-of-principle that, also away from LCS, the existence of the heterotic dual frame shown in~\cite{Hassfeld:2025uoy} can be turned into an actual duality between two concrete string theory constructions. In particular, it shows that not only the effective action but also the perturbative and non-perturbative BPS spectrum of the heterotic string theory fully reproduce the properties of the Type IIB compactifications. In particular, it provides a very transparent and simple example that the Distance and Emergent String Conjectures are indeed satisfied in their original formulation and also hold near type II$_0$ loci. However, here the duality does not provide much new insight, since the duality between Type IIB on $X_{3,3}$ and the heterotic string theory relies on rather standard constructions of the heterotic string. 

From this perspective, the heterotic theories dual to the $X_{4,4}$ and $X_{6,6}$ models are more interesting. As we show in this work, these correspond to heterotic compactifications on K3$\times T^2$ for which not only the $T^2$ moduli are fixed but the same is true for the overall volume modulus of the K3. For this reason, these models do not have decompactification limits to either eight or six dimensions and can be considered to be genuinely four-dimensional models. Interestingly, also the gauge theory data for these heterotic models are more exotic than for the $X_{3,3}$ model. Just from the heterotic perspective, it is not obvious that the resulting worldsheet theory is supersymmetric and anomaly free since the gauge theory data have some non-standard features. However, the latter are uniquely fixed by the requirement that the massless spectrums on the two sides of the duality have to match and moreover passes the (non-)perturbative duality tests. This provides very strong evidence that the worldsheet theory on the string associated with the corresponding type II degeneration of the Calabi--Yau threefold is indeed described by these data. On the Type IIB side this string arises as a BPS solution to Type IIB string theory which automatically ensures that the resulting worldsheet theory is anomaly free and supersymmetric. For the K-points in the $X_{4,4}$ and $X_{6,6}$ models, we thus use the Type IIB perspective to conclude that the proposed genuinely four-dimensional heterotic construction does indeed lead to a consistent, anomaly free and supersymmetric worldsheet theory. In particular, the Type IIB side automatically encodes the fully $\alpha'$-corrected supersymmetry conditions on the heterotic side and provides solutions to these.

Our final example is a two-parameter model that was first studied in~\cite{CANDELAS1994481,Kachru:1995wm,Kachru:1995fv} and more recently also in~\cite{Lee:2019oct,Bastian:2021eom,Monnee:2025msf,Castellano:2026bnx,Hattab:2026lho}. This model features a type II$_0$ locus for which the heterotic dual was proposed to be a non-perturbative compactification involving a spacetime-filling NS5-brane~\cite{Monnee:2025msf}. In this work, we demonstrate that such a heterotic configuration with a single spacetime-filling NS5-brane indeed exists and reproduces the massless supergravity spectrum. We further show that it passes our (non-)perturbative duality tests. The gauge theory data for this model is of similar kind to that of the heterotic dual to the $X_{4,4}$ and $X_{6,6}$ models in that also here the heterotic K3 is fixed to string-size volume. Importantly, as anticipated in~\cite{Monnee:2025msf}, this exotic-looking heterotic string is dual to a more standard heterotic configuration without NS5-branes~\cite{Kachru:1995fv} implying that in four dimensions there can be smooth transitions in which NS5-branes are nucleated -- a process whose analogue in six-dimensions would necessarily be singular. The results of this paper therefore shine new light on properties of the heterotic string that are special to four dimensions. 

The rest of this paper is structured as follows: in Section~\ref{sec:het} we review the results of~\cite{Hassfeld:2025uoy}, demonstrating the emergence of a heterotic string in the vicinity of type II singularities in the vector multiplet moduli space of Type IIB string theory compactifications over Calabi--Yau threefold. We further review the relevant aspects of the complex and K\"ahler structure moduli spaces of the various threefolds that we consider in this work. In Section~\ref{sec:HetDualTypeII0}, we introduce the general strategy to identify the heterotic dual theories to the type II$_0$ loci and introduce the general procedure behind the (non-)perturbative tests of the proposed dualities that we perform in this work. We also present the fully worked out heterotic dual of Type IIB on $X_{3,3}$. In Section~\ref{sec:4dHetVac} we then construct the genuinely four-dimensional heterotic backgrounds dual to the $X_{4,4}$ and $X_{6,6}$ models and present the details of the non-trivial tests of these dualities. Section~\ref{sec:twoparametermodelshet} then presents the details of the heterotic dual of type II$_0$ loci in a two-parameter example and we further discuss the properties of the transitions nucleating spacetime-filling NS5-branes in four dimensions. We conclude in Section~\ref{sec:conclusions}. The appendices contain some background informations on specific aspects of heterotic gauge bundles. 

\section{Calabi--Yau Degenerations and Heterotic Strings}\label{sec:het}

Type II--heterotic duality is by no means a new topic. Although the first instances of four-dimensional $\cN=2$ Type II string theory compactifications with a heterotic dual were identified already in~\cite{Kachru:1995wm}, these classic examples cover only a particular class of Type II theories. This is most transparent in Type IIA, where the four-dimensional duality can be viewed as descending from the six-dimensional duality between Type IIA on K3 and the heterotic string on $T^4$. The evidence for the six-dimensional duality is based on the fact that the moduli space of a conformal field theory (CFT) on $T^{20,4}$ agrees with the sigma model moduli space on K3. Furthermore, reducing the worldvolume theory of a Type IIA NS5-brane on K3 yields a 2d field theory that can be identified with the worldsheet theory of the heterotic string. 

Taking the heterotic $T^4=T^2\times T^2$, the standard Type IIA--heterotic duality in four dimensions with $\cN=2$ supersymmetry~\cite{Kachru:1995fv,Klemm:1995tj} is obtained by adiabatically fibering the six-dimensional one over $\mathbb{P}^1$, schematically
\begin{equation}
    {\rm Type\;IIA\;on\;} {\rm K3}\to \mathbb{P}^1 \quad \longleftrightarrow\quad  {\rm Heterotic\;on\;}T^2\times (T^2 \to \mathbb{P}^1)\,.
\end{equation}
The factor $T^2 \to \bbP^1$ on the heterotic side can moreover be identified with a K3. Thus, the duality identifies Type IIA string theory on a K3-fibered Calabi--Yau threefold with the heterotic string on K3$\times T^2$. This heavily relies on the adiabatic fibration and hence requires the volumes of the $\mathbb{P}^1$s on both sides of the duality to be large. This in turn means that also the volumes of the threefolds are large.

Via mirror symmetry, the duality between the heterotic string and Type IIA compactifications extends to Type IIB. Here, the standard duality is realized in the vicinity of the large complex structure (LCS) regime of the mirror threefold which, via mirror symmetry, maps to the large volume regime of Type IIA on the original Calabi--Yau threefold. However, the large volume/LCS regime is just a small region of the moduli space of 4d $\cN=2$ supergravity theories and in principle other loci can have a weakly-coupled heterotic dual description as well. From the Type IIA perspective, these loci do not correspond to a non-linear sigma model with a Calabi--Yau target space, such that establishing the heterotic dual theory cannot make use of the adiabatic fibration described above. For this reason, a different strategy was pursued in~\cite{Hassfeld:2025uoy}, working directly in the complex structure moduli space of Type IIB string theory compactified on a Calabi--Yau threefold. Following this strategy, the existence of a heterotic duality frame in certain corners of the moduli space is argued for by identifying string solutions that become weakly coupled and tensionless. As we review in the following, using the geometry of the underlying Calabi--Yau threefold, the worldsheet spectrum on these strings can then be shown to agree with the heterotic string worldsheet.

\subsection{Type II Degenerations of Calabi--Yau Threefolds}\label{sec:type_II_degenerat}
\begin{figure}[!t]
    \centering
    \includegraphics[width=0.75\linewidth]{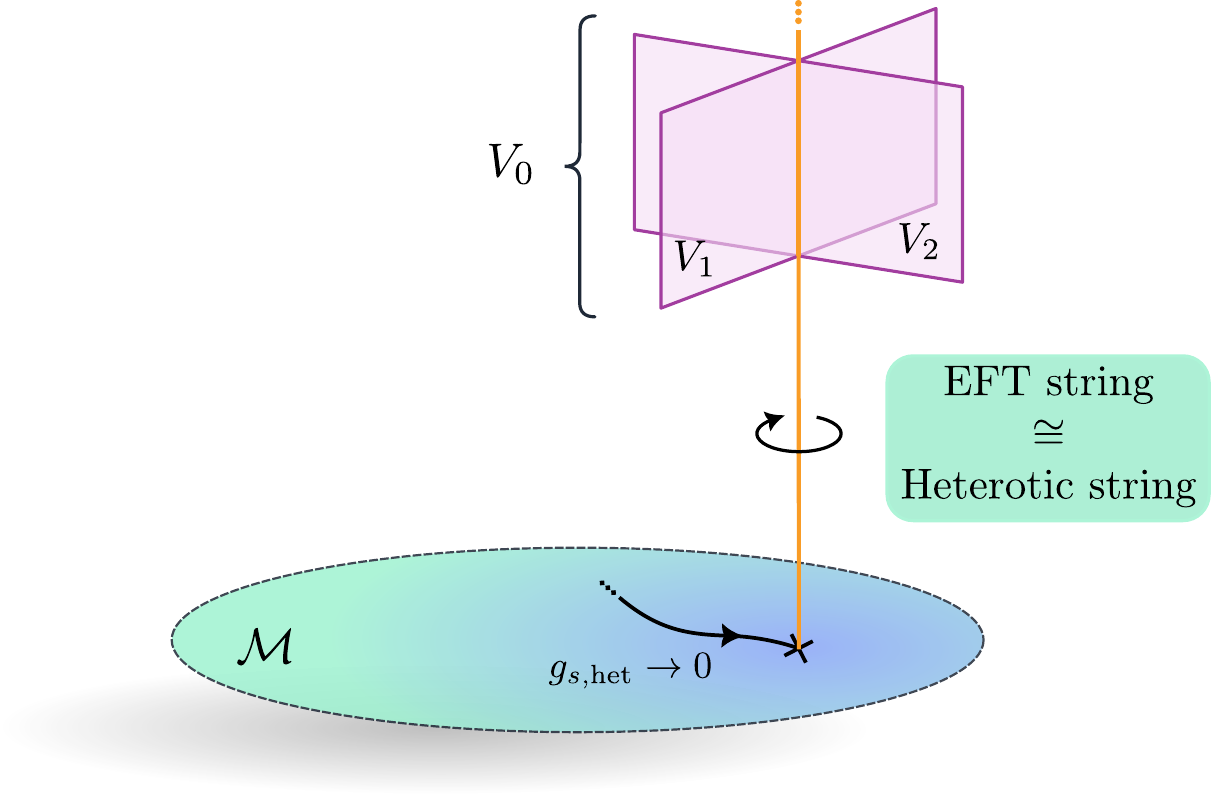}
    \caption{Sketch of the Tyurin--type degeneration near type II loci in the complex structure moduli space of a Calabi--Yau threefold. The supergravity EFT string that drives the moduli to this locus near its core is identified with a critical heterotic string becoming weakly coupled in the limit. }
    \label{fig:typeIIdeg}
\end{figure}

To study heterotic string solutions in Type IIB Calabi--Yau compactifications,~\cite{Hassfeld:2025uoy} focused on the so-called type~II loci in the complex structure moduli space of Calabi--Yau threefolds, in the nomenclature of~\cite{Grimm:2018ohb,Grimm:2019bey}. The simplest class of such type II loci correspond geometrically to Tyurin degenerations~\cite{tyurin2003fano}. Consider a Calabi--Yau threefold $V$ and a complex structure modulus $z$ such that $z\to 0$ realizes a Tyurin degeneration. There then exists a family $V_z$ of Calabi--Yau threefolds such that $V_0$ is singular and (possibly after base change and birational transformations) can be brought in semi-stable form 
\begin{equation}\label{eq:tyurin}
    V_0 = V_1 \cup_Z V_2 \,,\qquad Z\equiv V_1\cap V_2\,.
\end{equation}
Here $V_1$ and $V_2$ are two threefolds that intersect over a twofold $Z$. The Calabi--Yau condition for $V_{z\neq0}$ ensures that $Z$ itself is a K3 surface. The string solutions of the 4d effective supergravity theory that become tensionless in the limit $z\to 0$ can be described as fibrations 
\begin{equation}
    V \hookrightarrow \cV \to \mathbf{D}_z\,,
\end{equation}
where $\mathbf{D}_z$, with complex coordinate $z$, is interpreted as the spacetime coordinate transverse to the string in 4d. In particular, the fiber over $z=0\in \mathbf{D}_z$ is given by $V_0$ in~\eqref{eq:tyurin}. For the string solution to be supersymmetric, the total space $\cV$ of the fibration has to be a Calabi--Yau fourfold itself. This string is an instance of an EFT string introduced in~\cite{Lanza:2021udy} and the string configuration is illustrated in Figure~\ref{fig:typeIIdeg}.
The spectrum of massless states propagating on the string worldsheet is determined by the modes of the 10d Type IIB supergravity multiplet that in the reduction of Type IIB supergravity on $\cV$ localize to $\{z=0\}$. As it turns out, for the Tyurin-type degenerations~\eqref{eq:tyurin} the mode counting yields a string with left- and right-moving central charges $(c_L,c_R)=(24,12)$, which we identify with the heterotic string in light-cone gauge. For later convenience, let $t=\frac{\log(z)}{2\pi i}+\mathcal{O}(z)$ denote the asymptotic value of the complex structure parameter of $V$ that controls the degeneration. For $t\to i \infty$ the degeneration in~\eqref{eq:tyurin} is realized everywhere in the space transverse to the string. From the supergravity solution, one confirms that the heterotic string becomes tensionless and weakly coupled in the limit~\cite{Hassfeld:2025uoy}. 

The degeneration~\eqref{eq:tyurin} is thus compatible with the existence of a heterotic string. Since the Type IIB supergravity solution preserves four supercharges of the same chirality, the worldsheet theory on the string has $\cN=(0,4)$ right-moving supersymmetry. The weakly-coupled heterotic string arising in the limit $t\to i \infty$ is hence realized as a compactification on K3$\times T^2$ or, as we will see in the following, a non-geometric variant thereof. This, however, does not fully determine the heterotic dual, since the gauge theory data must still be specified or, equivalently, the actual worldsheet action has to be determined. Indeed, the geometry of the degeneration of $V$ on the Type IIB side only provides partial insight, as it only fixes the rank of the unbroken perturbative heterotic gauge group and the number of spacetime-filling NS5-branes wrapping the heterotic torus. Specifically, the rank of the heterotic gauge group is encoded in the polarization of the K3 surface $Z=V_1\cap V_2$: the components $V_1$ and $V_2$ of the normal crossing variety $V_0$ induce a polarization on the K3 encoded in the polarization lattice $\Lambda_{\rm pol}$ with signature 
\begin{equation}\label{eq:sgnLambdapol}
    {\rm sgn}\,\Lambda_{\rm pol} = (1,\rho)\,,\qquad 0\leq \rho\leq 19\,. 
\end{equation}
The orthogonal complement of $\Lambda_{\rm pol}$ in $H^2(Z,\mathbb{Z})$ is referred to as the transcendental lattice $\Lambda_{\rm trans}$ with signature
\begin{equation}
    {\rm sgn}\,\Lambda_{\rm trans} = (2, 19-\rho) \,. 
\end{equation}
As explained in~\cite{Hassfeld:2025uoy}, elements in $\Lambda_{\rm trans}$ correspond to cohomology classes in $H^3(V_{z\neq 0},\mathbb{Z})$. In Type IIB, they give rise to closed-string $U(1)$s that become weakly coupled in the $t\to i\infty$ limit. From the heterotic perspective, these are associated with the Cartan subgroup of the perturbative gauge group that is unbroken by the choice of gauge background.  

The signature of $\Lambda_{\rm trans}$ gives a finer classification of the singularity type. Suppose a type II limit realizes a degeneration of the form~\eqref{eq:tyurin}, inducing a polarization on $Z$ with lattice signature~\eqref{eq:sgnLambdapol}. This singularity then corresponds to a type II$_{b}$ degeneration, with $b=19-\rho$ in the notation of~\cite{Grimm:2018cpv}. In addition to the perturbative heterotic gauge group, the compactification of Type IIB string theory on $V$ can contain additional $U(1)$ factors that do not become weakly coupled in the limit $t\to i \infty$. According to the proposal in~\cite{Monnee:2025msf}, these correspond to spacetime-filling NS5-branes on the heterotic side which, if wrapped on the $T^2$ factor, yield additional $U(1)$ gauge factors for the effective 4d $\cN=2$ theory. 

The degeneration of the Calabi--Yau threefold on the Type IIB side thus determines the rank of the perturbative heterotic gauge group and the number of spacetime-filling NS5-branes wrapped on the $T^2$ factor of the heterotic model. However, it does not determine its exact gauge theory data. This should not come as a surprise since the degeneration of $V$ on the Type IIB side is intimately connected to the vector multiplet moduli space data of the 4d $\cN=2$ theory. On the heterotic side, the gauge data is instead associated with the hypermultiplet sector since, for example, the heterotic bundle moduli correspond to scalars in the hypermultiplets of the 4d $\cN=2$ effective theory. To actually construct the heterotic dual to type II loci in Type IIB vector multiplet moduli spaces, we hence have to also consider the hypermultiplet sector of both theories.

\subsection{One- and Two-Parameter Models}\label{sec:one-two_parameters_models}
The study of Type IIB--heterotic duality is particularly interesting for type II$_0$ loci in the complex structure moduli space of Calabi--Yau threefolds. For these, the perturbative heterotic gauge group arising from the left-moving sector is completely broken, leaving only the right-moving $U(1)^2$. Moreover, the heterotic torus moduli are expected to be fixed to special points in their moduli space~\cite{Hassfeld:2025uoy}, such that the heterotic theory has no 6d $\cN=(1,0)$ supergravity origin. Understanding the Type II--heterotic duality for these models in more detail is one of the goals of this work. 

The details of the heterotic dual construction associated with type II$_0$ loci depend crucially on the chosen threefold. In this work, we exclusively focus on one- and two-parameter Calabi--Yau threefolds. The reason is that on the Type IIB side the vector multiplet moduli space is well-known, since the full periods can be computed efficiently and explicitly in these low-dimensional moduli spaces. Importantly, the type II$_0$ loci, that are the main interest of this work, appear already in these cases with $n_V=1,2$. In this section, we give some basic details about the Calabi--Yau threefolds for which we construct the heterotic dual in the next sections. 
\subsubsection{One-Parameter Models}
For simplicity we focus on a particularly well-studied class of one-parameter Calabi--Yau threefolds for which the complex structure moduli space is equivalent to three-punctured sphere $\cM_{\rm c.s.}=\mathbb{P}^1\backslash\{0,1,\infty\}$. The Picard--Fuchs operator for these geometries is given by 
\begin{equation}
    L = \theta^4 - \mu z \prod_{i=1}^4 (\theta + a_i)\,,
\end{equation}
where $z$ is the coordinate on $\mathbb{P}^1\backslash\{0,1,\infty\}$ with logarithmic derivative $\theta = z \frac{d}{d z}$ and $\mu$ is a constant. The various Calabi--Yau threefolds with $\cM_{\rm c.s.}=\mathbb{P}^1\backslash\{0,1,\infty\}$ have been classified in~\cite{Doran:2005gu} and are distinguished by the values of the vector $(a_1,a_2,a_3,a_4)$. In total there are 14 such Calabi--Yau threefolds which are referred to as \emph{hypergeometric} Calabi--Yaus, since the solutions to the Picard--Fuchs equations are hypergeometric functions. For each of these Calabi--Yau threefolds there are three special points in $\cM_{\rm c.s.}$: the LCS point at $z=0$, a conifold point at $z=1/\mu$ and a third singularity at $z=\infty$ whose exact nature depends on the coefficients $(a_1,\dots, a_4)$. The third singularity can either be an orbifold singularity, another conifold point, a second LCS point or a K-point. In this work, we are interested in the last type of singularity since it corresponds to a type II$_0$ degeneration at $z=\infty$. There are three hypergeometric Calabi--Yau threefolds with K-points, which are commonly referred to as $X_{3,3}$, $X_{4,4}$ and $X_{6,6}$ and are characterized by~\cite{Doran:2005gu,vandeHeisteeg:2024lsa}
\begin{equation}
   (a_1, a_2, a_3, a_4) = \left\{\begin{matrix}
       (\frac13, \frac13, \frac23, \frac23) & \text{for}& X_{3,3} \\ &&\\ 
       (\frac14, \frac14, \frac34, \frac34) & \text{for}& X_{4,4} \\ &&\\
       (\frac16, \frac16, \frac56, \frac56) & \text{for}& X_{6,6} 
   \end{matrix}\right.\,. 
\end{equation}
The solutions at $z=0$ can be used to determine an integer basis of 3-cycles on the respective threefolds which, via mirror symmetry, is mapped to a basis of D$(2p)$-branes wrapping holomorphic cycles on the mirror threefold $\tilde{X}_3$. Under mirror symmetry, the point $z=0$ is mapped to the large-volume regime of $\tilde{X}_3$. From the solution of the Picard-Fuchs equations around $z=0$, the topological data of $\tilde{X}_3$ can be read off. Of particular interest for us is the Euler characteristic $\chi(\tilde{X}_3)=-\chi(X_3)$ since it determines the number of hypermultiplets, given by 
\begin{equation}\label{eq:number_hypers_K-points}
    n_H = h^{1,1}(X_3) + 1 =\frac12 \chi(X_3) +2 = \left\{\begin{matrix}
          74 & \text{for}& X_{3,3} \;\text{and} \;X_{4,4} \\
       62 & \text{for}& X_{6,6} 
          \end{matrix}\right.\,,
\end{equation}
where we used $\chi(X_3)= 2(h^{1,1}-h^{2,1})$ and $h^{2,1}(X_3)=1$. 

For the analysis in this paper, the models featuring a K-point are the most relevant. In particular, for later reference, we review some of their key aspects, focusing on their monodromy matrices. For later convenience, it is useful to introduce a new coordinate $s$, such that the K-point is located at $s=0$, and the monodromy matrix can be computed by encircling the singularity as $s\rightarrow e^{2\pi i}s$. The details about this coordinate redefinition, as well as the structure of the periods in the vicinity of this point and the couplings of the effective field theory derived from it have been analysed in detail in~\cite{Bastian:2023shf}. These models are characterized by quasi-unipotent monodromy matrices for which, given $n,d\in\bbZ$, we have:
\begin{equation}
    \left(T^n-\mathrm{Id}\right)^{d+1}=0,
\end{equation}
where $T$ represents the monodromy matrix and $\mathrm{Id}$ is the identity matrix. Moreover, $T$ can be decomposed into two commuting matrices as $T=T_{ss}T_u$, where the  semi-simple and unipotent parts are denoted by $T_{ss}$ and $T_u$, respectively, satisfying\footnote{As in~\cite{Bastian:2023shf}, the order is defined through the projectivization of $Sp(4,\bbZ)$. Indeed, as we will see in Section~\ref{ssec:hetDualII0}, these semi-simple factors correspond to stabilizer subgroups of $PSL(2,\bbZ)$ on the heterotic side. Although charges of BPS particles organize into $SL(2,\bbZ)$ orbits, the moduli space is instead sensitive to its projectivization.}
\begin{equation}
    (T_{ss})^n=\pm{\rm{Id}}\,,\quad (T_u)^n=e^N\,, \label{eq:monodromy_matrices_K-point_general}
\end{equation}
where we introduced the log-monodromy matrix $N$. In particular, $T\in Sp(4,\bbZ)$ is an integer valued matrix whereas in general $T_{ss}$ and $T_u$ can be rational valued, i.e. $T_{ss},T_u\in Sp(4,\bbQ)$. Moreover, $T_u$ and $T_{ss}$ are symplectic with respect to the pairing 
\begin{equation}
   \Sigma=\begin{pmatrix}0&\rm{Id}\\-\rm{Id}&0\end{pmatrix}\, .
\end{equation}
To describe instead the moduli space in the vicinity of a the K-points, it is more convenient to consider a finite cover of the moduli space. This is achieved via a redefinition of the coordinates of the form $s\mapsto s^n$ in such a way that the resulting monodromy matrix is unipotent. In this universal cover, the information about $T_{ss}$ is encoded in the periods due to the presence of a $\bbZ_n$ symmetry under $s\mapsto e^{2\pi i/n}s$. In all examples considered in this paper, this symmetry acts multiplicatively on the periods.

In order to obtain an explicit form of the monodromy matrices, consider the form of the periods near the K-point. As shown in~\cite{Bastian:2023shf}, it is simpler to obtain the asymptotic form of the periods in an integral symplectic basis via yet another coordinate redefinition. In this way, the periods can be expanded as $\Pi=\Pi_{(0)}+e^{-2\pi\text{Im}(t)}\Pi_{(1)}$, where $t$ is the coordinate in the new covering of the singularity, such that the K-point is at $t=i\infty$. The expression of the 0th order term of these periods is given by 
\begin{equation}\label{eq:asympt_exp_periods}
    \Pi_{(0)}=\begin{pmatrix}
        1\\\tau\\(\a+\b\tau)t+\delta+\gamma\tau\\(\b+\k\tau)t+\gamma
    \end{pmatrix},
\end{equation}
with $\tau=\frac{-\b+i\sqrt{\a\k-\b^2}}{\k}$, where $\a,\b,\k\in\bbZ$ and $\delta,\gamma\in\bbQ$ are parameters that define the Calabi--Yau threefold. In particular, we see that the K-point models are characterized by the fact that there is no vanishing cycle near the singularity. This is in contrast to a conifold singularity, for which we would have a cycle shrinking to zero volume near the singular point. From the explicit form of the periods, we can compute the log-monodromy matrix (by encircling the singularity at infinity via the transformation $t\mapsto t+1$), which takes the form:
\begin{equation}
    N=\left(
\begin{array}{cccc}
 0 & 0 & 0 & 0 \\
 0 & 0 & 0 & 0 \\
 \a & \b & 0 & 0 \\
 \b & \k & 0 & 0 \\
\end{array}
\right),
\end{equation}
and it is completely determined in terms of the data of the 3-cycles near the K-point. For the three models introduced above, the geometric data are given in Table~\ref{table:geometric_data_K-point}.
\begin{savenotes}
\begin{table}[t]
    \centering
    \renewcommand{\arraystretch}{2}
    \begin{tabular}{|c|c|c|c|c|}\hline
         & $\tau=\frac{-\b+i\sqrt{\a\k-\b^2}}{\k}$ & $(\a,\b,\k)$ & $\left(\delta,\gamma\right)$\footnote{These parameter are called extension data; they do not appear in the K\"ahler potential, but enter the periods as constant shifts, mixing the periods components as in~\eqref{eq:asympt_exp_periods}. In particular, their value is not unique, but can be determined up to some basis transformations; see~\cite{Bastian:2023shf} for an extended discussion.}&Order $n$\\ \hline
        $X_{3,3}$ & $e^{2\pi i/3}=-\frac12+i\frac{\sqrt3}2$ & $(2,1,2)$ & $\left(\frac23,\frac16\right)$&3 \\\hline
        $X_{4,4}$ & $e^{\pi i/2}=i$ & $(1,0,1)$ & $\left(\frac12,0\right)$&2 \\\hline
        $X_{6,6}$ & $e^{\pi i/3}=\frac12+i\frac{\sqrt3}2$ & $(2,-1,2)$ & $\left(\frac23,-\frac16\right)$&3\\\hline
    \end{tabular}
    \caption{Geometric data characterizing each K-point model.}
    \label{table:geometric_data_K-point}
\end{table}
\end{savenotes}

With this preparation, we can write down the explicit form of the monodromy matrices for the three hypergeometric Calabi--Yau threefolds introduced above. Here we give the expression for the full monodromy matrices (while the log-monodromy matrices can be recovered with the data in Table~\ref{table:geometric_data_K-point}) which take the form\footnote{Notice that the semi-simple part of the monodromies acts multiplicatively on the period~\eqref{eq:asympt_exp_periods}; in particular it does not act as $t\to t+1$. This is compatible with the fact that they correspond to automorphism of the limiting mixed Hodge structure~\cite{Bastian:2023shf}.}
\begin{equation}\label{eq:monodromy_matrices_K-points_explicit}\begin{aligned}
    &T_{3,3}=\left(
\begin{array}{cccc}
 0 & 1 & 0 & 0 \\
 -1 & -1 & 0 & 0 \\
 0 & 1 & -1 & 1 \\
 0 & 0 & -1 & 0 \\
\end{array}
\right)\,,\qquad     T_{4,4}=\left(
\begin{array}{cccc}
 0 & 1 & 0 & 0 \\
 -1 & 0 & 0 & 0 \\
 0 & 1 & 0 & 1 \\
 0 & 0 & -1 & 0 \\
\end{array}
\right)\,,\\    &\hspace{3.5cm}T_{6,6}=\left(
\begin{array}{cccc}
 0 & 1 & 0 & 0 \\
 -1 & 1 & 0 & 0 \\
 0 & 1 & 1 & 1 \\
 0 & 0 & -1 & 0 \\
\end{array}
\right).
\end{aligned}\end{equation}
Together with the monodromies around the conifold and the large complex structure point, these form the duality group $\Gamma \subset {\rm Sp}(4,\mathbb{Z})$ of the vector multiplet sector of the 4d $\cN=2$ effective theory of gravity. As the moduli space is the three-punctured sphere, we only have to specify two of the three monodromies to determine the full modular group. We take this second monodromy to be the monodromy around the conifold point which in the models of interest in this work takes the general form\footnote{This universal form of the conifold monodromy is recovered after an integral change of basis from the K-point basis that we use in this paper to the LCS/conifold basis~\cite{Bastian:2023shf}. The change of basis is such that the symplectic pairing used at the conifold is brought to the standard symplectic pairing $\Sigma$ we use here~\cite{Bastian:2023shf}.}
\begin{equation}
     T_{\rm coni}=\left(
\begin{array}{cccc}
 1 & 0 & 0 & 0 \\
 0 & 1 & 0 & 0 \\
 0 & 0 & 1 & 0 \\
 -1 & 0 & 0 & 1 \\
\end{array}\right)
\,. 
\end{equation}
As part of the discussion of the dual heterotic model we recover the duality group from the heterotic data. 

\subsubsection{Two-Parameter Model: \texorpdfstring{$X_{12}$}{X12}}\label{sec:p11226}
Apart from the one-parameter models discussed above, we also consider the heterotic dual to type II$_0$ loci in two-parameter models. A well-studied example of a two-parameter Calabi--Yau threefold for which the complex structure moduli space features a type II$_0$ locus is given by the degree-12 hypersurface in the weighted projective space $\mathbb{P}_{1,1,2,2,6}/(\mathbb{Z}_6^2\times \mathbb{Z}_2)$; this model will be denoted $X_{12}$ in the following, and was first discussed in~\cite{Candelas:1993dm, Kachru:1995fv} and more recently in~\cite{Lee:2019oct,Bastian:2021eom,Monnee:2025msf,Castellano:2026bnx,Hattab:2026lho}. The Calabi--Yau hypersurface in the weighted projective space is given by the zero locus of the polynomial 
\begin{equation}
    P = z_1^{12} + z_2^{12} +z_3^{6} + z_4^6 + z_5^2 -12 \psi z_1z_2 z_3 z_4 z_5 -2\phi z_1^6 z_2^6\,.
\end{equation}
Modding out the ambient space by the group $G=\mathbb{Z}_6^2\times \mathbb{Z}_2$ yields orbifold singularities such that, as it stands, the hypersurface $\{P=0\}\subset \mathbb{P}_{1,1,2,2,6}/(\mathbb{Z}_6^2\times \mathbb{Z}_2)$ is singular. The orbifold singularities can be resolved by blow-ups leading to a smooth Calabi--Yau threefold with $(h^{1,1},h^{2,1})=(128,2) $. In the following, we refer to this Calabi--Yau threefold as $X_{12}$. For the effective 4d $\cN=2$ supergravity action obtained by compactifying Type IIB string theory on the threefold, the dimensions of the vector multiplet moduli space $\cM_{\rm V}$ and the hypermultiplet moduli space $\cM_{\rm H}$ are respectively given by 
\begin{equation}\label{eq:numb_modulii_2_parameter}
    \text{dim}_{\mathbb{C}}(\cM_{\rm V})= 2 \,,\qquad \text{dim}_{\mathbb{H}}(\cM_{\rm H})=129\,, 
\end{equation}
where $\dim_\bbC$ and $\dim_\bbH$ denote respectively their complex and quaternionic dimensions. The two-dimensional vector multiplet moduli space is identified with the complex structure moduli space spanned by the parameters $(\psi, \phi)$. The limit $\phi\to \infty$ corresponds to a type II degeneration in which the Calabi--Yau threefold splits into the union of two threefolds
\begin{equation}
    X_{12} \to X_{12,0} = \{z_1^6=0\}\cup \{z_2^6=0\}\,,
\end{equation}
where the action of the discrete group $G=\mathbb{Z}_6^2 \times \mathbb{Z}_2$ ensures that both components are fully reduced. In terms of these coordinates, the singular divisors of $\cM_{\rm c.s.}(X_{12})$ are not normal crossing. To obtain normal-crossing divisors one introduces the coordinates $(\epsilon, u)$ such that 
\begin{equation}
    \phi = -\frac{1}{\epsilon} \,,\qquad \psi=-\frac{1}{\sqrt{3}} (2^5 \epsilon (1+\epsilon u))^{-1/6}\,. 
\end{equation}
In this way, one obtains the normal crossing divisors $\Delta_0=\{x_0=0\}$ and $\Delta_1 = \{x_1=0\}$ with $x_0= \epsilon u$ and $x_1 = u^{-2}$. The periods can be derived from the Picard-Fuchs operator~\cite{Lee:2019oct}, and in an integral\footnote{In this work, we refer to any basis related to the LCS basis by an integral (possibly non-symplectic) change of basis as \emph{integral}, owning to the basis vectors belonging to the integral charge lattice.} basis as a function of $(x_0,x_1)$ they read
\begin{equation}\label{eq:periods_two_parameter}
    \Pi = \begin{pmatrix}
        X^0\\X^1\\X^2 \\F_0 \\F_1 \\F_2
    \end{pmatrix} = \begin{pmatrix}
         (1 + \beta^2 x_0) \\
       i(1 - \beta^2 x_0) \\
        -2{\beta} \sqrt{x_0} \\
        \frac{\log(x_0^2 x_1) + \beta^2 (\log(x_0^2 x_1) - 4) x_0}{2\pi i}\\
        \frac{\log(x_0^2 x_1) - \beta^2 (\log(x_0^2 x_1) - 4) x_0}{2\pi} \\
        \frac{\beta}{\pi i} \log(x_1) \sqrt{x_0} + \frac{4\beta}{i\pi} \sqrt{x_0} 
    \end{pmatrix}\, ,
\end{equation}
with $\beta$ some constant data.\footnote{As noted in~\cite{Castellano:2026bnx}, the periods in~\cite{Monnee:2025msf} do not posses integral monodromies, and as such are not expressed in a suitable integral basis. One can perform a change of coordinates $M=\diag(1,1,2,4,4,2)$ to get to the integral basis of this paper. Moreover, we rescaled the periods by $\pi$ with respect to~\cite{Monnee:2025msf}. One can check that the periods in the main text are related to those of~\cite{Castellano:2026bnx} instead by
\begin{equation}
\renewcommand{\arraystretch}{0.85}
     M= \begin{pmatrix}
0 & 0 & 0 & 0 & 0 & 1 \\
0 & 2 & 0 & 0 & 0 & 0 \\
2 & 0 & 0 & 0 & 0 & -1 \\
0 & 0 & 2 & -1 & 0 & 0 \\
0 & 0 & 0 & 0 & -1 & 0 \\
0 & 0 & 0 & -1 & 0 & 0
\end{pmatrix}\, .
\end{equation}
\label{footnote:changeofbasis} 
Notice that, since $\det M = -8$, the transformation is integral but not symplectic.}
A general point along the divisor $\Delta_0$ corresponds to $\epsilon\to 0$ at finite $u$ which is hence a certain double-scaling limit for $\phi \simeq \psi^{1/6} \to \infty$. The singularities along the divisors $\Delta_0$ and $\Delta_1$ are of type II$_0$ and II$_1$, respectively. The monodromy around them acting on the integral basis of periods above are respectively~\cite{Castellano:2026bnx}
\begin{equation}
    T_{\Delta_0}=\left(
\begin{array}{cccccc}
 1 & 0 & 0 & 0&0&0 \\
 0 & 1 & 0 & 0&0&0 \\
 0 & 0 & -1 & 0&0&0  \\
 2 & 0 & 0 & 1 & 0& 0  \\
 0&2& 0& 0& 1 &0\\
 0&0&0&0&0&-1
\end{array}\right) \ , \quad  T_{\Delta_1}=\left(
\begin{array}{cccccc}
 1 & 0 & 0 & 0&0&0 \\
 0 & 1 & 0 & 0&0&0 \\
 0 & 0 & 1 & 0&0&0  \\
 1 & 0 & 0 & 1 & 0& 0  \\
 0&1& 0& 0& 1 &0\\
 0&0&-1&0&0&1
\end{array}\right) \ .
\end{equation}
Notice that the monodromy around $\Delta_0$ can be split into a finite order $\bbZ_2$ monodromy $T_{\Delta_0,\bbZ_2}$, due to the $\sqrt{x_0}$ branch-cut, and a unipotent contribution $T_{\Delta_0,u}$ reading
\begin{equation}\label{eq:2-param-monodromy}
    T_{\Delta_0,u}=\left(
\begin{array}{cccccc}
 1 & 0 & 0 & 0&0&0 \\
 0 & 1 & 0 & 0&0&0 \\
 0 & 0 & 1 & 0&0&0  \\
 2 & 0 & 0 & 1 & 0& 0  \\
 0&2& 0& 0& 1 &0\\
 0&0&0&0&0&1
\end{array}\right) \ , \quad T_{\Delta_0,\bbZ_2}=\left(
\begin{array}{cccccc}
 1 & 0 & 0 & 0&0&0 \\
 0 & 1 & 0 & 0&0&0 \\
 0 & 0 & -1 & 0&0&0  \\
 0 & 0 & 0 & 1 & 0& 0  \\
 0&0& 0& 0& 1 &0\\
 0&0&0&0&0&-1
\end{array}\right)\,.
\end{equation}
As remarked in~\cite{Lee:2019oct}, the finite-order part is associated with the Weyl group of the underlying $SU(2)$ Seiberg--Witten rigid theory~\cite{Seiberg:1994rs} with periods $(X^2,F_2)$.\footnote{In particular, the SW monodromy matrix is given in the above conventions by $T_{\rm SW}=-T_{\Delta_0}^{-1}T_{\Delta_1}^{2}$ ~\cite{Lee:2019oct}.} For this reason, the symplectic pairing in this basis adapted to the field-theory sector is $\hat \Sigma=-2\Sigma$, compatibly with~\cite{Castellano:2026bnx}. Indeed, one can verify that the matrix associated to the integer change of basis (see  Footnote~\ref{footnote:changeofbasis}) obeys $M^T\Sigma M=-2\Sigma$. As a consequence, the states associated to $(X^I,F_I)$ are not electromagnetic-dual pairs, since their Dirac pairing is non-minimal in the chosen basis. Finally, let us stress that the above monodromies are associated with the shift $x_0 \to e^{2\pi i}x_0$. There could be other semi-simple parts of the monodromy that are associated to automorphisms of the limiting mixed Hodge structure (LMHS)~\cite{Schmid1973,Cattani1982,Cattani1986,Grimm:2018ohb,Grimm:2018cpv}\footnote{Although we will not make directly use of asymptotic Hodge theory in this work, the interested reader can consult~\cite{Monnee:2024gsq} for a review in the string theory context.} and act multiplicatively on the period. We will see how to derive these from the duality with the heterotic theory in Section~\ref{sec:twoparametermodelshet}.

\section{Testing Type IIB--Heterotic Duality}\label{sec:HetDualTypeII0}
\begin{figure}[!t]
    \centering
    \includegraphics[width=1\linewidth]{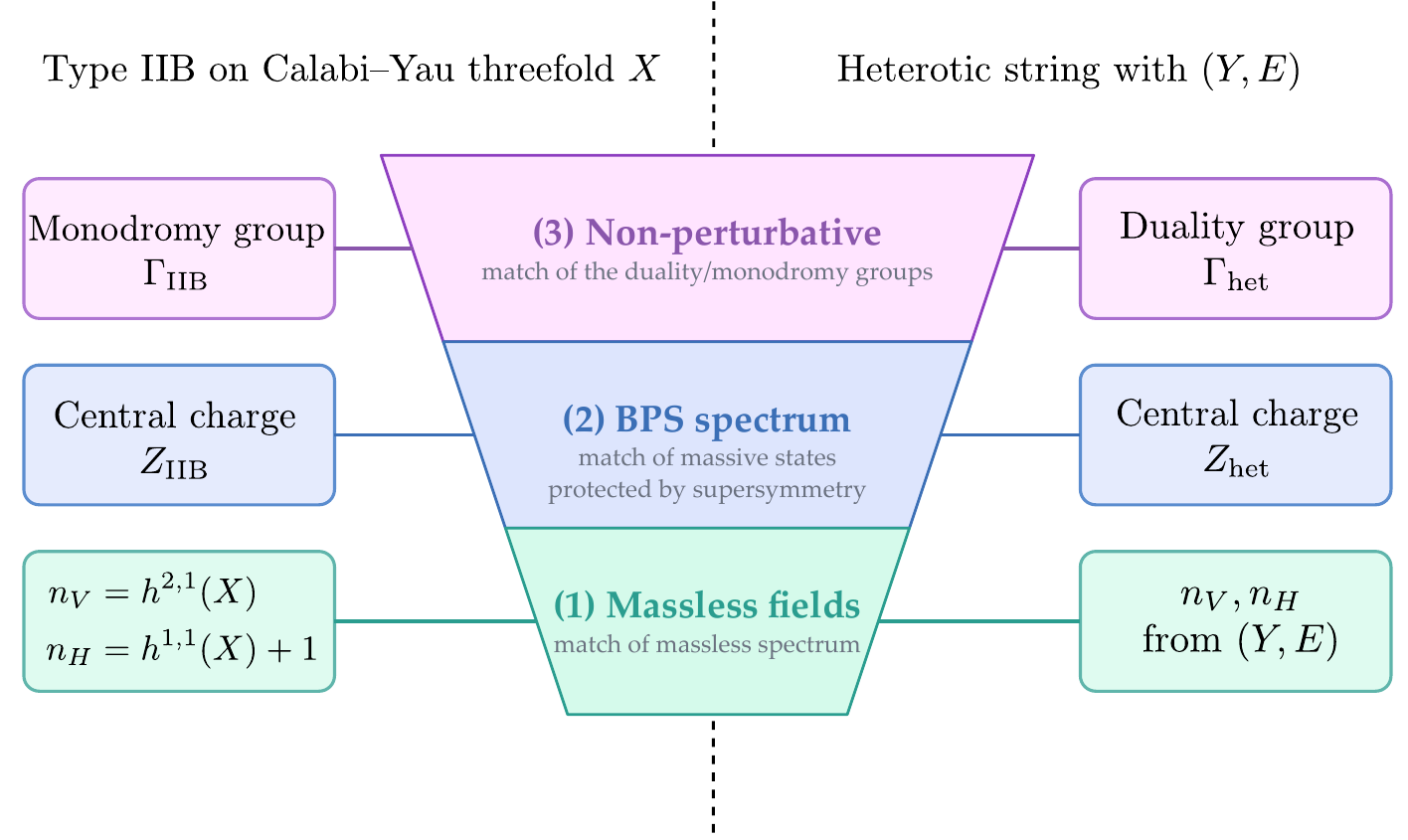}
    \caption{Depiction of the three levels of evidence for the duality between Type IIB on a Calabi--Yau threefold and the heterotic string on $Y$ with gauge theory data $E$. For the models of interest, we moreover either check directly or argue through duality with Type IIB that the heterotic worldsheet preserves $\cN=(0,4)$ supersymmetry and is anomaly-free.}
    \label{fig:level_duality}
\end{figure}

As reviewed in the previous section, by studying the geometry of Calabi--Yau degenerations realized at type II loci in the complex structure moduli space,~\cite{Hassfeld:2025uoy} identified a solitonic string solution to the 4d $\mathcal{N}=2$ supergravity equations of motion obtained from Type IIB compactified on a Calabi--Yau threefold. This string solution has two important properties:
\begin{itemize}
    \item[$\circ$] The string becomes tensionless if the vacuum expectation value for the scalar fields in the vector multiplets of the 4d $\cN=2$ effective action are tuned to the vicinity of the type II locus. 
    \item[$\circ$] The degrees of freedom of the worldsheet theory correspond to the degrees of freedom of a critical heterotic string with target-space sigma model K3$\times T^2$. 
\end{itemize}
The results of~\cite{Hassfeld:2025uoy} hence imply that there should exist a duality between Type IIB compactified on a Calabi--Yau threefold whose complex structure moduli space contains a type II locus and suitable supersymmetric configurations of the heterotic string. However, little is known about the heterotic dual, since its existence has been argued for without an explicit construction. In this work, we fill this gap by identifying the heterotic dual of a certain class of type II loci, demonstrating that the duality proposed in~\cite{Hassfeld:2025uoy,Monnee:2025msf} is indeed realized in string theory. Note that, since we lack a non-perturbative formulation of string theory, dualities are typically argued for at several levels rather than established by matching two full quantum theories (see Figure~\ref{fig:level_duality} for a schematic representation). 
\begin{enumerate}
    \item[$(1)$] The most basic requirement for two string constructions to be dual is that they lead to the same low-energy effective theory. Here, we consider supersymmetry-preserving compactifications of both Type IIB and heterotic string theory, such that the low-energy effective action is a 4d $\mathcal{N}=2$ supergravity theory. Apart from the supergravity multiplet, all massless fields then form vector multiplets or hypermultiplets, both of which contain scalar fields. At the massless level, the duality therefore requires the dimensions of the vector and hypermultiplet moduli spaces to agree. 
    \item[$(2)$] Whereas the massless spectrum can be compared at the supergravity level, the duality must extend to the massive states as well. Even though computing the full spectrum is out of reach for most string theories, supersymmetric theories allow for BPS states which can be reliably described in perturbative string theory. To test the duality beyond the massless level, we can hence match the BPS spectrum and the corresponding masses and central charges. The direct matching of the BPS states is in practice only possible if either we consider perturbative sectors or if non-renormalization theorems ensure the absence of $g_s$ corrections to their central charge.
    \item[$(3)$] The BPS spectrum is not directly sensitive to non-perturbative physics. Yet, the latter is crucial for extending the duality to the full quantum theory. The self-duality or modular groups of a theory offer a particularly tractable window into non-perturbative physics, since they map between the perturbative and non-perturbative sectors. Matching the modular groups of two string constructions therefore provides a non-perturbative test of the duality. 
\end{enumerate}

The goal of this chapter is to provide a general analysis of how the above three different levels of tests for Type IIB--heterotic duality are performed in practice. In particular, we focus on the specific class of type II$_0$ loci on the Type IIB side. For these examples, we show how to construct the heterotic dual theory based on the first requirement that their effective supergravity actions match. We then demonstrate how to perform the perturbative test for the duality by matching the perturbative BPS spectrum, and the third test by identifying the modular group from the heterotic string perspective and show that it matches the Type IIB results. 

\subsection{Matching the Low-Energy Spectrum}\label{sec:consist_cond_het_bundles}
As mentioned above, the basic requirement for identifying two string theories is that their low-energy effective actions agree. Since Type IIB compactified on a Calabi--Yau threefold yields a 4d $\mathcal{N}=2$ theory of supergravity in the low-energy approximation, we are interested in heterotic configurations that similarly reduce to 4d $\mathcal{N}=2$ theories in the IR. For a Type IIB compactification on a  Calabi--Yau threefold $X$, the massless spectrum is entirely determined by the topology of $X$. In particular, the theory contains $n_V=h^{2,1}(X)$ vector multiplets and $n_H=h^{1,1}(X)+1$ hypermultiplets.

In contrast, the counting of massless degrees of freedom on the heterotic side is more involved. As it is well-known~\cite{Hull:1985jv}, the worldsheet theory of the critical heterotic string in ten dimensions consists of a right-moving supersymmetric SCFT with central charge $c_R=15$ and a left-moving CFT with central charge $c_L=26$. Here, we are interested in heterotic string theories with four macroscopic dimensions and preserving eight supercharges. The $\cN=(0,4)$ supersymmetric worldsheet theory thus splits into a free part describing the extended directions and an interacting part, respectively 
\begin{equation}\label{eq:centralchargesY}
    (c_L, c_R) = (26,15)=(4,6) \oplus (22,9)\, .
\end{equation}
More generally, the worldsheet theory of a four-dimensional heterotic vacuum can be described by a gauged linear $\sigma$-model (GLSM) with $(0,2)$ supersymmetry~\cite{Witten:1993yc, Distler:1993mk}. In the simplest case, such a theory can also be obtained from the low-energy limit of the ten-dimensional heterotic string compactified on a six-dimensional manifold $Y$. To preserve 4d $\mathcal{N}=2$ supersymmetry, this manifold has to be a product of K3\footnote{For a review of the geometric and topological properties of K3, the interested reader can consult~\cite{Aspinwall:1996mn}.} and $T^2$.\footnote{More generally supersymmetry allows it to be a complex \emph{balanced} manifold. That is, conformally K\"ahler. These manifolds have been discussed in, e.g.,~\cite{Garcia-Fernandez:2016azr} in relation to heterotic construction with $H$-flux. In the mathematical literature, the resulting equations of motion are denoted Strominger--Hull systems, from the works~\cite{Strominger:1986uh, Hull:1986kz}.} Furthermore, the left-moving CFT of the heterotic string contains an internal part with $c_L=16$ describing the current algebra of the spacetime gauge group $G=E_8\times E_8$ (HE) or $G=SO(32)$ (HO) in the supersymmetric case.\footnote{In this work, our analysis is insensitive to the global form of the HO gauge group and we will conflate Spin(32)/$\bbZ_2$ with $SO(32)$.} Thus, unlike for compactifications of Type II string theory, in addition to $Y$ one has to specify a background for the 10d gauge theory. In terms of the CFT, this means that an interacting $(16,0)$ CFT has to be chosen. In the simplest case, the gauge background can be described by a vector bundle $V$ which is constrained by anomaly cancellation, supersymmetry condition and has to satisfy the hermitean Yang-Mills equations of motion.

Instead of compactifications of the ten-dimensional heterotic string, we can also consider genuinely lower-dimensional constructions. In this case, the interacting CFT with $(c_L,c_R)=(22,9)$ cannot be interpreted as, e.g., ten-dimensional supergravity on K3$\times T^2$ with a vector bundle. As we will see, the heterotic duals of all the Type IIB models introduced in Section~\ref{sec:type_II_degenerat} are of this type and correspond to either of the following constructions:
\begin{itemize}
    \item[$\circ$] Starting from the heterotic string in ten dimensions, we compactify on a torus whose volume is fixed to a string-sized value. This theory hence cannot be smoothly connected to the ten-dimensional heterotic theories by decompactifying the torus and therefore is a genuinely eight-dimensional gravitational theory. To get to a four-dimensional $\cN=2$ theory, this eight-dimensional theory has to be further compactified on a K3 with suitable gauge bundles. Importantly, in this case the volume of the K3 is a modulus, and the four-dimensional theory allows for a decompactification limit to eight dimensions. This is the case for the dual of the $X_{3,3}$ model discussed in Section~\ref{sec:x33}.
    \item[$\circ$] Instead of considering compactifications for which only the torus volume is fixed to string size, also the K3 volume can be fixed to string-sized values. In this case, there is no decompactification limit for the four-dimensional theory to eight dimensions. Moreover, since the K3 does not possess a large-volume regime, this construction cannot be viewed as a geometric compactification with vector bundles. One should instead use a direct worldsheet approach to describe the internal CFT. This is the case for the duals of the $X_{4,4}$, $X_{6,6}$ models discussed respectively in Section~\ref{sec:x44},~\ref{ssec:x66} and for the dual of $X_{12}$ discussed in Section~\ref{ssec:x12}.
\end{itemize}
As we review below, for K3 the stringy regime can still be reliably described using topological properties of the manifold. Hence, even without a geometric K3 background, we can still use the K3 language to describe the four-dimensional background. 

The heterotic string comes with a further complication, since, in addition to the perturbative states of the heterotic worldsheet, the background can include spacetime-filling NS5-branes wrapping holomorphic cycles in $Y$~\cite{Lukas:1998hk}. These contribute additional massless degrees of freedom to the effective action which have to be taken into account when matching the massless spectrum in the duality with Type IIB on a Calabi--Yau threefold. Suppose we have $n_5$ spacetime-filling NS5-branes wrapping curves $C_i$, $i=1, \dots, n_5$. Then each of these NS5-branes contributes a hypermultiplet to the four-dimensional effective action inherited from their six-dimensional worldvolume theory. The scalars in this hypermultiplet are identified as the coordinates transverse to $C_i$ inside $Y$. Furthermore, their presence also implies additional vector multiplets. For the NS5-branes in the HE theory, these arise from the tensor multiplet on the six-dimensional worldvolume theory. The resulting number of vector multiplets is
\begin{equation}
    n_{V}^{5,i} = h^{1,0}(C_i)\,. 
\end{equation}
Instead, for NS5-branes in the HO string, the vector multiplets in the 4d effective action arise from the dimensional reduction of the vector multiplets in the 6d worldvolume theory.

For a heterotic $E_8\times E_8$ compactification with NS5-branes, the number of vector and hypermultiplets is hence given by 
\begin{equation}\label{eq:nV_and_nH}
    n_V = n_V^{\rm pert} + \sum_i h^{1,0}(C_i) \,,\qquad n_H = n_H^{\rm pert} + n_5\,. 
\end{equation}
Here, the superscript ${}^{\rm pert}$ indicates the massless degrees of freedom associated with the perturbative heterotic string itself. 

To determine the number of massless degrees of freedom, we first need to specify a consistent background. In this section, we focus on geometric compactifications of the heterotic string. We instead defer the discussion of non-geometric compactifications in Section~\ref{sec:4dHetVac}. We take the background to be given by the pair $(X,V)$ with $X$ being a Calabi--Yau threefold or K3$\times T^2$ and $V$ a vector bundle. The condition that the resulting theory satisfies the equations of motion and is anomaly free constrains the possible vector bundles in the following way. 
\paragraph{Anomaly cancellation.}

Compactifying the heterotic string on a curved background such as a Calabi--Yau threefold induces gravitational anomalies. These can be cancelled by a background flux or spacetime-filling NS5-branes wrapping curves $C_i$. The anomaly cancellation condition for a geometric heterotic compactification is encoded in the Bianchi identity for the three-form $H$, which can be expressed as
\begin{equation}\label{eq:tadpole}
    \ch_2(T_X) =\ch_2(V) - \sum_{i=1}^{n_5}[C_i] \,,
\end{equation}
in terms of the Chern character of the tangent bundle $T_X$ and the vector bundle $V$. The presence of NS5-branes modifies the perturbative heterotic Bianchi identity and is central to the discussion of the heterotic dual of type II$_0$ loci in multi-parameter models in Section~\ref{sec:twoparametermodelshet}.

\paragraph{Stability and DUY.} 

For the pair $(X,V)$ to give a well defined heterotic vacuum, both the Einstein and the super Yang--Mills equations of motion have to be solved. For the Einstein equations, this is ensured if $X$ has special holonomy. For the bundle, the hermitian Yang--Mills (HYM) equations 
\begin{equation}
    J^{2} \wedge c_1(V)  = \mu(V) \mathrm{vol}(X) \mathrm{Id} 
\end{equation}
have to be satisfied. Here $\mathrm{vol}(X)$ is the volume form on $X$, $J$ its K\"ahler form and $c_1(V)$ is the first Chern class of $V$. At tree-level, the solutions to the hermitian Yang--Mills equations correspond to holomorphic, stable bundles. The stability condition on the bundle $V$ can then be recast in terms of the $\mu$-slope of the bundle, defined as 
\begin{equation}\label{eq:muslope}
    \mu(V) = \frac{1}{\text{vol}(X)\, \rk(V)}\int_X J^{n-1} \wedge c_1(V) \ .
\end{equation}
Indeed, it has been proven by Donaldson~\cite{donaldson} and Uhlenbeck-Yau~\cite{uhlenbeckyau} that the hermitian Yang-Mills equations have a solution if and only if the bundle is $\mu$-stable, i.e., if for any sub-bundle $V'$ with $\rk(V')<\rk(V)$, one has
\begin{equation}
    \mu(V') < \mu(V) \ .
\end{equation}
This condition can be relaxed to include equality, in which case the bundle is said to be semi-stable. Notice that the $\mu$-slope strongly depends on the K\"ahler form. The K\"ahler cone thus naturally splits in chambers of stability separated by walls~\cite{Sharpe:1998zu}: as the K\"ahler moduli vary, some stable bundles can become semi-stable, then unstable, and vice-versa. Besides stability, supersymmetry imposes that the slope vanishes
\begin{equation}\label{eq:muvalue}
    \mu(V) = 0\ ,
\end{equation}
which, from the EFT viewpoint, manifests as satisfying a D-term equation~\cite{Blumenhagen:2005ga}. The above integrability condition is denoted as the Donaldson-Uhlenbeck-Yau (DUY) condition.

\paragraph{Bundle spinors.} 

The above conditions ensure that the pair $(X,V)$ defines a good background for 10d heterotic supergravity, as spacetime anomalies are cancelled, the equations of motion are satisfied and some amount of supersymmetry is preserved. However, to obtain a consistent UV completion we have to ensure that the worldsheet theory is well-defined. A final consistency condition is therefore that the bundle admits spinors. At the level of the worldsheet, this ensures the absence of worldsheet anomalies~\cite{Witten:1985mj,Freed:1986zx}.\footnote{From the dual Type I perspective, this condition reflects instead a K-theory constraint on the non-BPS D7 brane~\cite{Uranga:2000xp}.} For the bundle $V$ to admit spinors, its second Stiefel--Whitney class has to vanish, which is ensured if the first Chern class of the bundle $V$ is even, i.e.
\begin{equation}
    c_1(V) \in H^2(X,2\bbZ) \ .
\end{equation}

\subsubsection{Massless Vector Multiplets}
Given the above constraints on the gauge sector, we can now discuss the counting of the perturbative vector and hypermultiplets. In a 4d $\mathcal{N}=2$ effective action the dimension of the former can be read off from the rank of the perturbative gauge group as $$n_V^{\rm pert}=\rk\, G_{\rm pert} -1\,,$$ where the $-1$ accounts for the graviphoton $U(1)$. Before turning on any gauge background, at a generic point in the moduli space, the perturbative gauge group of the heterotic string compactified on $Y$ is thus given by 
\begin{equation}
    G_{\rm pert} = G_E \times U(1)^{2 b_1(Y)} \stackrel{Y=K3\times T^2}{\longrightarrow } G_E \times U(1)^4
\end{equation}
Here, $G_E$ is the 10d gauge group of the HE or HO theories, and the $U(1)^4$ factor arises from the torus isometries. If we now turn on a non-Abelian vector bundle breaking $G_E \to G_0$, the number of perturbative vector multiplets is reduced to $n_V^{\rm pert}= \rk\,G_0 + 3$. At generic points in the torus moduli space, the number of perturbative vector multiplets is hence $n_V^{\rm pert}\geq 3$. However, the Type IIB models discussed in Section~\ref{sec:type_II_degenerat} have a small number of complex structure deformations, requiring $n_V<3$. These models can hence not be dual to heterotic string compactifications at generic points in the torus moduli space, as was already discussed in~\cite{Kachru:1995wm}. To this end, let us denote the complexified K\"ahler and complex structure parameters of the torus by $T,U$. The torus moduli space takes the simple form 
\begin{equation}
    \cM_{T,U} = \left(SL(2,\mathbb{Z}) \Big \backslash \frac{SL(2,\mathbb{R})}{U(1)} \right)^2\,.
\end{equation}
At special points in $\cM_{T,U}$, the perturbative gauge groups enhances. More precisely, we have the following possible gauge enhancements
\begin{equation}\label{eq:gauge_group_enhanc_torus}
\begin{split}
    T=U  &\implies  G=G_E \times SU(2) \times U(1)^3 \, , \\
    T=U=i &\implies G=G_E \times SU(2)^2 \times U(1)^2 \, , \\
    T=U=\{e^{2\pi i/3},e^{\pi i/3}\} &\implies G=G_E \times SU(3)\times U(1)^2 \ .
\end{split}
\end{equation}
In all cases, the $U(1)$ gauge factors that are enhanced to a non-Abelian factor arise as a linear combination of KK and winding momenta associated with the 1-cycles of $T^2$. From the worldsheet perspective, these spacetime $U(1)$s are associated with a purely left-moving algebra. For this reason, we will occasionally refer to the enhanced $U(1)$s at special loci in the $\cM_{T,U}$ as left-moving gauge group.

If we fix the torus moduli to one of the gauge enhancement loci, the perturbative gauge group can be further broken by turning on a gauge background for the additional non-Abelian gauge factors. To that end, we now consider vector bundles on K3 with structure group $G_V \subseteq G$, for which the Chern character is constrained by~\eqref{eq:tadpole} to be\footnote{In the following, we often abuse terminology by identifying Chern classes of top-dimension and their integrated counterparts.} $$-\ch_2(V)=-\ch_2(T_{K3})-n_5=24-n_5\,.$$ Decomposing the adjoint of $G$ into representations of $G_V$, each representation defines an associated vector bundle $U$ to which the Dirac operator couples. The massless spectrum is then determined by the dimensions of the cohomology groups valued in these bundles~\cite{Weigand:2006wyl}. 

Vector multiplets must transform in the adjoint of the unbroken gauge group, such that at a generic point in the vector multiplet moduli space the total number of massless vector multiplets~\eqref{eq:nV_and_nH} reads explicitly 
\begin{equation}
    n_V = \rk(G_V^\perp\subset G) + \sum_i h^{1,0}(C_i) -1 \,,
\end{equation}
where $G_V^\perp\subset G$ is the commutant of $G_V$ in $G$. As an aside, notice that here we describe heterotic compactifications to 4d on K3$\times T^2$ by means of an intermediate 8d description compactifying on the torus factor and further compactifying on K3 with bundles. In particular, to completely break the left-moving gauge theory, the bundles have to include part of the 8d gauge group coming from the reduction on $T^2$. The 10d perspective on these types of compactifications has been discussed in e.g.~\cite{Melnikov:2012cv,Melnikov:2014ywa}. This is reviewed for completeness in Appendix~\ref{app:A}, and holds \emph{mutatis mutandis} for the non-geometric constructions considered in Section~\ref{sec:4dHetVac}.

\subsubsection{Massless Hypermultiplets}
In addition to the vector multiplets, the theory contains massless hypermultiplets. Apart from the hypermultiplets associated with spacetime-filling NS5-branes discussed above, there are perturbative heterotic hypermultiplets of two origins~\cite{Louis:2011aa}. First, there are the degrees of freedom corresponding to the metric on K3. Indeed, a K3 has 20 metric moduli, such that the maximal number of hypermultiplets originating from the higher-dimensional metric is $n_{H, K3}=20$. Let us stress that for the genuinely four-dimensional constructions of the heterotic string considered in Section~\ref{sec:4dHetVac}, the number of the massless modes associated with the K3 deformation can be smaller than 20. 

Apart from the K3 moduli, the gauge bundle also contributes to the hypermultiplet sector. Decomposing the adjoint ${\rm ad}\,G$ into representations of $G_V \oplus G_V^{\perp}$, the number of hypermultiplets that are sections of the bundle $U$ associated to each of the representations is counted by
\begin{equation}
n_{H}^{\rm pert}(U) = h^1(X,U)\,. 
\end{equation}
In this context, the Hirzebruch-Riemann-Roch index theorem is an important tool that allows for the computation of the number of neutral hypermultiplets, which reads 
\begin{equation}\label{eq:indexthm}
    \chi(U) \equiv \sum_i (-1)^i \dim H^i(X,U) = \int_X \ch (U) \td(X) \ ,
\end{equation}
with $\td(X)$ the Todd class of $X$. Moreover, if the bundle is stable,\footnote{This also applies to the generalized notions of stability discussed in Section~\ref{sec:4dHetVac}.} then $H^0(X,\ad V) = H^2(X,\ad V)=1$, where the equality follows by Serre duality on K3 and self-conjugacy of the adjoint bundle~\cite{bridgeland2006stabilityconditionsk3surfaces}. 

To determine the dimension of the hypermultiplet moduli space, we have to focus on neutral hypermultiplets. First, there are the neutral hypermultiplets that transform in the adjoint of $G_V$. These are counted by $h^1(X, {\rm End}\,V)$. For bundles on K3, the number of such moduli can be computed just from the topological data of $V$. More generally, the gauge sector can be described by an object $E\in D^b({\rm Coh}(X))$, that is to say, in the bounded derived category of coherent sheaves on $X$~\cite{Distler:1996tj,Aspinwall:1998he}. This is the description we use in Section~\ref{sec:4dHetVac} for the genuinely 4d constructions. The relevant information of this object is encoded in its Mukai vector~\cite{mukai1987moduli,bridgeland2006stabilityconditionsk3surfaces}
\begin{equation}\label{def:mukaivector}
    \vec v = {\rm ch}(E) \sqrt{{\rm Td}(X)} =(r,c,s) \in  H^*(X,\bbZ)\, ,
\end{equation}
where $r$ is the rank of the the coherent sheaf, $c \in H^2(X,\bbZ)$ and $s \in H^4(X,\bbZ)$. In case a large volume for the surface $X$ is available and the coherent sheaf $E$ is given by a smooth rank-$r$ $U(n)$ vector bundle $V$,\footnote{That is, a coherent locally-free sheaf.} the Mukai vector can be expressed through the characteristic classes of $V$ as
\begin{equation}\label{eq:mukai_geom}
    \vec v(V)  = \left(r, c_1(V), r + \frac{c_1(V)^2}{2}-c_2(V)\right) \, .
\end{equation}
The number of neutral bundle moduli associated with the gauge theory sector is then simply given by
\begin{equation}\label{eq:mukai_dimension}
    h^1(X, {\rm End}\,V) = \frac12 \left(\vec v\,^2 +2 \right) = \frac12 \left[c^2 -2rs +2 \right]\, , 
\end{equation}
where $\vec v^2$ denotes the Mukai self-pairing.\footnote{The Mukai pairing is defined more generally as the intersection form 
\begin{equation}
    \vec v \cdot \vec v' = cc' - rs' - sr'\, ,
\end{equation}
in $H^*(X,\bbZ)$. In particular, the product $cc'$ is computed with the intersection form in $H^2(X,\bbZ)$.} 
For $U(n)$ bundles, we take $V$ to be defined by the fundamental representation. This correctly yields the dimension of the adjoint bundle through~\eqref{eq:mukai_dimension}. Indeed, in this case, the self-pairing of the Mukai vector is associated with the Euler characteristic as $\chi(U)=-\langle\vec{v}(U),\vec{v}(U)\rangle$ and, by exploiting~\eqref{eq:indexthm}, we recover the above identity.  For $E_8$ or $SO(32)$, the self-pairing of the Mukai vector does not yield the correct dimensions of moduli space, as the adjoint representation cannot be constructed from the fundamental as the traceless part of $V\otimes V$. For these cases one can simply use the known expression in terms of the dual Coxeter number $h(G)$~\cite{Klemm:1995tj}
\begin{equation}\label{eq:coxeterformula}
    \dim_{\bbH} \cM_{\ad G}= c_2(V) h(G) - \dim(G)  \ ,
\end{equation}
with $\dim(G)$ the dimension of the algebra and $c_2(V)$ the integrated second Chern class. 

In addition, there can be massless neutral hypermultiplet arising from states transforming under $U(1)$ factors in $G_V^\perp$ that become massive through a St\"uckelberg mechanism, as described in Section \ref{sec:4dHetVac}. These hypermultiplets are not associated with the adjoint representation of $G_V$, such that one has to compute their number by other means. This is relevant for some of the rank-2 gauge backgrounds considered in Section~\ref{sec:4dHetVac} and~\ref{sec:twoparametermodelshet}. In this case, we have to construct the corresponding bundles explicitly to compute the dimension of the relevant cohomology groups. For the bundles appearing in the later analysis, this can be achieved by exploiting the Heartshorne--Serre correspondence, reviewed in Appendix~\ref{app:B}, which allows us to construct $\mu$-stable bundles of rank two, which suffices for our purposes. Let us denote the resulting hypermultiplets that couple to the St\"uckelberg $U(1)$s by $n_{H,U(1)}^{\rm pert}$. The dimension of the hypermultiplet moduli space of a heterotic compactification is then given by 
\begin{equation}
    \dim_{\mathbb{H}}(\mathcal{M}_{\rm H}) = n_{\rm H, K3} +  h^1(X, {\rm End}\,V) + n_5 + n_{\rm H, U(1)}^{\rm pert} \,,
\end{equation}
where $n_{\rm H, K3}\leq 20$ and the other quantities are only indirectly constrained by the anomaly cancellation and bundle stability condition. 

With the above discussion at hand, we can hence construct heterotic configurations that satisfy the first requirement for the duality to the Type IIB configurations discussed in the previous section. In particular, we have to find consistent bundles that satisfy
\begin{equation}\begin{aligned}\label{identificationhodgenumbers}
    h^{1,1}(X_3) +1 &= n_{\rm H, K3} +  h^1(X, {\rm End}\,V)+ n_5 + n_{\rm H, U(1)}^{\rm pert}\,,\\
    h^{2,1}(X_3) &=\rk(G_V^\perp\subset G) + \sum_i h^{1,0}(C_i) -1 \,. 
\end{aligned}    
\end{equation}
As discussed in Section~\ref{sec:type_II_degenerat}, the results of~\cite{Hassfeld:2025uoy} imply that the rank of of $G_V^\perp$ can be read-off from the subindex of the type II$_b$ singularity on the Type IIB side as 
\begin{equation}
    \rk(G_V^\perp\subset G) = 2+b\,. 
\end{equation}
Identifying a heterotic configuration for which the dimension of the moduli spaces matches with the Hodge numbers of the Type IIB Calabi--Yau threefold tells us that the spectrum of the low-energy 4d $\mathcal{N}=2$ effective supergravity theories can be identified. We hence can use the first level of the duality test outlined above to identify candidates for string dualities that can then be further tested on the perturbative and non-perturbative quantum gravity level which are discussed below. \\

Even though the matching of the low-energy spectrum of the heterotic and the Type IIB theories focuses on the supergravity approximation of the respective string theories, let us stress that already at this level the existence of a UV completion, especially on the Type IIB side, is important. To appreciate this, we notice that the classification of asymptotic limits in the 4d $\cN=2$  vector multiplet moduli space using asymptotic Hodge structure does not rely on the UV completion of the supergravity theory~\cite{Compratore:2026rcn}. Similarly, the fact that the scaling of the couplings in the vector multiplet sector in the vicinity of a type II degeneration are consistent with the interpretation of the limit as an emergent string limit follows just from the properties of supergravity~\cite{Compratore:2026rcn}. Instead, as also stressed in~\cite{Hassfeld:2025uoy,Monnee:2025msf}, to conclude that the EFT string corresponds to a critical heterotic string requires the detailed geometry of the Calabi--Yau threefold, which is quantum gravitational information on the Type IIB side. Similarly, just from the supergravity perspective, a type II limit can arise in any 4d $\cN=2$ supergravity theory regardless of the number of hypermultiplets. The landscape of 4d $\cN=2$ supergravity theories with a Type IIB UV completion is more constraining, since the Calabi--Yau threefolds which have a type II locus in their complex structure moduli space do not have arbitrary $h^{1,1}$. In particular, not all hypergeometric Calabi--Yau threefolds have a K-point; for example, the mirror quintic with $(h^{2,1},h^{1,1})=(1,101)$ does not have a K-point, implying that there is no perturbative heterotic model with $n_H=102$ hypermultiplets. On the Type IIB side, the existence of a type II locus in a given complex structure moduli space is global information that is not directly accessible from the supergravity perspective but requires the full UV completion and hence constrains the possible heterotic dual constructions. 

It would be interesting to understand the exact constraints on a 4d $\cN=2$ supergravity theory to allow for a type II locus in the vector multiplet moduli space and hence for a heterotic or Type II dual description. In particular, this should manifest as a constraint on the possible number of vector and hypermultiplets. It would hence be similar to the constraint on a 4d $\cN=2$ supergravity theory to have a type III locus in the vector multiplet moduli space, which requires the number of hyper- and vector multiplet to be such that the gravitational anomaly in the emerging six-dimensional $\cN=(1,0)$ theory is cancelled,  
\begin{equation}
    (n_H-n_V) = 273-29n_T \,,
\end{equation}
for $n_H$, $n_V$ and $n_T$ the number of hyper-, vector and tensor multiplets of the 6d theory, respectively.

\subsection{Matching the Perturbative BPS Spectrum}\label{ssec:hetDualII0}
The second level of the duality test goes beyond the supergravity approximations and checks whether (part of) the massive spectrum agrees on both sides of the proposed duality. Although the full spectrum is in general tremendously difficult to compute, BPS states in 4d $\mathcal{N}=2$ theories are characterized by a central charge that can be reliably computed in perturbative string theory and matched on both sides of the duality. In what follows, we relate the the BPS spectra on the Type IIB side, associated with the light D3-brane spectrum, with the corresponding perturbative heterotic states by matching the charge lattice and their associated masses.

On the heterotic side, the accessible part of the BPS spectrum consists of the excitations of the heterotic string. At weak coupling, their masses follow from the standard mass formula, including those of the BPS states, for which the latter is associated with the central charge. Away from the weak coupling regime, the central charge receives perturbative and non-perturbative corrections in the heterotic string coupling. These are suppressed in the weak-coupling regime, which the proposed duality identifies with the type II locus in the Type IIB moduli space.

Instead, there are no BPS states in the perturbative spectrum of the Type IIB string. Here, the relevant BPS states arise instead from D3-branes wrapping special Lagrangian 3-cycles, whose central charge is simply given by their calibrated volume. Since for Type IIB Calabi--Yau threefold compactifications the string coupling is part of the hypermultiplet sector, the central charge of the D3-branes does not receive any $g_s$ corrections and is hence perturbatively exact. For this reason, the Type IIB D3-branes can be used for the second duality test which requires the perturbative BPS spectrum to match. 

In this regard, it was already noticed in~\cite{Hassfeld:2025uoy} that, along type II loci in the complex structure moduli space of Calabi--Yau threefolds, there exist certain special Lagrangian 3-cycles for which the associated volume vanish. More precisely, for a type II$_b$ locus there is a $b+2$-dimensional lattice of such 3-cycles. Due to their prominent role in this work and for simplicity of exposition, we focus in the following on type II$_0$ singularities for which this lattice is two-dimensional. By our previous discussion, this implies that the perturbative gauge group of the dual heterotic string to type II$_0$ loci is always broken to $U(1)^2$. Hence, when considering type II$_0$ loci in multi-parameter models with $h^{2,1}(X_3)>1$, the heterotic dual necessarily involves NS5-branes. 

\paragraph{Type IIB BPS states.} Let us start by studying the BPS states on the Type IIB side corresponding to D3-branes wrapping 3-cycles of the Calabi--Yau threefold. Their central charge is computed in terms of the period vector $\Pi$ as\footnote{Following~\cite{Hattab:2026lho}, the prefactor of $2\sqrt{2\pi}$ is fixed by the normalization of the graviphoton flux; however, we depart from their conventions by omitting the factor of $-i$ in the central charge, which is anyway irrelevant for the BPS masses.} 
\begin{equation}\label{eq:BPS_states_from_periods}
    Z_\mathbf{q}=2\sqrt{2\pi}e^{K/2}\left(\mathbf{q}\cdot\Sigma\cdot\Pi \right),
\end{equation}
where $K$ is the K\"ahler potential, $\mathbf{q}$ is the electromagnetic charge vector, and $\Sigma$ is the symplectic pairing.\footnote{We use conventions such that in a symplectic basis it is given by $\Sigma=\begin{pmatrix}0&\rm{Id}\\-\rm{Id}&0\end{pmatrix}$ as in Section~\ref{sec:type_II_degenerat}.} The mass of these BPS states is then given by
\begin{equation}
    m_{\mathbf{q}}=|Z_{\mathbf{q}}|.
\end{equation}
To restrict to the BPS states that become light near the K-point, we first recall that the Calabi--Yau threefold undergoes a stable degeneration limit near a type II$_0$ locus, as reviewed in Section~\ref{sec:het}. The degenerate variety can be brought into normal-crossing form corresponding to a collection of threefolds that intersect over a K3 surface $Z$ with transcendental lattice having signature $(2,0)$. As shown in~\cite{Hassfeld:2025uoy}, there exist a collection of 3-cycles $\Gamma \subset X_3$ that can be viewed as circle fibrations 
\begin{equation}
    S^1 \hookrightarrow \Gamma \to C \,, \quad C\in H_2(Z)\,,
\end{equation}
and where the $S^1$ fiber shrinks to zero size over $Z$ as the variety degenerates towards the singular locus. Here the curves $C$ span a two-dimensional lattice and all have non-negative self-intersection inside $Z$. These properties implies that the $\Gamma$ are special-Lagrangian, such that there exists a BPS state for each lattice site obtained by multi-wrapping a D3-brane on the corresponding cycle. This establishes the existence of a tower of BPS states becoming light near the type II$_0$ locus.

The exact mass of these states can be computed from the explicit form of the periods of the Calabi--Yau threefold $X_3$. Their asymptotic expansion near a type II$_0$ locus in one-parameter models was discussed around~\eqref{eq:asympt_exp_periods}. As explained above, for type II$_0$ loci in higher-dimensional moduli spaces, the main difference is the presence of NS5-branes in the heterotic dual description. Nonetheless, the periods associated with BPS excitations of the perturbative heterotic string take a similar form in both cases. We therefore focus on one-parameter models in what follows. In the vicinity of the type II$_0$ locus, the K\"ahler potential can be expanded as
\begin{equation}\label{Kahlerpotential}
    K=-\text{log}\left[K_{(0)}+e^{-2\pi\text{Im}(t)}K_{(1)}+\dots\right]\, .
\end{equation}
The leading order term in the singular limit $t \to i\infty$ is given by~\cite{Bastian:2023shf}
\begin{align}
    K_{(0)}=4|\k|(\tau_2)^2\,\text{Im}(t)\, .
\end{align}
The leading order expression for the central charge then takes the form:
\begin{equation}\begin{aligned}\label{eq:centr_charge_BPS_IIB}
    Z_{\mathbf{q}}=\frac{\sqrt{2\pi}}{\sqrt{|\k|(\tau_2)^2\,\Im(t)}}&\bigg\{p^1\left[(\a+\b\tau)\Im(t)+\delta+\gamma\,\tau\right] \\ &+p^2\left[(\b+\k\tau)\Im(t)+\gamma\right]-q_1-q_2\tau\bigg\},
\end{aligned}\end{equation}
where we used $\mathbf{q}\equiv\left(p^1,p^2,q_1,q_2\right)$, with $p^I$ the magnetic charges and $q_I$ the electric ones.

From~\eqref{eq:centr_charge_BPS_IIB}, we see that the only divergent terms in the expression are the ones proportional to ${\rm Im}(t)$. Using the fact that $\a\k-\b^2>0$ and $|2\b|\leq\a\leq\k$, in order for the complex structure parameter to be in the fundamental domain, the light BPS states are then characterized by $p^1=p^2=0$ and span a 2d charge sublattice given by 
\begin{equation}
\label{eq:lattice_BPS_IIB}
    \begin{pmatrix}
        \Re Z\\ \Im Z
    \end{pmatrix}_{\mathbf{q}=(0,0,q_1,q_2)}=-\frac{\sqrt{2\pi}}{\sqrt{|\k|(\tau_2)^2\,\text{Im}(t)}}\begin{pmatrix}
        1&\tau_1\\0&\tau_2
    \end{pmatrix}\begin{pmatrix}
        q_1\\q_2
    \end{pmatrix}.
\end{equation}
In particular, we see that their central charge and mass are defined only in terms of geometric data of the degenerating threefold. 

\paragraph{Perturbative heterotic BPS states.} The above expressions for the central charge and hence the mass of the BPS states becoming light at the type II$_0$ singularity can now be compared to the perturbative BPS states of the heterotic string. Given a compactification of the heterotic string on $T^2$, the left- and right-moving momentum are given by~\cite{Antoniadis:1995ct}
\begin{align}\label{eq:het_momenta}
    \boldsymbol{p}_R &= \frac{-i}{\sqrt{2\Im T\Im U}}\big(n_1U-n_2+w_1 T+w_2 TU \big) \,,  \\
    \boldsymbol{p}_L &= \frac{-i}{\sqrt{2\Im T\Im U}}\big(n_1 U-n_2+w_1 \bar{T}+w_2 \bar{T}U \big) \,.
\end{align}
Here $U,T$ are, respectively, the complex structure and the complexified K\"ahler structure of the torus, taking values in $(\bbH/{\rm SL}(2,\bbZ))^2$;\footnote{More precisely, in addition to the ${\rm SL}(2,\bbZ)$ transformations acting on the single parameters $U$ and $T$, there are two additional symmetries, namely mirror symmetry and spatial inversion, and the resulting moduli space of the torus is given by:
\begin{equation}
    \mathcal{T}^2=\left(\bbH_U/{\rm SL}(2,\bbZ)_U\times\bbH_T/{\rm SL}(2,\bbZ)_T\right)/(\bbZ_2^M\times\bbZ_2^I).
\end{equation}
\label{footnote:moduli_space_torus}}
instead, $n_1,n_2\in\bbZ$ are the momentum and $w_1,w_2\in\bbZ$ the winding numbers along the two circles. In particular, given a duality transformation acting on the complex structure of the torus as\footnote{This convention also induces an action on the period $\begin{pmatrix}
    1\\\tau
\end{pmatrix}$ of the form
\begin{equation}\label{eq:duality_transf_basis_1_U}
    {\rm SL}(2,\bbZ)_U\ni\begin{pmatrix}
        a&b\\c&d
    \end{pmatrix}:\begin{pmatrix}
    1\\\tau
\end{pmatrix}\mapsto\begin{pmatrix}
        a&b\\c&d
    \end{pmatrix}\begin{pmatrix}
    1\\\tau
\end{pmatrix}.
\end{equation}
We choose this convention in order to match with~\cite{Bastian:2023shf}.
}
\begin{equation}\label{eq:duality_transf_torus}
    \begin{pmatrix}
        a&b\\c&d
    \end{pmatrix}\in{\rm SL}(2,\bbZ)_U:\;U\,\longrightarrow\,U'=\frac{c+dU}{a+bU},
\end{equation}
this induces an action on the momentum and winding lattice of the form
\begin{equation}\label{eq:duality_transf_mom_latt_torus}
    \begin{pmatrix}
        n_1 \\ n_2 \\ w_1 \\ w_2
    \end{pmatrix}\,\xrightarrow[]{\;{\rm SL}(2,\bbZ)_U\;}\,\begin{pmatrix}
        d&-b&0&0\\-c&a&0&0\\0&0&a&c\\0&0&b&d
    \end{pmatrix}\begin{pmatrix}
         n_1 \\ n_2 \\ w_1 \\ w_2
    \end{pmatrix},
\end{equation}
and similarly for ${\rm SL}(2,\bbZ)_T$. In particular, these act as duality transformations on the momentum lattice and, in the case of ${\rm SL}(2,\bbZ)_U$, they do not mix the left- and right-moving components of the momentum. Let us stress a subtle but important difference between the action of the duality group on the torus moduli $U,\,T$ and the one on the momentum and winding lattice just presented. As is clear from \eqref{eq:duality_transf_torus}, the torus moduli are not affected by the action of $-{\rm Id}$, since the $SL(2,\bbZ)$ duality group acts in terms of a quotient on them, implying that the effective duality group is $SL(2,\bbZ)/\{\pm{\rm Id}\}=PSL(2,\bbZ)$. This can also be seen from the periods' perspective, since their duality group is, in principle, the full $SL(2,\bbZ)$. However, we always have the freedom to fix one of them to unit value, such that the effective action is again given by $PSL(2,\bbZ)$. Now, this is not true in the case of the charge lattice just presented, for which the effective duality group is the full $SL(2,\bbZ)$. This distinction plays a role in the following. Finally, notice that we chose $\boldsymbol{p}_{L,R}$ to be dimensionless, in such a way that they span the even uni-modular lattice $\Gamma^{2,2}$. With this convention, the mass-formula and level matching condition take the form\footnote{Since it is not relevant in the discussion, we ignore the extra left-moving momenta associated with the $E_8\times E_8$/$SO(32)$ lattice.}
\begin{equation}
    \frac{\alpha'}{4}M^2 = \frac12 p_R^2 + (\tilde N_R-1)= \frac{1}{2}p_L^2 +(N_L-1) \, ,
\end{equation}
where $N_L$ and $\tilde N_R$ are the oscillator numbers.\footnote{For notational convenience, we define the right-moving oscillator number $\tilde N_R$ to reabsorb zero-point energies in the R and NS sectors, namely $\tilde N_R=N_R$ in the R sector and $\tilde N_R=N_R-1/2$ in the NS one, with respect to e.g.~\cite[eq. (10.102)]{Blumenhagen:2013fgp}, such that the formula is symmetric with respect to the exchange $L\leftrightarrow R$.}

As we reviewed above, the surviving gauge group of the heterotic theory must be $U(1)^2_R$, and the BPS states are given in terms of the right moving momentum only. In particular, these states have to satisfy $\boldsymbol{p}_L=0$, together with 
\begin{equation}
    N_L-\tilde N_R=(n_1w_1+n_2w_2)\, ,\quad \tilde N_R=1\, .
\end{equation}
The mass of the BPS states then takes the form:
\begin{equation}\label{eq:MBPS_PR}
     M_{\rm BPS}=\sqrt{2} |\boldsymbol{p}_R|\, M_{\rm het} \, ,
\end{equation}
where $M_{\rm het}$ is defined as $M_{\rm het}^{-2}=\alpha'$.

As we show below, the relevant tori for our discussion are those characterized by $U=T$ at rational points in their moduli space. At these special loci $T$ and $U$ take the form\footnote{Notice that this is the same expression as the one considered around~\eqref{eq:asympt_exp_periods}. Indeed, the rational points are defined as solutions to:
\begin{equation}
    \k U^2+2\b U+\a=0,
\end{equation}for $\a,\b,\k\in\bbZ.$}
\begin{equation}
    U=T=\frac{-\b+i\sqrt{\a\k-\b^2}}{\k} \, , \quad \mathrm{with} \ \ \a,\b,\k\in\bbZ\, .
\end{equation} 
For the purposes of this work, we can further restrict ourselves to the subset $\alpha=\kappa$. By explicitly solving $\boldsymbol{p}_L=0$, we then get:
\begin{equation}\label{eq:2d_lattice_BPS_heter}
    \boldsymbol{p}_R
    =\begin{pmatrix}\Re p_R\\ \Im p_R\end{pmatrix}=\sqrt2\begin{pmatrix}
        1&U
    \end{pmatrix}\begin{pmatrix}
        n_1\\w_2
    \end{pmatrix} =\sqrt2\begin{pmatrix}
        1&U_1\\0&U_2
    \end{pmatrix}\begin{pmatrix}
        n_1\\w_2
    \end{pmatrix}\equiv L \cdot \boldsymbol{q}_R\, .
\end{equation}

\paragraph{Matching the two sides.} As we pointed out above, the two-dimensional lattice of states \eqref{eq:2d_lattice_BPS_heter} that become light in the weak-coupling limit of the heterotic theory should be identified with the corresponding light states \eqref{eq:lattice_BPS_IIB} near the type II$_0$ locus in the Type IIB vector multiplet moduli space. In order to show this matching, the masses of the heterotic string states and the masses of the wrapped D3-brane in Type IIB string theory have to be computed in the same units. In particular, notice that that the central charge in~\eqref{eq:centr_charge_BPS_IIB} gives the mass of the BPS states in 4d Planck units, whereas $\mathbf{p}_R$ determines the mass of the BPS excitations of the heterotic string via~\eqref{eq:MBPS_PR} in units of the heterotic string scale. To compare the mass of the states in the heterotic and the Type IIB duality frame, we then have to relate the Planck and the heterotic string scale. To do so, we recall that for an EFT string magnetically charged under an axion with saxion $\text{Im}\,t$, the tension $T$ is given by~\cite{Lanza:2021udy}
\begin{equation}\label{eq:tension}
    \frac{T}{M_{\rm pl}^2} =- \frac12 \frac{\partial K}{\partial \text{Im}\,t}\,. 
\end{equation}
However, for this relation to be correct, the axion $\text{Re}\,t$ must have the correct periodicity of $2\pi$. The periodicity of the axion differs from $2\pi$ if there is an orbifold singularity at the type II locus in the moduli space. Indeed, if $n$ is the order of the orbifold singularity, the periodicity of the axion is $2\pi/n$. For $n>1$, the elementary axion charge lattice is finer by the order $n$ of the type II locus. This means that the dual EFT string lattice is coarser, such that~\eqref{eq:tension} computes the tension of a string with effective charge $n$. Instead of working with the axions with non-standard periodicity, one could also work on the $n$-fold cover such that the axion has standard periodicity and also the elementary EFT string has unit charge. In any event, the tension computed by~\eqref{eq:tension} is $n$ times the tension of the fundamental heterotic string. By looking at the explicit form of the K\"ahler potential for type II boundaries given in~\eqref{Kahlerpotential}, we find
\begin{equation}\label{stringtension-n}
     \frac{T_{\rm het}}{M_{\rm pl}^2} = \frac{M_{\rm het}^2}{2\pi M_{\rm pl}^2}=\frac{1}{2n \Im(t)}\,.
\end{equation}
Plugging this into~\eqref{eq:MBPS_PR} we obtain 
\begin{equation}\label{eq:momwind_hetunits}
    \frac{M_{n_1, w_2}}{M_{\rm pl}} = \sqrt{\frac{2\pi}{n\Im(t)} }|\boldsymbol{p}_R|=\frac{2\sqrt\pi}{\sqrt {n \Im(t)}}|n_1+ w_2U|\, .
\end{equation}

As we have already pointed out, for the K-point models $X_{3,3}$, $X_{4,4}$ and $X_{6,6}$ the integer $n$ is given by the finite order of the monodromy matrix. Taking this into account, the lattice of the BPS states on the heterotic side clearly matches with the BPS states in the Type IIB description given in~\eqref{eq:lattice_BPS_IIB} once we identify the complex structure of the torus, $U$, with the the period of the K3 surface $Z$ at the type II degeneration, $\tau$, as $U=\tau$. Consequently, the heterotic momentum and winding charges $(n_1,w_2)$ are identified with the integral lattice of electric 3-cycles, spanned by $(q_0,q_1)$ in a canonical symplectic basis. Moreover, notice that the order of the monodromy, which at first sight appears to be accessible only from the Type IIB side of the duality, is available also in the heterotic theory. Indeed, the heterotic torus is fixed at special values in the moduli space determined by the periods of the degenerating K3 surface from Table~\ref{table:geometric_data_K-point}; more precisely, these points are the ones fixed under some specific duality transformations, as reported in Table~\ref{table:geometric_data_K-point_heterotic}. It turns out that the order of the monodromy around the K-point is precisely the order of the respective stabilizer in~$PSL(2,\bbZ)$. In fact, while the order of $(ST)^2$, the stabilizer of $X_{3,3}$, is 3 in both $SL(2,\bbZ)$ and $PSL(2,\bbZ)$, this is not true for the other two, for which the order in the latter are respectively 2 and 3, instead of 4 and 6, matching Table~\ref{table:geometric_data_K-point}. Finally, notice that, while the matching of the massless modes is not necessarily sensitive to the different enhancement points in the torus moduli space, the same does not hold for the matching of the perturbative and, as we will see in the next section, the non-perturbative BPS states.

To summarize, the BPS states corresponding to D3-branes wrapping 3-cycles of the three-fold are mapped to the BPS states in the right-moving momentum lattice of the heterotic string. Concretely, for the three hypergeometric models featuring a K-point presented above, we then have that the dual description is given as the heterotic string compactified over K3$\times T^2_{U=T=\tau}$~\cite{Hassfeld:2025uoy}, where the complex structure values are given in Table~\ref{table:geometric_data_K-point}.

\subsection{Non-Perturbative Test of the Duality and Monodromy Group}\label{ssec:nonperttest}
By the above discussion, the spectrum of light BPS states in the heterotic theory matches with the Type IIB's one. However, this part of the spectrum is only sensitive to the perturbative UV completion of the supergravity theory on the heterotic side. For the full quantum theories to be equivalent, we should also test the duality at the non-perturbative level, which is, of course, not possible in full generality, since the entire non-perturbative spectrum of the heterotic string is not even known in the BPS sector.\footnote{See \cite{Alvarez-Garcia:2024vnr} for recent novel aspects of non-perturbative effects in heterotic theories.}

Nonetheless, we know that there can be states corresponding to linear combinations of KK-monopoles and NS5-branes wrapping the K3 factor and a 1-cycle of the torus. Due to their non-perturbative nature, their central charge scales like~\cite{Ceresole:1995jg, Antoniadis:1995ct}
\begin{equation}\label{Znp}
    Z_{\rm n.p} \sim S \times Z_{\rm n.p.}^{(0)}\,,
\end{equation}
where we factored out the dependence of the complexified heterotic 4d dilaton $S$. To test the duality between Type IIB and the heterotic string at non-perturbative level, we now show that it is possible to re-derive the monodromy group of the Type IIB complex structure moduli space from the heterotic string. 

\paragraph{Unipotent monodromies.} The monodromy around the type II locus in the Type IIB moduli space maps on the heterotic side to a monodromy around the weak coupling locus, induced by sending $S\to S+1$.\footnote{This fact is well-known from the more-established Type II--heterotic dualities that have a large volume Type IIA realization.} 
Let the states obtained from KK-monopoles and NS5-branes have charges 
\begin{equation}\label{eq:non-pert-charges}
    \mathbf{q}_{\rm n.p} = \begin{pmatrix}
        q_{{\rm n.p}}^1\\q_{{\rm n.p.}}^2
    \end{pmatrix}\, .
\end{equation}
At the rational points in the torus moduli space, these charges form an integer lattice dual to the one spanned by the BPS right-moving perturbative states.\footnote{This fact can be understood as follows. Heterotic compactifications on K3 give rise to two sets of magnetically dual strings in six-dimensions: the perturbative heterotic string itself and NS5-branes wrapping the K3 factor. One set of (perturbative) BPS states is given by a linear combination of momentum and winding modes along a 1-cycle $\gamma$ of the torus, while the other (non-perturbative) set can be chosen to be given by NS5-branes/KK-monopole states along the dual magnetic cycle $\gamma'$.} Together with the perturbative charge vector $\boldsymbol{q}_R$, these non-perturbative states thus form a $Sp(4,\bbZ)$-vector of electric and magnetic charges in the four-dimensional theory. The central charges of these states are given by~\cite{Ceresole:1995jg}
\begin{equation}\label{eq:het_BPS_charge}
    Z_{\bf q} = 2\sqrt{2\pi}e^{K/2}\left(-q_{R,I} X^I+q_{\rm n.p.}^I S \eta_{IJ}X^J \right)\, .
\end{equation}
Comparing with~\eqref{Znp} we thus see that $Z_{\rm n.p.}^{(0)}$ is simply proportional to the central charge of the perturbative heterotic BPS states. Indeed, since the non-perturbative states come from KK-monopoles and NS5-branes wrapping the torus, their central charge has to be proportional to the central charge of the BPS states in the perturbative sector. In the above expression, the $X^I$ are symplectic sections and $\eta_{IJ}$ the intersection form on the magnetic lattice given by $\boldsymbol{q}_{\rm n.p}$. The total charge can be assembled in the symplectic vector $\boldsymbol{q}=(p^I,q_I)=(q_{\rm n.p}^I,q_{R,I})$. In a canonical symplectic basis, the lattice of perturbative and non-perturbative heterotic states is hence identified respectively with the integral lattice of electric and magnetic 3-cycles $H_3(X,\bbZ)$. At the elliptic points in the $T^2$ moduli space and in the basis of winding and momenta, the intersection form $\eta^{IJ}$ on the lattice of perturbative electric states is proportional to the Gram matrix\footnote{This can be understood as the fact that the intersection form in the basis given by $\boldsymbol{p}_R$ is simply the identity. Changing basis through~\eqref{eq:2d_lattice_BPS_heter} gets us to the intersection form on the electric lattice being given by $L^TL \equiv G$. }
\begin{equation}\label{eq:gram}
    \eta^{IJ} = \frac1n G^{IJ}\equiv \frac{1}{n}\frac{2T_2}{U_2} \begin{pmatrix}
        1& U_1\\U_1&|U|^2 
    \end{pmatrix}\,,
\end{equation}
where the presence of the orbifold order $n$ comes from switching to the finite covering to match the Type IIB description, following the discussion around \eqref{eq:tension}. This, in turn, determines which perturbative charges are picked up by $\mathbf{q}_{\rm n.p.}$ under the unipotent monodromy $S \to S+1$ around the weak coupling locus, in virtue of the former spanning the dual lattice. Notice indeed that defining the magnetic periods as $F_I = S\eta_{IJ}X^J$, with $\eta_{IJ}$ the inverse of $\eta^{IJ}$, the associated monodromy matrix acting on $(X^I,F_I)$ has the form
\begin{equation}\label{eq:inverse_Gram_matrix_monodromy}
    T_u^n = \begin{pmatrix}
        \mathrm{Id}_2 & 0\\  nG^{-1} & \mathrm{Id}_2
    \end{pmatrix}\,,
\end{equation}
where the subscript denotes the dimension of the identity matrices. Note that we identified the above monodromy with the $n$-th power of $T_u$ in order to account for passing to the $n$-fold cover of moduli space.
Moreover, the identification of $F_I$ above allows us to recover the precise expression of the periods from the heterotic perspective. 

\paragraph{Semi-simple monodromies.} The matrix $T_u$ captures the unipotent part of the monodromy matrix but is in general not an integer matrix. The reason is that the monodromy around $S=i\infty$ also has a semi-simple part, which was removed above by going to the finite cover of moduli space. The latter encodes that the perturbative charge lattice is itself invariant under certain duality transformations of the torus.
Concretely, since the torus moduli are fixed at a self-dual locus, the charge lattice of the perturbative heterotic string is invariant under the action of the corresponding stabilizer in $SL(2,\mathbb{Z})$ that leaves the respective point in moduli space invariant. The stabilizers are listed in Table~\ref{table:geometric_data_K-point_heterotic} in terms of the generators 
\begin{equation}\label{S_T_matrices_convention}
            S=\begin{pmatrix}
                0&1\\-1&0
            \end{pmatrix}:U\mapsto- U^{-1} \;\; {\rm and}\;\; T=\begin{pmatrix}
                1&0\\1&1
            \end{pmatrix}:U\mapsto U+1
\end{equation}
of the $SL(2,\mathbb{Z})$ duality group of the torus.  The semi-simple part of the monodromy on the heterotic side hence takes the form\footnote{The block diagonal entries of $T_{ss}$ are obtained from the generators of the lattices of BPS states. Indeed, the perturbative states are spanned by $X^I=\begin{pmatrix}
    1&U
\end{pmatrix}^T$, which transforms as~\eqref{eq:duality_transf_basis_1_U} $X\longrightarrow AX$ for $A\in SL(2,\bbZ)$. Via direct computation, instead, the magnetic dual periods associated with the non-perturbative charges are $\eta_{IJ}X^J\sim\begin{pmatrix}
    U&-1
\end{pmatrix}^T$ and transform as $\eta\,X\longrightarrow (A^{-1,T})\,\eta\,X$. It is possible to notice that these are strictly related to the diagonal blocks of the duality transformations defined in~\eqref{eq:duality_transf_mom_latt_torus}, where the diagonal blocks get exchanged.} 
\begin{equation}\label{eq:Tss_het}
    T_{ss} = \begin{pmatrix}
        a&b&0&0\\c&d&0&0 \\\multicolumn{2}{c}{\multirow{2}{*}{$H$}} & d&-c \\ && -b&a 
    \end{pmatrix}\,,
\end{equation}
where, as far as the action of the stabilizer group is concerned, the non-perturbative charge-lattice takes over the role of the left-moving sector due to the symplectic condition $T^T \Sigma T = \Sigma$ imposed on $T_{ss}$. The lower block-triangular form of $T_{ss}$ follows from the fact that the monodromy around the weak-coupling point $S=i \infty$ cannot induce a non-perturbative charge on states only charged under the perturbative heterotic gauge group. The lower-left $2 $-by-$2$ matrix $H$ is moreover required to make the full monodromy matrix $T= T_{ss}T_u$ integral and to ensure $T_{ss}^n=\mathrm{Id}_4$, where $n$ is the order of the stabilizer in Table~\ref{table:geometric_data_K-point_heterotic}.

The semi-simple part of the monodromy hence encodes additional information about the heterotic construction, namely the order of the stabilizer subgroup $G_{\rm stab}\subset PSL(2,\mathbb{Z})$ of the elliptic points at which the moduli of the torus are fixed. As we pointed out above, differently from the torus moduli, which give the order of the monodromy around the K-point as the order of the stabilizer in $PSL(2,\bbZ)$, the charge lattice is instead sensitive to the whole $SL(2,\bbZ)$. For this reason, the $2\times2$ blocks that appear on the diagonal of the semi-simple part are given by the stabilizers of the torus moduli in the full $SL(2,\bbZ)$ duality group.

\paragraph{Monodromies of non-gravitational sectors.} In the presence of spacetime-filling NS5-branes there are additional non-gravitational sectors which can have additional discrete stabilizer groups $G_{\rm n.p.}$. These can also contribute a semi-simple part to the monodromy that acts on the charges in the NS5-brane sector. The NS5-brane theory describes non-gravitational sectors which, in the simplest case of a single NS5-brane, is given by the 4d $\mathcal{N}=2$ super-Yang-Mills gauge theory with gauge group $SU(2)$. As explained in~\cite{Monnee:2025msf}, the gauge coupling of such a Seiberg--Witten theory decoupling at a Type II$_0$ locus is free to vary along $\Delta_0$, such that the stabilizer of non-gravitational sector for a generic coupling is in this case simply a $\mathbb{Z}_2\subset SL(2,\mathbb{Z})$ symmetry. Here the $SL(2,\mathbb{Z})$ symmetry acts on the charges of the non-gravitational sector and is not to be confused with the $SL(2,\mathbb{Z})$ action on the heterotic torus.
The non-gravitational sector is invariant under the monodromies associated with the perturbative sector.\footnote{Indeed, this has been formulated as a criterion for the consistent decoupling of field theory sectors in~\cite{Castellano:2026bnx}.} Conversely, this means that, when computing the string tension, the $n$ appearing in~\eqref{stringtension-n} is simply given by the order of the stabilizer of the torus appearing for the perturbative string. 
Putting all these ingredients together, the total monodromy around the $S=i\infty$ in the heterotic moduli space turns out to be
\begin{equation}
    T_S = \tilde T_u\, \tilde T_{ss} \,\tilde T_{\rm n.p.} \,,
\end{equation}
with each factor reading (after a possible reordering of the periods)
\begin{equation}\label{eq:monodromy_factorization}
    \tilde T_u = \begin{pmatrix}
        T_u &\mathbf{0}\\ \mathbf{0} & \mathrm{Id}_{2n_5} 
    \end{pmatrix}\,,\quad \tilde{T}_{ss} = \begin{pmatrix}
        T_{ss} &\mathbf{0}\\ \mathbf{0} & \mathrm{Id}_{2n_5} 
    \end{pmatrix}\,,\quad \tilde{T}_{\rm n.g.} = \begin{pmatrix}
        \mathrm{Id}_4 &\mathbf{0}\\ \mathbf{0} & T_{\rm n.g.} 
    \end{pmatrix}\,,
\end{equation}
and where $T_{\rm n.g.}$ is the stabilizer group acting on the non-gravitational sector. In terms of the order $n$, the log-monodromy around the type II locus is given by
\begin{equation}\label{eq:N_het}
N=\log (T_u^n) = \begin{pmatrix}
        \mathbf{0}_{2} & \mathbf{0}_2 \\  nG^{-1} & \mathbf{0}_2
    \end{pmatrix}\,,
\end{equation}
which is ensured to be an integral matrix. \\

\begin{table}[!t]
    \centering
    \renewcommand{\arraystretch}{2}
    \begin{tabular}{|c|c|c|c|}\hline
         & $U=\frac{-\b+i\sqrt{4\a\k-\b^2}}{2\k}$ & Gauge group & Stabilizer \\ \hline
        $X_{3,3}$ & $e^{2\pi i/3}=-\frac12+i\frac{\sqrt3}2$ & $SU(3)\times U(1)^2$ & {\renewcommand{\arraystretch}{1}$(ST)^2=\begin{pmatrix}
         0 & 1\\
         -1 & -1
        \end{pmatrix}$}\footnote{One could equivalently identify the stabilizer of $U=e^{2\pi i/3}$ to be $ST$. However, with the convention for $S$ and $T$ given in~\eqref{S_T_matrices_convention}, $(ST)^3=-1$ would lead to a stabilizer of order $6$. Indeed, even though $(ST)^3$ and $(ST)^6$ are identified as elements of $PSL(2,\bbZ)$, when we take their representation in the bigger duality group of the complex torus $O(2,2,\bbZ)$ the two differ, since the latter is a representation of $SL(2,\bbZ)$ (see Footnote~\ref{footnote:moduli_space_torus}). For this reason we are forced to consider the element $(ST)^2$ as generator of the stabilizer group; notice that, when taking into account this subtlety related with the topology of the duality group, this matrix then coincides with top left diagonal block of~\eqref{eq:monodromy_matrices_K-points_explicit}.} \\\hline
        $X_{4,4}$ & $e^{\pi i/2}=i$ & $SU(2)\times SU(2)\times U(1)^2$ & {\renewcommand{\arraystretch}{1}$S=\begin{pmatrix}
         0 & 1\\
         -1 & 0
        \end{pmatrix}$} \\\hline
        $X_{6,6}$ & $e^{\pi i/3}=\frac12+i\frac{\sqrt3}2$ & $SU(3)\times U(1)^2$ & {\renewcommand{\arraystretch}{1}$TS=\begin{pmatrix}
         0 & 1\\
         -1 & 1
         \end{pmatrix}$}\\\hline
    \end{tabular}
    \caption{Geometric data characterizing the heterotic dual of each K-point model.}
    \label{table:geometric_data_K-point_heterotic}
\end{table}

To summarize, we have given a prescription that allows us to compute the full monodromy matrix around the weak-coupling point in the heterotic moduli space and which encodes both perturbative and non-perturbative information. This can then be compared to the monodromy around the type II locus in the Type IIB moduli space and be used to test the proposed duality also at the non-perturbative level. 

\subsection{The Heterotic Dual of \texorpdfstring{$X_{3,3}$}{X33}}\label{sec:x33}
The above discussion gives a general three-step procedure to construct and test the heterotic theory dual to the type II$_0$ loci in the Type IIB moduli space. We now show how this can be implemented in a concrete setup. We start by considering the bicubic Calabi--Yau threefold $X_{3,3}$, which has also been the main example of the analysis in~\cite{Hattab:2025aok}.

\paragraph{(1) Matching of massless spectrum.} We first have to find a heterotic bundle that reproduces the number of vector multiplets ($n_V=1$) and hypermultiplets ($n_H=74$) of $X_{3,3}$. As explained above this requires to tune the torus moduli to one of the elliptic points in the $SL(2,\mathbb{Z})$ fundamental domain such that the 8d perturbative gauge group is either $G=G_E \times SU(3)\times U(1)^2$ or $G=G_E \times SU(2)\times SU(2) \times U(1)^2$. Either of these gauge groups has to be broken to $U(1)^2$ to match the massless field content of Type IIB on $X_{3,3}$. This is achieved by considering a bundle
\begin{equation}
    V = V_G \oplus V_{SU} \, ,
\end{equation}
with $V_G$ an $E_8 \times E_8$ or $SO(32)$ bundle and $V_{SU}$ an $SU(3)$ or $SU(2)^2$ bundle, both embedded in the corresponding factors of the gauge group. Furthermore, the model cannot contain any spacetime-filling NS5-branes. This choice of structure group ensures that $c_1(V)=0$, so that the bundle is ensured to be stable and of vanishing slope, hence leading to a supersymmetric gauge background.

The neutral hypermultiplets arising as bundle moduli of the adjoint $SO(32)$ (or $E_8\times E_8$) and $SU(3)$ representations can be computed using the index theorem~\eqref{eq:indexthm}, which yields \eqref{eq:coxeterformula}. We thus have to find a bundle that $i)$ breaks the left-moving gauge group entirely, $ii)$ exactly cancels the curvature induced contribution to the Bianchi identity and $iii)$ reproduces the correct number of bundle moduli $\text{dim}_\mathbb{H}\cM_{\rm bundle}=54$, in addition to the 20 geometric moduli of K3. It turns out that this is only possible for $G=G_E \times SU(3) \times U(1)^2$ with $G_E=SO(32)$. Following the discussion in Section~\ref{sec:consist_cond_het_bundles}, we choose 
\begin{equation}\label{eq:choiceX33}
    c_2(V_G) = 18  \quad {\rm and} \quad c_2(V_{SU}) = 6 \,,
\end{equation}
which cancels the anomaly~\eqref{eq:tadpole} and gives the required dimension of the hypermultiplet moduli space. Indeed, the first condition can be trivially checked, while using~\eqref{eq:indexthm} we get $\text{dim}_\mathbb{H} \, \cM_{V_{SU}} =10$ and $\text{dim}_\mathbb{H} \, \cM_{V_G} =44$ which, together with the 20 K3 moduli, match the dimension of the hypermultiplet moduli space of the Type IIB side.

This naive counting, however, still has to be supplemented by a concrete realization of the bundle. For instance, if $V_G$ on a geometric K3 were embedded into $E_8 \times E_8$ with instanton numbers $(9,9)$, the four-dimensional theory would retain an $SU(3)^2$ gauge symmetry. Indeed, as explained in~\cite{Duff:1996rs}, $G_E$ can not be completely Higgsed in this case even though the naive dimension $h^1(K3, V_G)$ is positive. This can also be seen from the Type IIA or F-theory dual, which is available for a geometric K3~\cite{Friedman:1997yq}. The same obstruction is present for $G_E=SO(32)$ with $c_2(V_G)=18$ for a bundle with vector structure, as the index theorem leaves an unbroken $SO(14)$ gauge group with no charged hypermultiplets available to Higgs it further~\cite{Berkooz:1996iz}. However, it is still possible to achieve the complete breaking of the left-moving gauge group via a bundle with this topological data by considering a less standard realization of the latter. To this end, we start by considering an initial rank-16 bundle $E$ with structure group $U(16)$ and Chern classes
\begin{equation}
    c_2(E) = 138 \, , \quad c_1(E)  = 8H \,.
\end{equation}
From this we define a bundle $V_G$ \emph{without vector structure}, following the procedure outlined in~\cite{Witten:1997bs, Berkooz:1996iz}. To do so, we formally twist $E$ by the inverse of a line bundle $\cL$ with $\ch(\cL) = e^{H/2}$, that is, with half-integral first Chern class. Although $\cL$ does not exist as a genuine line bundle, the twisted object $V_G = E \otimes \mathcal{L}^{-1}$ defines an honest principal bundle with structure group $G_V=SU(16)/\bbZ_2$. This twist removes the first Chern class as 
\begin{equation}\label{eq:bundleSo32}
    c_1(V_G) = 8H - 16 (H/2) = 0\, , \quad c_2(V_G) = 18  \quad \implies \quad v(V_G) = (16,0,-2)\, .
\end{equation}
At the same time it introduces a non-trivial obstruction $\tilde w_2 = H \,\mathrm{mod} \, 2 \ne 0$ to the existence of the vector representation. As a consequence, the index--theorem argument, that forces an unbroken $SO(32-k)$ for bundles with vector structure, no longer applies. Moreover, since $c_1(V_G) = 0$, the bundle automatically satisfies the classical DUY condition~\eqref{eq:muvalue}.
Decomposing the adjoint of $G_E$ into representations of the maximal subgroup $ SU(16) \times U(1)$, one has\footnote{In particular, the fundamental of $SO(32)$ decomposes into $SU(16)\times U(1)$ representations as ${\bf 32} = {\bf 16}_1 \oplus {\bf \overline{16}}_{-1}$, with charge $1$ on the $\bf 16$ and $-1$ on the $\overline{\bf 16}$. Using that the adjoint of $SO(32)$ is the antisymmetric square of the fundamental, the above decomposition follows.}  
\begin{equation}
    {\bf 496} \to {\bf 255}_0 + {\bf 1}_0 +{\bf 120}_{+2} + \overline{\bf 120}_{-2}\, ,
\end{equation}
where the subscript denotes the charge with respect to the commutant $G_V^\perp =U(1)$ of $SU(16)\times U(1)$ inside $SO(32)$. The $\mathbf{255}$ is the adjoint of $SU(16)$, associated with the bundle $\mathrm{End}\,V_G$. The singlet, associated with the trivial bundle $\cO$, corresponds to the $U(1)$ gauge field, which remains massless since $c_1(V_G) = 0$.\footnote{For more details on this point, see the discussion in Section \ref{sec:4dHetVac}.} The charged representations are associated instead with the antisymmetrized tensor product $\Lambda^2 V_G$ and its dual. Applying the index theorem, with $\ch(\Lambda^2V_G)\equiv (\ch_0,\ch_1,\ch_2) = (120,0,-252)$ one finally has
\begin{equation}
    \dim \cM_{\ad G_V} = \frac{1}{2}(v^2(V_G) +2) = 33\,, \quad \dim \cM_{\Lambda^2V_G \otimes\overline{\Lambda^2V_G} } =  -\chi(\Lambda^2 V_G) = 12 \, .
\end{equation}
Using one of the charged hypermultiplets to Higgs the remaining $U(1)$ vector multiplet, the total quaternionic dimension of the hypermultiplet moduli space for this bundle is
\begin{equation}
    \dim \cM_{V_G} = 33+12- 1 = 44 \, ,
\end{equation}
as required.
All in all, this choice of bundles completely breaks the non-abelian factors of the gauge group, leaving only a perturbative $U(1)^2$ at low-energy, thereby reproducing the dimension of the vector multiplet moduli space. Together with the hypermultiplet counting, this precisely matches the four-dimensional low-energy spectrum.

\paragraph{(2) Matching the perturbative BPS spectrum.}
We can now test the proposed duality between Type IIB on $X_{3,3}$ and the heterotic construction specified above at the level of the massive perturbative spectrum. We start by noticing that the K-point in $X_{3,3}$ has an order-3 semi-simple part, such that we have $n=3$ in~\eqref{eq:momwind_hetunits}. Furthermore, the boundary data of the hypergeometric models have been computed in~\cite{Bastian:2023shf} and are given in Table~\ref{table:geometric_data_K-point}; in particular, for the $X_{3,3}$ case we have
\begin{equation}
    (\alpha, \beta, \gamma, \delta, \kappa)= \left(2,1,\frac23,\frac16,2\right)\,. 
\end{equation}
Using these values and comparing~\eqref{eq:centr_charge_BPS_IIB} with~\eqref{eq:momwind_hetunits}, we hence find that the heterotic model is realized at $T=U=e^{2\pi i/3}$, consistently with the enhanced perturbative gauge group $G$ obtained above. Thus, for this choice of the torus moduli, the perturbative BPS spectrum of the heterotic compactification on K3$\times T^2$, with bundle data as specified in~\eqref{eq:choiceX33}, matches the one obtained on the Type II side. 

\paragraph{(3) Matching the monodromy group.} As we have mentioned above, the value of the complex structure of the torus on the heterotic side determines the semi-simple part of the monodromy matrix~\eqref{eq:Tss_het}, which for the heterotic model considered here takes the form 
\begin{equation}
    T_{ss} = 
\left(
\begin{array}{cccc}
 0 & 1 & 0 & 0 \\
 -1 & -1 & 0 & 0 \\
 p & q & -1 & 1 \\
 r & s & -1 & 0 \\
\end{array}
\right)\,,
\end{equation}
where $p,q,r,s \in \bbQ$ are parameters which are not fixed by the geometric data. In order to fix them, we have to consider the unipotent part arising from the non-perturbative mixing described in Section~\ref{ssec:nonperttest}. At the point $U=T=e^{2\pi i/3}$, the Gram matrix~\eqref{eq:gram} determining the log-monodromy $N$ in~\eqref{eq:N_het} is
\begin{equation}
    N=\log(T_u^3)=\begin{pmatrix}
        0 & 0&0&0 \\ 0&0&0&0\\  2&1 &  0&0\\ 1 & 2 &0&0
    \end{pmatrix}\,,
\end{equation}
where we used that the order of the semi-simple part is $n=3$. In this way, the full heterotic duality matrix $T_{\rm het}\equiv T_u T_{ss}$ takes the form:
\begin{equation}
    T_{\rm het} =\left(
\begin{array}{cccc}
 0 & 1 & 0 & 0 \\
 -1 & -1 & 0 & 0 \\
 p-\frac{1}{3} & q+\frac{1}{3} & -1 & 1 \\
 r-\frac{2}{3} & s-\frac{1}{3} & -1 & 0 \\
\end{array}
\right) \, .
\end{equation}
In particular, this has to satisfy $T_{\rm het} \in Sp(4,\bbZ)$, which, together with the condition $T_{ss}^3 = 1$ imposed by the order-3 orbifold, gives
\begin{equation}\label{eq:extension_data_het_side}
    q=r \,,\quad  s =-p+q\,, \quad p\in \frac13 + \bbZ\,, \quad  q\in \frac{2}{3}+\bbZ\,.
\end{equation}
As we can see, these two conditions fix the bottom-left block of $T_{\rm het}$ in~\eqref{eq:Tss_het} only up to integer shifts. This, however, is in agreement with the extension data $\delta,\,\gamma$ computed in~\cite{Bastian:2023shf} on the Type IIB side and summarized in~\eqref{table:geometric_data_K-point}, which characterize the semi-simple part of the monodromy matrix. Indeed, as extensively discussed in~\cite{Bastian:2023shf}, this extension data is also determined only up to changes of basis of the periods. In particular, we can see that by considering the minimal value for $p,\,q$ in~\eqref{eq:extension_data_het_side} we recover the correct value of the Type IIB description, in such a way that the monodromy matrices on the two sides match.

Thus, following the discussion in Section~\ref{ssec:nonperttest}, the action of the monodromy around the weak coupling point of the heterotic string matches the monodromy around the K-point in $X_{3,3}$. Together with the matching of the perturbative BPS spectrum, this confirms the proposed heterotic dual construction with bundle data given in~\eqref{eq:choiceX33} for the $X_{3,3}$ model. The above duality clearly shows that, indeed, the EFT string realizing the K-point limit in $X_{3,3}$ is a heterotic string with a stable gauge configuration flowing to a $\cN=(0,4)$ worldsheet SCFT in the IR. For the dual of the $X_{3,3}$, we identified a geometric model for which bundle stability, worldsheet anomaly cancellation and D-flatness can be straight-forwardly established. This model hence serves as a proof-of-principle that, also for type II loci away from the LCS point, the existence of a critical tensionless string indeed implies the existence of a dual heterotic theory that can be explicitly constructed. For the examples that we discuss in the next section, the supersymmetry condition and absence of worldsheet anomalies will be more intricate. In these cases, given that the proof-of-principle in this section shows that the EFT string associated with general type II limits has indeed a consistent heterotic realization, we will use the Type IIB realization of the heterotic string as a BPS EFT string in a UV complete theory to argue for worldsheet supersymmetry and anomaly cancellation.

\section{Genuinely 4d Type IIB--Heterotic Dualities}\label{sec:4dHetVac}
The bundles and constructions considered up to now relied on a large-volume, geometric description of the heterotic K3. In particular, we showed explicitly that the Type IIB EFT string realizing the K-point limit in $X_{3,3}$ corresponds to a heterotic string with a simple gauge bundle. This confirms that the worldsheet theory of an EFT string realizing a type II$_0$ limit indeed corresponds to a supersymmetric heterotic background. 

In this section, we turn to the heterotic duals of the K-points in $X_{4,4}$ and $X_{6,6}$. As it turns out, these constructions do not admit a large-volume description with gauge data defined by suitable bundles. Instead, the heterotic duals are realized at the Gepner point in the K3 moduli space, with the gauge sector being determined generically by an object $E$ in the derived category of coherent sheaves on K3, $E\in D^b({\rm Coh}(K3))$~\cite{Aspinwall:1998he, Douglas:2000ah}. For these models, the supersymmetry condition on the worldsheet, as well as the absence of anomalies, cannot be inferred from the large volume consistency conditions discussed in~\ref{sec:consist_cond_het_bundles}
\begin{equation}
    \int_{\rm K3} c_1(V)\wedge J= 0 \quad {\rm and}\quad c_1(V)\in H^2({\rm K3},2\mathbb{Z})\,. 
\end{equation}
Instead of deriving the Type IIB--heterotic duality from the worldsheet theory, here we reverse the logic: on the Type IIB side, the EFT string corresponds to a solution of the 4d $\mathcal{N}=2$ supergravity equations of motion, such that the worldsheet theory is manifestly supersymmetric and anomaly free.\footnote{The latter is due to the expectation that probe objects in a UV-complete description are free of anomalies (including potential bulk inflow).} In addition, the central charges of the worldsheet theory identify this string as a critical heterotic string, which is confirmed by the explicit proof-of-principle in the previous section. Indeed, suppose now we identify a stable heterotic background that passes the perturbative and non-perturbative tests outlined in Sections~\ref{ssec:hetDualII0} and~\ref{ssec:nonperttest} for the $X_{4,4}$ and $X_{6,6}$ models. This establishes this heterotic configuration as the the worldsheet theory on the Type IIB EFT string probing the K-point limit which, from the above discussion, is manifestly supersymmetric and anomaly free. In this way, we use the Type IIB--heterotic duality to identify new supersymmetric, anomaly-free heterotic worldsheet theories without having to directly check for supersymmetry and absence of anomalies in the heterotic duality frame.

\subsection{Non-Geometric Heterotic Backgrounds}\label{ssec:nonGeomComp}
From the worldsheet perspective, the heterotic models that we are interested in correspond to a CFT with $\mathcal{N}=(0,4)$ supersymmetry and left- and right-moving central charges $(c_L, c_R) = (26,15)$, as we discussed in a previous section.
In the following, it will be useful to split this CFT as a product of a free CFT$_f$ and an interacting CFT$_i$ with
\begin{equation}
    (c_L,c_R)_f = (4,9) \,\quad (c_L,c_R)_i = (22,6)\,.
\end{equation}
The free CFT$_f$ describes the motion in the four extended spacetime directions and the free right-moving scalars of the torus. Notice that this is different from the general threefold case in~\eqref{eq:centralchargesY} since for the K3$\times T^2$ background, the torus contributes to the free part of the right-moving SCFT. The choice of interacting CFT$_i$ corresponds to the K3 model with a non-trivial gauge background for the torus isometries and the 10d gauge group, as briefly described in Appendix \ref{app:A}.

We recall from the discussion in Section~\ref{sec:HetDualTypeII0} that the heterotic gauge sector is generically determined by an object in the derived category of coherent sheaves $E\in D^b({\rm Coh}({\rm K3}))$. This object is characterized by a choice of invariants encoded in the Mukai vector, as given in~\eqref{def:mukaivector}.
In case a large-volume regime for the K3 is available and the coherent sheaf $E$ is given by a smooth rank-$r$ $U(n)$ vector bundle $V$, the Mukai vector can be expressed through the characteristic classes of $V$, as in~\eqref{eq:mukai_geom}.
Even though we will be interested in more general objects describing the gauge sector, having in mind this geometric realization of the Mukai vector on K3 will be helpful. For the gauge theory sector to give rise to a good heterotic worldsheet theory, three conditions have to be ensured:
\begin{enumerate}
    \item[$\circ$] The gauge sector has to be described by a poly-stable object $E\in D^b({\rm Coh}({\rm K3})$,
    \item[$\circ$] The worldsheet theory has to flow to a CFT with $\mathcal{N}=(0,4)$ supersymmetry,
    \item[$\circ$] The worldsheet theory has to be anomaly free. 
\end{enumerate}
As we explained in Section~\ref{sec:HetDualTypeII0}, at large-volume, the first condition is satisfied if the gauge data is described by a vector bundle $V$ that is $\mu$-stable, i.e. for every sub-bundle $A\subset V$ 
\begin{equation}
    \mu(A)\leq \mu(V)\,,
\end{equation}
with the $\mu$-slope defined in~\eqref{eq:muslope}. Moreover, supersymmetry requires that the phase of the central charge of $V$ aligns with the heterotic background. Using the duality to Type I string theory~\cite{Polchinski:1995df}, one can show that for geometric compactifications of the heterotic string this phase vanishes, such that the supersymmetry requirement is satisfied if
\begin{equation}
    \mu(V)=0\,.
\end{equation}
As already mentioned, from the spacetime perspective this can be viewed as the vanishing of a D-term of the supersymmetric gauge theory. Classically the D-term is hence given by~\cite{Blumenhagen:2005ga}
\begin{equation}\label{eq:D-term}
    D= \mu(V)\,. 
\end{equation}
Finally, absence of anomalies requires $c_1(V)\in H^2({\rm K3},2\mathbb{Z})$ for a vector bundle in a geometric compactification. 

Away from the large volume regime, the conditions above receive corrections. In principle these could be both corrections in the string coupling and in $\alpha'$. However, here we are interested in 4d $\mathcal{N}=2$ configurations of the heterotic string for which the string coupling is part of the vector multiplet moduli space while the bundle data and the moduli of the K3 are, instead, associated with the hypermultiplet sector. Accordingly, the $g_s$ corrections to the $\mu$-stability condition as computed in~\cite{Blumenhagen:2005ga,Blumenhagen:2005pm}\footnote{For related references discussing loop-corrected D-terms in string theory, and in particular the generation of Fayet-Iliopolous terms in the effective supergravity theory, see~\cite{Atick:1987gy,Dine:1987xk,Dine:1987gj}. For an account of how these get modified by the presence of five-branes, see instead~\cite{Moore:2000fs}.} vanish identically on K3. On the other hand, there can be both perturbative and non-perturbative $\alpha'$ corrections to the consistency condition of the heterotic gauge sector away from the large volume limit; we will focus on these in the following.

\paragraph{Stability condition.} The classical stability condition is expressed in terms of the $\mu$-slope. At large volume, the $\mu$-slope is nothing but the phase of the central charge of the vector bundle $V$. This classical central charge is given by~\cite{Blumenhagen:2005pm}
\begin{equation}\label{Zclass}
    Z_{\rm cl.}(V) \equiv \int_{\rm K3} e^{iJ} \wedge {\rm ch}(V) \sqrt{{\rm Td}({\rm K3})} \,, 
\end{equation}
with phase 
\begin{equation}
    \pi(V) = \frac{1}{\rk(V)}\Im(Z_{\rm cl}(V))=\mu(V)\,.
\end{equation}
The classical central charge is expressed in terms of the classical K\"ahler form $J$ and hence depends on the classical volume of the K3 surface, which receives $\alpha'$ corrections away from the strict large volume regime. 

To compute these, it is easier to use the Type I dual of the $SO(32)$ heterotic string. We therefore recall that the heterotic $SO(32)$ string is dual to the Type I string once we identify the heterotic B-field with the $C_2$ field and we invert the string coupling. Since the string coupling is part of a vector multiplet, for the following discussion we simply fix it to a constant finite value. At large volume, we identify the central charge $Z_{\rm cl.}(V)$ with the central charge of a D9-brane with suitable induced lower-dimensional brane charge in Type I string theory. To compute this, we use that Type I is obtained from Type IIB string theory by modding out worldsheet parity. The central charge of a brane is then inherited from the Type IIB parent theory, where it is given by 
\begin{equation}\label{centralchargeIIB}
        Z_{\rm cl., IIB}(V) \equiv \int_{\rm K3} e^{iJ_{\rm IIB}} \wedge {\rm ch}(V) \sqrt{{\rm Td}({\rm K3})} \,, 
\end{equation}
where $J_{\rm IIB}$ is the complexified Type IIB K\"ahler form 
\begin{equation}
    J_{\rm IIB} = J - i B_2\,. 
\end{equation}
In Type IIB compactifications on K3, we can extend the notion of classical central charge beyond the large volume regime. To do so, we employ mirror symmetry, mapping Type IIB on a K3 surface $S$ to Type IIA on its mirror $\hat{S}$~\cite{Aspinwall:1994rg}. The central charge of $V$ can then be expressed in terms of the period vector $\widehat{\Pi}$ of $\hat{S}$ as 
\begin{equation}\label{eq:centralcharge}
    Z(E) = \widehat{\Pi} \cdot \mathrm{ch}(E) \sqrt{\mathrm{Td}({\rm K3})} \, .
\end{equation}
The period vector of $\hat{S}$ is expressed in terms of local coordinates $\psi_i$ on the complex structure moduli space of $\hat{S}$. The large complex structure point $\psi^i\to \infty$ of $\hat{S}$ is identified with the large volume point for $S$, and the K\"ahler moduli of $S$ are given by 
\begin{equation}\label{clmirrormap}
    z^i = \int_{C_i} B_2 + iJ \to \frac{i}{2\pi}\log \psi^i\,. 
\end{equation}
Here $\{C_i\}$ is a basis of holomorphic curves on $S$. Using this identification, the central charge in~\eqref{eq:centralcharge} reduces to its classical expression in~\eqref{centralchargeIIB} near the large complex structure point. However, the expression in~\eqref{eq:centralcharge} is valid beyond this regime and computes the central charge beyond the $\psi^i\to \infty$ limit. 

To get back to Type I/Heterotic SO(32) string theory, we have to implement the action of worldsheet parity on the Type IIB string theory, which effectively amounts to freezing the real part of the K\"ahler moduli $z^i$. Given the classical mirror map~\eqref{clmirrormap}, this restricts the complex structure parameters $\psi^i$ of $\hat{S}$ to 
\begin{equation}
    {\rm Type\;I:}\quad \psi^i\in \mathbb{R}\,.
\end{equation}
The central charge of the D9-brane in Type I string theory is hence given by 
\begin{equation}
    Z(E) = \widehat{\Pi}(\psi)|_{\psi \in \mathbb{R}}\cdot \mathrm{ch}(E) \sqrt{\mathrm{Td}({\rm K3})}
\end{equation}
In terms of this central charge, we can now analyse the stability condition. For D-branes, $\mu$-stability is replaced, away from the large volume regime, by the more general notion of Bridgeland stability, in particular $\Pi$-stability~\cite{Douglas:2000ah,Enger:2003ue,Aspinwall:2004mb}. For the central charge $Z(E)$, the $\Pi$-slope is defined in general as
\begin{equation}\label{phislope}
    \phi(E) = \frac{1}{\pi} \arg Z(E)\,. 
\end{equation}
For a semi-stable object $E\in D^b({\rm Coh}({\rm K3}))$ we thus require 
\begin{equation}
    \phi(A)\leq \phi(E)\,,
\end{equation}
for each sub-object $A\subset E$ that embeds into $E$ such that there exists a map $A\hookrightarrow E$. In general, testing for the stability of an object on a general manifold $X$ is rather difficult. However, here we are interested in objects supported on K3 for which the analysis simplifies considerably. Specifically, we are interested in Mukai vectors $v\in H^*({\rm K3},\mathbb{Z})$ and asking whether there exist stable objects for this choice of $v$. For K3, corollary 6.9 of~\cite{bayer2013projectivitybirationalgeometrybridgeland} ensures that for a primitive vector $v\in H^*({\rm K3},\mathbb{Z})$ with $v^2\geq -2$ the moduli space is non-empty, of dimension
\begin{equation}\label{dimv}
    \dim{v} = \frac12 (v^2 +2) = \frac12 (c^2-2rs +2 )\,,
\end{equation}
and consists of stable objects with respect to a generic stability condition.\footnote{If the object described by the Mukai vector is required to have structure, e.g., it has to preserve a quadratic form, the dimension of the moduli space for this object differs from \eqref{dimv}, see~\eqref{eq:coxeterformula} for the $E_8$ and $SO(32)$ gauge backgrounds.} We will use this notion of stability in the following.

Let us briefly comment on one point: the relation between the periods of $\hat{S}$ and the central charge of the object describing the gauge theory sector for the heterotic string, or Type I, compactified on $S$ makes use of K3 mirror symmetry; strictly speaking, the latter assumes a $(4,4)$ supersymmetry as opposed to the $(0,4)$ supersymmetric worldsheet theory of the heterotic string. However, here we are just interested in the stability condition for the heterotic vacuum and merely use mirror symmetry as a computational tool to evaluate the central charge; unlike the usual application of mirror symmetry for Type II compactifications, we do not identify the heterotic theory on $S$ with another theory on $\hat{S}$. Consequently, exploiting mirror symmetry to understand the stability condition of heterotic gauge bundles is similar to the application of mirror symmetry to understand the supersymmetry conditions of branes or flux superpotentials in Type II orientifolds. It would be interesting to see whether there is any issue arising with this notion of stability or whether it is indeed the correct generalization of $\mu$-stability in the non-geometric regime.\footnote{In the context of the heterotic theory, $\Pi$-stability has also been considered as a generalized stability condition in~\cite{He:2012iy}. For what concerns its application in the non-geometric regime, and in particular in the vicinity of the Landau-Ginzburg point, see~\cite{Walcher:2004tx} instead.}

\paragraph{Supersymmetry and worldsheet anomalies.} 
The existence of a stable object $E$ at a given point in the moduli space of the K3 surface $S$ does not yet ensure that the pair $(S,E)$ is a good heterotic configuration leading to a 4d $\mathcal{N}=2$ theory of supergravity. Indeed, we still have to ensure that the pair $(S,E)$ yields a $\mathcal{N}=(0,4)$ supersymmetric worldsheet theory that is free of anomalies. Let us stress that this is automatically satisfied for the dual of $X_{3,3}$ discussed in Section \ref{sec:x33}, as the K3 allows for a large-volume regime. 

Let us first discuss the supersymmetry condition. Given the classical D-term of the heterotic theory in~\eqref{eq:D-term}, one might be tempted to simply replace the $\mu$-slope in the D-term with the slope $\phi(E)$ in~\eqref{phislope}. However, even though the $\phi$-slope captures the (non-)perturbative corrections in $\alpha'$ to the central charge, there can be additional $\alpha'$ corrections to the heterotic D-term condition. Schematically, we have
\begin{equation}\label{eq:D-term-corrections}
    D =\underbrace{D_0}_{=\phi(E)} + \alpha' D_1 + \dots + \sum_{\mathbf{n}} \mathcal{D}_\mathbf{n} e^{-n_it^i/\alpha' }\,, 
\end{equation}
where $D_i$ are perturbative and $\mathcal{D}_{\mathbf{n}}$ non-perturbative corrections to the classical D-term. Notice that the corrections can be different for each factor of the gauge group. In the following, we are interested in the heterotic setups that are realized deep in the interior of the K3 moduli space, where these corrections can be sizeable such that there is no reason why $\phi(E)=0$ corresponds to the vanishing locus for the D-term.\footnote{In this work, we motivated the notion of stability through the duality with Type I. If one had access to the corrections in~\eqref{eq:D-term-corrections}, one could instead start from these to define a corrected slope, whose value would then vanish by construction on the supersymmetric loci. This is what is pursued in the context of $\cN=1$ perturbative corrections to the HYM equations in~\cite{Blumenhagen:2005ga,Blumenhagen:2005pm}.} In fact, using the duality to the Type IIB setups, we will argue that for the models that we consider the supersymmetry condition chooses a phase $\phi(E)\neq 0$. Let us stress that, even though in the interior of the K\"ahler moduli space of $S$ the supersymmetry condition differs from the classical one, for the K\"ahler deformations of $S$ to survive as moduli of the theory, the pair $(S,E)$ has to satisfy the ordinary zero $\mu$-slope condition at large volume. If this is not the case, the worldsheet theory cannot have a marginal deformation associated with the K\"ahler moduli of $S$. 

A similar analysis applies to the anomaly cancellation condition. If the heterotic worldsheet theory is realized as a non-linear sigma model on a geometric K3 surface $S$ with some vector bundle $V$, the worldsheet anomaly cancellation requires $c_1(V)\in H^2(S,2\mathbb{Z})$, as discussed above. In the interior of the K\"ahler moduli space of $S$ the stable object $E\in D^b({\rm Coh}(S))$ does not necessarily allow for a description in terms of a vector bundle. This means that the worldsheet anomalies are also not captured by the second entry of the Mukai vector mod 2.\footnote{For example, if the Type I dual to these vacua are given by gauge sectors without a vector structure~\cite{Witten:1997bs} and with discrete B-field, as it happens in Type I Gepner points~\cite{Bachas:2008jv}, then the evenness condition is shifted by multiples of $B$.} In the following, we will indeed consider models that have Mukai vectors $v$ with $c\in H^2_{\rm alg}(S, 2\mathbb{Z}+1)$, the Picard group of $S$. In this case, anomaly cancellation tells us that the stable object $E$ with Mukai vector $v$ cannot be a vector bundle. Instead, it has to have another description and the worldsheet theory must have an obstruction to deforming the K\"ahler structure to large volume, where for this choice of $v$ we would obtain an anomalous non-linear sigma model. 

In the 4d effective action, the K\"ahler modulus of $S$ is a scalar in a hypermultiplet. For this reason, if the K\"ahler deformation is obstructed, an entire hypermultiplet must be lifted from the massless spectrum. That this is indeed the case, can be seen already at the geometric level. Indeed, notice that for each real K\"ahler modulus $t^i={\rm Im}\,z^i$ that is fixed by the condition $\phi(E)=\phi_0$, its real part $b^i=\text{Re}\,z^i$ takes over the role of the axion in the St\"uckleberg mechanism for the determinant $U(1)$s associated with $c$.\footnote{This four-dimensional mechanism descends from the ten-dimensional Green--Schwarz anomaly cancellation~\cite{Ibanez:1986xy}, which grants these $U(1)$s the name of \emph{anomalous} $U(1)$s. For a similar treatment of the six-dimensional case, see~\cite{Honecker:2006dt}.} In this work, we consider models with a single $U(1)$ and a single K\"ahler modulus. In this case, the mass term for the St\"uckelberg $U(1)$ is simply given by~\cite{Blumenhagen:2005ga}
\begin{equation}\label{eq:massu1}
    M_{U(1)} = {c_1(E)}|_J \, ,
\end{equation}
where the subscript denotes the component along $J$. The masses of all fields removed from the effective theory end up being of the order of the string scale.

Finally, let us mention that, even though the supersymmetry condition and worldsheet anomaly cancellation are affected by quantum corrections, the spacetime anomaly-cancellation condition encoded in the Bianchi identity is not affected. This means that, if the gauge sector is described by multiple objects $E_i \in D^b({\rm Coh}(S))$ with Mukai vectors 
\begin{equation}
    v(E_i) = (r_i, c_i,s_i) \,,
\end{equation}
we still have 
\begin{equation}
    24 = \sum_i s_i +n_5\,. 
\end{equation}
The anomaly cancellation condition is hence still a topological information about the admissible Mukai charges.

\paragraph{Quartic K3 and Landau-Ginzburg models.} The heterotic models that we consider in these sections are characterized by pairs $(S\times T^2\,;E_1\oplus E_2)$ where $E_1$ describes the gauge theory data of the $SO(32)$ or $E_8 \times E_8$ gauge groups and $E_2$ the gauge theory data for the enhanced gauge symmetry coming from the torus isometries. As it turns out, the models that allow for a dual Type IIB description require Mukai vectors for $E_2$ of the form 
\begin{equation}
    v(E_2) = (2, c, s)\,, \quad \text{with} \quad 0<c^2 \in 4 \mathbb{Z} \;{\rm and}\; c \in H^2_{\rm alg}(S,2\mathbb{Z}+1)\,.  
\end{equation}
As mentioned above, since $c$ is odd, worldsheet anomaly cancellation then tells us that the heterotic models cannot be realized in the large-volume regime. This is consistent with the requirement of supersymmetry. In the large volume phase, the Mukai vector $v(E_2)$ would have $\mu$-slope
\begin{equation}
    \mu(E_2) = \frac{1}{2} \int_S J\wedge c\,. 
\end{equation}
Since $c^2 >0$, its Poincar\'e dual is a movable curve in $S$ and $\mu(E_2)$ is simply the volume of this curve. The classical D-term condition then requires the volume of this curve to vanish. This, in turn, requires the overall volume of $S$ to vanish, such that a supersymmetric locus can at best exist in a non-geometric phase of the theory. The theories that we consider in this section are hence realized at the Gepner/Landau-Ginzburg point of the surface $S$. Moreover, since the Mukai vector breaks supersymmetry and leads to an anomalous worldsheet theory at large volume, the resulting worldsheet theory cannot have a marginal deformation associated with the overall volume of $S$. 

In fact, we can give a simple realization of the surface $S$. Since $c^2\in 4\mathbb{Z}$, the surface $S$ can be described as the mirror quartic in $\mathbb{P}^4$. Here we briefly describe it following~\cite{Aspinwall:1996mn}. Recall that a K3 surface is a hyper-K\"ahler manifold that is equipped with three complex structures $(\omega_1,\omega_2,\omega_3)$. The hyper-K\"ahler metric on K3 has 20 deformation moduli associated with elements of $H^{1,1}({\rm K3},\mathbb{R})$. The three complex structures of K3 can split into the K\"ahler form $J=\omega_3$ and the holomorphic $(2,0)$ form $\Omega=\omega_1 + i \omega_2$. The choice of K\"ahler form is not unique since hyper-K\"ahler rotations can exchange $\Omega$ and $J$. The intersection pairing on a K3 surface is given by the lattice 
\begin{equation}
    \Gamma_{3,19} = U^{\oplus 3} \oplus (-\Gamma_{E_8}) \oplus (-\Gamma_{E_8}) \,,
\end{equation}
where $U$ is the hyperbolic plane and $\Gamma_{E_8}$ the root lattice of $E_8$. To distinguish K\"ahler and complex structure deformations, one considers polarized K3 surfaces. To that end, one defines the polarization or Picard lattice 
\begin{equation}
    \Lambda_{\rm pol} = H^2({\rm K3},\mathbb{Z}) \cap H^{1,1}({\rm K3})\,. 
\end{equation}
The K\"ahler form $J$ then takes values in $\Lambda_{\rm pol}\otimes \mathbb{R}$. The polarization lattice of a K3 surfaces has signature $(1,\rho-1)$ for $0\leq \rho \leq 20$, such that, for a polarized K3 surface, the number of K\"ahler moduli is given by $\rho$. The remaining moduli correspond to complex structure deformations that leave the polarization lattice invariant and are associated with the transcendental lattice $\Lambda_{\rm trans}$ with 
\begin{equation}
    {\rm sgn}\,\Lambda_{\rm trans} = (2,20-\rho)\,. 
\end{equation}
The quartic surface corresponds to a polarized K3 surface with Picard number $\rho=1$, for which the K\"ahler moduli space is hence one-dimensional. Concretely, we focus on the quartic K3 surface that can be described as a quartic hypersurface $S\subset \mathbb{P}^3$
\begin{equation}
    x_0^4 + x_1^4 + x_2^4 +x_3^4 -\frac{4}\psi x_0x_1x_2x_3 + {\rm 19\;additional\;monomials}= 0 \ .
\end{equation}
The only curve class in the Picard lattice is given by the hyperplane class $H$ which $S$ inherits from the ambient space $\mathbb{P}^4$. The hyperplane class has self-intersection $H\cdot H=4$ such that $c\in H^2_{\rm alg}(S,\mathbb{Z})$ can be taken to be a multiple of $H$. To compute $Z(E)$ as defined in~\eqref{eq:centralcharge}, we consider the mirror quartic $\hat{S}$ given by the hypersurface 
\begin{equation}\label{quarticequation} 
    \hat{S}=\{ x_0^4 + x_1^4 + x_2^4 +x_3^4 -\frac{4}{\psi} x_0x_1x_2x_3 =0 \}\subset \mathbb{P}^3/\mathbb{Z}_4^2\,,
\end{equation}
and compute its periods. The periods of $\hat{S}$ are defined as
\begin{equation}
 \Pi_{\hat{S}}^a(\psi) = \int_{\gamma_a} \Omega\,,
\end{equation}
for ${\gamma_a}$ a basis of primitive 2-cycles on $\hat{S}$ satisfying $\hat{J}|_{\gamma_a}=0$ for $\hat{J}$ the K\"ahler form on $\hat{S}$. The periods are solutions to the Picard-Fuchs equations that are derived from the hypersurface equations~\eqref{quarticequation} as in~\cite{Hartmann_2012}. At large complex structure, i.e.~$\psi\to 0$, the period vector $\mathbf{\Pi}(\psi)$ for the mirror quartic reads 
\begin{equation}
    \mathbf{\Pi}(\psi) = \begin{pmatrix} \Pi_0 \\ \Pi_1 \\\Pi_2\end{pmatrix}= \begin{pmatrix}
        1 \\ \frac{1}{2\pi i} \log \psi \\ -\frac{1}{2\pi} (\log \psi)^2 
    \end{pmatrix}+\mathcal{O}(\psi)\,,
\end{equation}
which translates into 
\begin{equation}
    z=\int_H (B_2 + iJ) = \frac{1}{2\pi i}\log \psi\,,
\end{equation}
for the complexified K\"ahler modulus of the quartic. Away from $\psi=0$, $\hat{S}$ encodes the quantum volume of $S$. The phase structure of the one-dimensional complex structure moduli space of $\hat{S}$ is very simple. Apart from the LCS singularity at $\psi=0$, the quartic has two additional singularities at $\psi^4=1$ and at $\psi=\infty$. The singularity at $\psi^4=1$ is a conifold singularity, which signals the transition from the geometric phase into a non-geometric phase. The second phase is the Landau-Ginzburg/Gepner one, associated with the finite order singularity at $\psi=\infty$. The periods at the Gepner point can be explicitly evaluated to be
\begin{equation}\label{LGperiods}
    \Pi(\psi) \stackrel{\psi\to \infty}{\rightarrow } C \psi^{-1/4} \begin{pmatrix}
        1 \\ \frac{1+i}{2} \\ i
    \end{pmatrix}\,,
\end{equation}
where the exact normalization $C$ is irrelevant for our purposes. We hence notice that, even though all periods vanish as $\psi\to \infty$, the mirror map identifies the quantum volume of the hyperplane class $H$ of the mirror $S$ as 
\begin{equation}
    z=\int_H (B_2 + iJ) =\frac{\Pi_1}{\Pi_0} = \frac{1+i}{2}\,,
\end{equation}
which is hence non-zero at the Gepner point. In particular, we also have to turn on a non-zero B-field at the Gepner point. Importantly, unlike what one might have expected classically, the string theory is perfectly smooth as $\psi\to\infty$. This is hence an example where the classical supergravity analysis does not capture correctly the stringy physics.

\subsection{The Heterotic Dual of \texorpdfstring{$X_{4,4}$}{X44}}\label{sec:x44}
We now turn to discuss the heterotic dual of Type IIB on $X_{4,4}$ which has the same number of vector multiplets and hypermultiplets as Type IIB on $X_{3,3}$, namely $n_V=1$, $n_H=74$. As in Section~\ref{sec:HetDualTypeII0}, we proceed in three steps to argue for the duality. We first propose a heterotic background that matches the massless spectrum of the Type IIB supergravity. Let us stress that, unlike for the heterotic background dual to the $X_{3,3}$ model, here we first establish that the gauge configuration is stable. We then show that the heterotic string with this background yields the perturbative BPS spectrum that matches the light D3-brane spectrum in the vicinity of the K-point in the moduli space of $X_{4,4}$, and that the non-perturbative heterotic information encoded in the duality group matches the Type IIB side. We thus collect very strong evidence that the stable configuration on the heterotic side indeed describes the worldsheet theory of the Type IIB EFT string realizing the K-point in $X_{4,4}$. Using that the worldsheet theory on the EFT string is anomaly-free and supersymmetric, we then infer that the heterotic configuration indeed leads to a supersymmetric and anomaly-free theory. 

\paragraph{(1) Matching of massless spectrum.} As Type IIB on $X_{4,4}$ has the same number of massless hypermultiplets and vector multiplets as Type IIB on $X_{3,3}$, it is natural to consider its heterotic dual to be realized on the other special locus on the $T^2$ moduli space, namely the $SU(2)^2$ enhancement locus. This will be later confirmed by the matching of the BPS spectrum and monodromies. There are, however, no $SU(2)^2$ bundles matching the number of moduli. Thus, as anticipated throughout this section, we have to turn to unitary bundles with non-trivial $c_1(V)$.\footnote{As we will show below via the perturbative and non-perturbative checks, the $SU(2)^2$ enhancement point is indeed the correct one for the heterotic dual of $X_{4,4}$. For this reason, we can not just embed the $SU(3)$ bundle~\eqref{eq:choiceX33} inside $SU(2)^2$, since it would lead to a residual $U(1)$ commuting group, which could not be further broken due to the trivial second entry of the Mukai vector $v(V_{SU})$.} Given the perturbative heterotic gauge group $G=G_E \times SU(2)^2 \times U(1)^2$, and the required Higgsing to $U(1)^2$, this suggests a $U(2)$-bundle embedded in the $SU(2)^2$ factor. All in all, in order to match the number of vector multiplets and hypermultiplets, we choose
\begin{equation}
    E= E_{G} \oplus E_U \, ,
\end{equation}
with 
\begin{equation}
    v(E_G) = (r_G, 0, s_G) \,,\qquad v(E_U) = (2, c_U, s_U) \,.
\end{equation}
For $E_G$, we can use the same construction of the $SO(32)$ bundle $V_G$ as discussed around \eqref{eq:bundleSo32}, with 
\begin{equation}
    c_2(V_G) = 18 \,,\quad \dim(\cM_{V_G}) = 44\,. 
\end{equation}
The dimension of the hypermultiplet moduli space is then
\begin{equation}
    \dim_\mathbb{H} \cM_H = 20-\delta + 44+ \frac12\left( v^2(E_U) +2 \right)\,,
\end{equation}
where $\delta=0$ if $c_U=0$ and $\delta=1$ for $c_U \neq 0$, to account for the volume modulus that can be fixed by the supersymmetry condition. It turns out that the only option for a $U(2)$ bundle that breaks the left-moving algebra entirely and yields $\dim_\mathbb{H}(\cM_H) =74$ is 
\begin{equation}\label{MukaivectorX44}
    v(E_U) = (2,c,-4) \,,\;\;\text{with} \;\; c^2 =4\,. 
\end{equation}
Notice that here, as for the $X_{3,3}$ case, we realize the gauge background $E_G$ geometrically, which requires $G_E=SO(32)$.\footnote{If instead we choose $G_E=E_8\times E_8$, we would require a gauge background with $c_2(E_{E_8})=9$ in each factor. As discussed below~\eqref{eq:choiceX33}, for geometric K3 compactifications of the heterotic string there are no known realizations of such a bundle that breaks $E_8$ completely. However, since here the CFT on the string worldsheet does not have a geometric interpretation, it is conceivable that the aforementioned geometric constraint can be evaded; at least, this would not lead to contradiction with Type IIA-- or F-theory--heterotic duality. Moreover, unlike the $X_{3,3}$ model, the heterotic dual of the $X_{4,4}$ model is realized at the self-T-dual point $T=U=i$, enforcing the expectation for the heterotic string to have a T-dual $E_8\times E_8$ realization with $c_2(E_{E_8})=9$.\label{fn:E8discussion}} Since $c^2=4$, we can focus on the quartic K3 surface $X$ described above and identify $c=H$, where $H$ is the hyperplane class on $S$. 

Due to the non-zero $c$ and, moreover, $c\notin H^2_{\rm alg}(S,2\mathbb{Z})$, the heterotic configuration cannot be realized as a large volume non-linear sigma model with gauge bundles. Instead, we have to consider the non-geometric Landau-Ginzburg phase to which we are confined, since our choice of gauge data does not satisfy the  supersymmetry and anomaly-cancellation conditions at large volume. As mentioned above, we do not show directly from the heterotic perspective that the gauge data leads to a supersymmetric and anomaly-free theory; instead, below we will make use of the Type IIB dual to infer this. For now, we have to show that the heterotic configuration is poly-stable. We focus on the stability of $E_U$. Using the periods of the mirror quartic at $\psi=\infty$ as given in~\eqref{LGperiods}, we find the central charge at the Gepner point to be
\begin{equation}
    Z(E_U) = 2+4 i \,. 
\end{equation}
We immediately notice that the Mukai vector $v(E_U)$ given in~\eqref{MukaivectorX44} is not primitive. We hence cannot immediately apply the result on the stability of~\cite{bayer2013projectivitybirationalgeometrybridgeland}. Indeed, we can write 
\begin{equation}
    v(E_U) = (2,H,-4) = (2,0,-1) \oplus (0,H,-3) \,,
\end{equation}
satisfying
\begin{equation}
    Z(2,0,-1) = Z(0,H,-3) = \frac{1}{2} Z(E_U)\,. 
\end{equation}
Hence, for $E_U$ with Mukai vector given in~\eqref{MukaivectorX44}, the Gepner point lies on a marginal stability wall. However, we have 
\begin{equation}
    v^2(2,0,-1) = 12 \,,\quad v^2(0,H,-3) =4\,,
\end{equation}
which both satisfy $v^2\geq -2$, hence both components into which $E_U$ splits at the Gepner point are primitive and have a non-zero moduli space consisting of stable objects. Thus, even though $E_U$ is not necessarily stable, it is poly-stable, which is sufficient for our heterotic application. 

Consider now the embedding of the structure group of $E_U$ into $SU(2)_1\times SU(2)_2$. Identifying $SU(2)_1$ with the $\mathfrak{su}(2)$ subalgebra of $\mathfrak{u}(2) \simeq \mathfrak{su}(2) \oplus \mathfrak{u}(1)$, and  picking the Cartan inside $SU(2)_2$ as the extra generator, the commutant subgroup left unbroken is $U(1)$.\footnote{This also corresponds to the standard embedding of $U(2)$ inside $SO(4)$.} Hence, the adjoint representation of the original group splits as
\begin{equation}\label{eq:u2decompSU}
    \ad[SU(2)_1\times SU(2)_2] \longrightarrow  \ad U(2) \oplus {\rm  det}_2 \oplus \overline \det_{-2} \ ,
\end{equation}
where the subscript denotes the $U(1)$ charge, and $\det$ indicates the determinant representation to which we associate the determinant $\Lambda^2 E_U$.\footnote{Given the fundamental representation ${\bf 2}_1$ of $\mathfrak{u}(2)$, which conventionally has charge 1 under the $\mathfrak{u}(1)$ subalgebra, the determinant representation is given by the antisymmetrization $\det = \Lambda^2 {\bf 2}_1$, and thus has charge 2. The same is true for the antifundamental and the inverse determinant representations. Since all charges are even numbers, all representations are allowed by the global form of $U(2)$. 
} 
As discussed in Section~\ref{ssec:nonGeomComp}, however, due to the non-zero $c_U$, the supersymmetry condition confines the model to the Gepner point. This in turn implies that the single $U(1)$ gauge field in the commutant subgroup of $U(2)$ inside $SU(2)^2$ absorbs the $B$-field axion $\text{Re}\,z$ and becomes massive due to the Green--Schwarz mechanism~\eqref{eq:massu1}. This reduces the perturbative gauge group to $U(1)^2$, matching $n_V=1$ on the Type IIB side. 

As far as the massless hypermultiplet sector is concerned, the gauge theory data is engineered exactly in such a way that the bundle moduli and the surviving K3 moduli exactly reproduce the number of massless hypermultiplets of Type IIB on $X_{4,4}$. Still, we have to ensure that no extra hypermultiplet state charged under the $U(1)$ is generated due to the decomposition~\eqref{eq:u2decompSU}, as they would become neutral hypermultiplets after St\"uckelberging the $U(1)$. To see whether there is any hypermultiplet charged under the $U(1)$, we have to consider the determinant representation of $E_U$. Even though in the heterotic theory $E_U$ is not realized as a vector bundle, we can still use a vector bundle representation of the Mukai vector $v(E_U)$ to check for additional neutral hypermultiplets. Therefore, consider a vector bundle $V$ with $v(V)=v(E_U)$. The number of hypermultiplets charged under the St\"uckelberg $U(1)$ is then given by $h^1(S, \Lambda^2 V)$. The determinant bundle $\Lambda^2 V$ is nothing but the line bundle associated with the curve class $H$, and as such is ample. The Kodaira vanishing theorem~\cite{Esnault2004LecturesOV} thus applies, determining that $h^1(S, \Lambda^2 V) =0$, meaning that there are no additional hypermultiplets charged under the St\"uckelberg $U(1)$, that would be hence neutral under the two massless vectors of the effective theory. We thus conclude that the proposed heterotic configuration is stable and exactly reproduces the low-energy spectrum of Type IIB compactified on $X_{4,4}$. 

\paragraph{(2) Matching the perturbative BPS spectrum.} As for the $X_{3,3}$ case, we can test the proposal for the duality between Type IIB on $X_{4,4}$ and the heterotic configuration described above beyond the massless level. We therefore consider again the perturbative heterotic BPS spectrum, which, as we already saw, has to be identified with the BPS states obtained from wrapped D3-branes on the Type IIB side. Starting from the latter, the central charge of the D3-branes that become light in the vicinity of the K-point in the moduli space of $X_{4,4}$ can be computed using~\eqref{eq:centr_charge_BPS_IIB}, with the boundary data for $X_{4,4}$ computed in~\cite{Bastian:2023shf} (see also Table~\ref{table:geometric_data_K-point})
\begin{equation}
    (\alpha, \beta, \gamma, \delta, \kappa)= \left(1,0,\frac12,0,1\right)\,.
\end{equation}
Comparing with~\eqref{eq:momwind_hetunits}, we immediately see that the heterotic torus moduli are confined to the elliptic point $T=U=i$, which corresponds to the fixed point under the order-2 monodromy $S$ in $PSL(2,\mathbb{Z})$. This confirms the enhancement of the perturbative gauge group to the $SU(2)^2$ gauge factor considered in the matching of the supergravity spectrum above. We conclude that the perturbative BPS spectrum of the heterotic configuration introduced above reproduces the light D3-brane spectrum of Type IIB on $X_{4,4}$ in the vicinity of the K-point. 

\paragraph{(3) Matching the monodromy group.} Finally, we can also test the proposed duality at the non-perturbative level by reproducing the monodromy matrix~\eqref{eq:monodromy_matrices_K-points_explicit} from the heterotic perspective. Following the discussion in Section~\ref{ssec:nonperttest}, the monodromies mixing the perturbative and non-perturbative heterotic states split into a unipotent part $T_u$, as in~\eqref{eq:inverse_Gram_matrix_monodromy}, and a semi-simple part $T_{ss}$, as in~\eqref{eq:Tss_het}. Using the fact that the stabilizer subgroup of $T=U=i$ has order $n=2$ in $PSL(2,\mathbb{Z})$, we find
\begin{equation}\label{eq:x44_het_monod}
     N=\log(T_u^2)=\begin{pmatrix}
        0 & 0&0&0 \\ 0&0&0&0\\  1&0 &  0&0\\ 0 & 1 &0&0
    \end{pmatrix}\, , \quad T_{ss} = \begin{pmatrix}
        0 & 1&0&0 \\ -1&0&0&0\\  p&q &  0&1\\ r& s &-1&0
    \end{pmatrix}\,.
\end{equation}
Imposing that $T_{\rm het}\equiv T_u T_{ss} \in Sp(4,\bbZ)$ and $T_{ss}^2=\pm\mathrm{Id}$, we get
\begin{equation}\label{eq:boundary_data_X44}
    r=q\, , \quad s=-p\, , \quad p \in \bbZ\, ,\quad q\in \bbZ+\frac{1}{2}\, ,
\end{equation}
and we reproduce the monodromy matrix in~\eqref{eq:monodromy_matrices_K-points_explicit} with the choice $p=0\,\ q=1/2$. Once again, these free parameters and their quantization condition can be identified with the choice of boundary data in~\cite{Bastian:2023shf}. The proposed duality between the heterotic configuration and Type IIB on $X_{4,4}$ hence also passes the non-perturbative test. 

\paragraph{(4) Heterotic worldsheet consistency from Type IIB.} Our above arguments show that the heterotic configuration, specified by the Mukai vectors~\eqref{eq:bundleSo32} and~\eqref{MukaivectorX44}, reproduces the low-energy effective action of Type IIB on $X_{4,4}$ and passes the perturbative and non-perturbative checks, probing the validity of the duality at the quantum level. We therefore have very strong evidence that the worldsheet theory on the EFT string in Type IIB on $X_{4,4}$ probing the K-point is given by the heterotic string on K3$\times T^2$, with the K3 volume fixed at a string-size volume, the torus moduli fixed at $T=U=i$ and the above gauge data. 

Recall that, on the heterotic side, we showed that there exists in general a stable object with this Mukai vector; however, we did not show that the heterotic configuration describes an anomaly-free worldsheet SCFT with $\mathcal{N}=(0,4)$ supersymmetry. To achieve this, we can exploit the duality to Type IIB on $X_{4,4}$. Indeed, the EFT string on the Type IIB side is constructed as a supersymmetric solution of the equations of motion, such that the worldsheet theory on this string is a supersymmetric, anomaly free 2d CFT. Given our evidence that the worldsheet on the EFT string is described by the heterotic configuration specified in this section, the Type IIB realization of the string thus ensures that this heterotic configuration indeed gives rise to an anomaly-free worldsheet theory with $\cN=(0,4)$ supersymmetry. Let us stress again that this makes use of the Type IIB duality and we do not infer this directly from the heterotic string setup, as it would require to compute the $\alpha'$-exact D-term and anomaly cancellation condition for the heterotic string. In other words, here we use the classical description of one side of the duality (Type IIB on a Calabi--Yau threefold) to infer the effect of $\alpha'$-corrections on the other side of the duality (the heterotic configuration). Finally, we recall that this last step was not necessary for the heterotic dual of the $X_{3,3}$ model, where supersymmetry and anomaly-cancellation on the worldsheet were guaranteed also on the heterotic side, thanks to the large K3 volume realization of the latter.

\subsection{The Heterotic Dual of \texorpdfstring{$X_{6,6}$}{X66}}\label{ssec:x66}
We now turn to the last one-parameter hypergeometric model featuring a K-point in its moduli space, namely $X_{6,6}$. Recall that Type IIB compactified on the latter leads to a 4d $\cN=2$ theory with $n_V=1, \, n_H = 62$. For the heterotic dual of this theory, we will see that the torus moduli will be fixed to the third root of unity, $T=U=e^{\pi i/3}$, which follows from the matching of the monodromies and BPS states with the Type IIB dual theory. The perturbative gauge group before Higgsing thus shows an enhancement to $SU(3)$. 

\paragraph{(1) Matching of massless spectrum.} Due to the low number of hypermultiplets, the choice of gauge theory data is quite restricted, and, as for the $X_{4,4}$ case, no $SU(3)$ or $SU(2)^2$ bundle can match the massless spectrum of Type IIB on $X_{6,6}$. For the gauge theory data, we then again make the ansatz
\begin{equation}
    E= E_{G} \oplus E_U \, ,
\end{equation}
with 
\begin{equation}\label{eq:bundle_choice_x66}
    v(E_G) = (r_G, 0, s_G) \,,\qquad v(E_U) = (2, c_U, s_U) \,,
\end{equation}
For $E_G$, one can consider again an $SO(32)$ bundle $V_G$, with a similar construction as the one discussed around~\eqref{eq:bundleSo32}, with
\begin{equation}\label{eq:topdataVG}
    c_2(V_G)= 17\,,\quad \dim(\cM_{V_G})=14\,. 
\end{equation}
Notice, however, that taking literally the same construction as in~\eqref{eq:bundleSo32} and just adjusting the relevant topological data will not lead to a complete Higgsing of the $SO(32)$ gauge group. Instead, we should slightly modify the construction and consider more involved tensor products of sub-bundles. Since conceptually this works as before, though more involved, we do not discuss this here in detail and assume that we have a gauge background $E_G$ satisfying~\eqref{eq:topdataVG} and completely breaking the $SO(32)$ gauge group.\footnote{To explicitly construct such a gauge background, it might be necessary to take into account the non-geometric nature of the heterotic K3, similarly to the considerations raised for the possibility of an $E_8\times E_8$ realization of the heterotic dual of $X_{4,4}$, see
Footnote~\ref{fn:E8discussion}. It would be interesting to investigate this in more detail.} For what concerns $E_U$, consider a gauge background with $U(2)$ structure group embedded in the enhanced gauge group obtained from the torus isometries, namely $SU(3)$. Imposing that the gauge configuration reproduces $n_H=62$ hypermultiplets and satisfies the Bianchi identity, this leads to
\begin{equation}\label{MukaivectorX66}
    v(E_G) = (16, 0,-1) \,,\qquad v(E_U) = (2,c,-12) \,,\;\;\text{with} \;\; c^2 =36\,. 
\end{equation}
This can again be realized for a K3 surface $S$ given by the quartic and $c=3H\in H^2_{\rm alg}(S,\mathbb{Z})$. To match the dimension of the hypermultiplet moduli space, we again used that, due to $c^2>0$ the overall volume of $S$ is not a marginal deformation of the worldsheet theory. Notice that the 10d gauge group has to be again $SO(32)$, since the above gauge data do not allow for a configuration that completely Higgses an $E_8\times E_8$ gauge group.

As before, we have to argue that the configuration~\eqref{MukaivectorX66} admits a stable representative. To do so, let us consider only the $E_U$ component. We notice that, at the Gepner point of the quartic, the central charge is given by 
\begin{equation}
    Z(v(E_U)) = 2(3+4i)\,. 
\end{equation}
Similarly to the $X_{4,4}$ case, the two Mukai vectors 
\begin{equation} 
v(E_1) = (2,H,-5) \,,\qquad v(E_2) = (0,2H,-7) \,, 
\end{equation}
satisfy 
\begin{equation}
    v(E_U) = v(E_1) + v(E_2) \,,\quad Z(v(E_U))= 2 Z(v(E_1)) = 2 Z(v(E_2))\,.
\end{equation}
Since also 
\begin{equation}  v^2(E_1),v^2(E_2)>-2\,,
\end{equation}
following the same reasoning as for the dual of the $X_{4,4}$ model, we conclude that the Mukai vector $v(E_U)$ in~\eqref{MukaivectorX66} admits a semi-stable representative. Moreover, as in the previous construction, the gauge background configuration with $U(2)$ structure group, when embedded in $SU(3)$, leaves the $U(1)$ commutant unbroken. Indeed, the embedding can be realized by identifying the $\mathfrak{su}(2)$ subalgebra of $\mathfrak{u}(2)$ with one of the two Cartan generators of $SU(3)$, and the additional $\mathfrak{u}(1)$ with the second one, leading to an unbroken $U(1)$. The latter is then Higgsed via the Green--Schwarz mechanism due to the non trivial $c_U$ of $E_U$, as described in~\eqref{eq:massu1}.

The above counting of the massless hypermultiplets takes into account the gauge background deformations as well as the surviving K3 moduli. As for the heterotic model dual to Type IIB on $X_{4,4}$, we must check that the $U(2) \hookrightarrow SU(3)$ embedding does not generate extra charged multiplets which become neutral after St\"uckelberging the $U(1)$ symmetry. To do so and for computational simplicity, we again consider a vector bundle $V$ as representative of the Mukai vector $v(E_U)$, even though, due to the non-geometric regime of the compactification, the actual object describing the heterotic background may be more general. As described above, the embedding can be chosen to be the canonical embedding of $SU(N)\times U(1)^M \subset SU(N+M)$, see e.g.~\cite{Blumenhagen:2005ga, Weigand:2006wyl}, in which the adjoint for $N=2,\, M=1$ decomposes in term of $U(2)$ representations as 
\begin{equation}
    \ad SU(3) \longrightarrow  \ad U(2) \oplus (V\otimes{\rm  det} V)_3 \oplus (\overline V\otimes \overline {{\rm  det} V})_{-3} \ ,
\end{equation}
where once again the subscript denotes the $U(1)$ charge. The $U(2)$ bundle can be constructed through the Heartshorne--Serre correspondence, reviewed in Appendix~\ref{app:B}. The dimension of $h^1(K3, V \otimes \det V)$ relevant for the representation $V\otimes{\rm  det} V$ has been computed, and vanishes. Thus, the $U(2)$ gauge background presented above has the required properties. Since the K3 volume is fixed to the Landau-Ginzburg phase and the heterotic theory is hence genuinely four-dimensional, we stress that the actual gauge theory data are not described by a vector bundle but by a more general object $E_U\in D^b({\rm Coh}(S))$ with the specified Mukai vector. 

\paragraph{(2) Matching of the perturbative BPS spectrum.} We can now match the perturbative heterotic spectrum with the BPS states of wrapped D3-branes. For the $X_{6,6}$ case, the boundary data computed in~\cite{Bastian:2023shf} and summarized in Table~\ref{table:geometric_data_K-point} are given by
\begin{equation}
    (\alpha, \beta, \gamma, \delta, \kappa)= \left(2,-1,\frac23,-\frac16,2\right)\,.
\end{equation}
The Type IIB BPS charge lattice, computed using~\eqref{eq:centr_charge_BPS_IIB}, then matches the perturbative heterotic charge lattice and central charges~\eqref{eq:momwind_hetunits} once the torus moduli are fixed to the elliptic point $T=U=e^{\pi i/3}$, with order-3 stabilizer in $PSL(2,\mathbb{Z})$. The model thus realizes an enhanced perturbative gauge group with an $SU(3)$ factor, and justifies the gauge choice~\eqref{eq:bundle_choice_x66} described above.

\paragraph{(3) Matching the monodromy group.} As a last step, we repeat the procedure outlined in Section~\ref{ssec:nonperttest} in order to reconstruct the Type IIB monodromy matrix from the heterotic perspective. The order $n=3$ stabilizer action identified in Table~\ref{table:geometric_data_K-point_heterotic} and~\eqref{eq:Tss_het} leads to the semi-simple part of the monodromy
\begin{equation}
    T_{ss}=\begin{pmatrix}\begin{array}{cccc}
 0 & 1 & 0 & 0 \\
 -1 & 1 & 0 & 0 \\
 p & q & 1 & 1 \\
 r & s & -1 & 0 \\
\end{array}\end{pmatrix}\,,
\end{equation}
where once again the parameters $p,q,r,s$ are determined by requiring an integer symplectic action of the full monodromy matrix $T_{\rm het}$ and $T_{ss}^3=\pm\mathrm{Id}$. To this end, using~\eqref{eq:gram} and~\eqref{eq:N_het}, one gets
\begin{equation}
    T_{\rm het}=T_{ss}T_{u}=\left(
\begin{array}{cccc}
 0 & 1 & 0 & 0 \\
 -1 & 1 & 0 & 0 \\
 p+\frac{1}{3} & q+\frac{1}{3} & 1 & 1 \\
 r-\frac{2}{3} & s+\frac{1}{3} & -1 & 0 \\
\end{array}
\right) \, .
\end{equation}
The constraints above require the following conditions
\begin{equation}
    r=q\,, \quad s=-p-q\,, \quad p \in -\frac13 + \bbZ \,,\quad q \in \frac23+ \bbZ \,,
\end{equation}
which we can once again match the extension data characterizing the semi-simple part of the monodromy matrix, as discussed in Section~\ref{sec:one-two_parameters_models} and in~\cite{Bastian:2023shf}. The resulting full monodromy matrix matches the Type IIB one in~\eqref{eq:monodromy_matrices_K-points_explicit}.

\paragraph{(4) Heterotic worldsheet consistency from Type IIB.} Let us stress that, similar to the $X_{4,4}$ case, we have not shown that the configuration of gauge data with Mukai vectors given in~\eqref{MukaivectorX66} leads to a supersymmetric and anomaly-free worldsheet theory. However, given that this configuration satisfies all the above tests to qualify as the heterotic theory arising from the Type IIB EFT string near the K-point, we can use the same arguments as in the $X_{4,4}$ case and exploit the Type IIB realization of the EFT string to argue that the above configuration leads to an anomaly-free Gepner worldsheet theory with $\cN=(0,4)$ supersymmetry. 

\section{Dualities with Fivebranes and Heterotic Transitions}\label{sec:twoparametermodelshet}

In the previous sections, we have discussed the dual heterotic theories emerging in the vicinity of K-points in one-parameter compactifications of Type IIB string theory, focusing on the different consistency conditions that must be imposed on the geometry and gauge theory data of the theory. In particular, we saw that a non-trivial first Chern class of the gauge bundle forces to consider non-perturbative corrections to the consistency conditions and leads to a genuinely 4d duality, fixing the volume of the K3 compactification manifold on the heterotic side to string-scale. 

However, K-points in one-parameter moduli spaces are just a specific example of type II$_0$ degenerations. In fact, these degenerations can also arise in higher-dimensional moduli spaces and, similarly, are expected to be dual to a weakly coupled heterotic string compactified on K3$\times T^2$~\cite{Hassfeld:2025uoy, Monnee:2025msf} satisfying the same consistency conditions discussed in the previous sections. The main difference with the one-parameter case, however, is that these models also have additional $U(1)$ symmetries which, on the heterotic side, are associated with the presence of spacetime-filling NS5-branes.

In this section, we analyse these more involved models and, in particular, how the presence of spacetime-filling NS5-branes affects the physics of the heterotic theory. To do so, we will focus on the specific Calabi--Yau threefold $X_{12}$ introduced in Section~\ref{sec:p11226}, given by the degree-12 hypersurface in $\bbP_{1,1,2,2,6}/(\mathbb{Z}_6^2\times \mathbb{Z}_2)$, which has been already extensively studied in the literature~\cite{Kachru:1995fv, Monnee:2025msf, Hattab:2026lho}.

\subsection{The Heterotic Dual of \texorpdfstring{$X_{12}$}{X12}}\label{ssec:x12}

Recall that Type IIB on $X_{12}$ leads to a 4d $\cN=2$ supergravity theory with $n_H=129$ and $n_V=2$. As reviewed in Section~\ref{sec:p11226}, the complex structure moduli space of $X_{12}$ contains two type II loci that intersect normally. The singular locus $\Delta_1$ is a type II$_1$ singularity and is connected to the LCS point. The heterotic dual associated with this locus has already been identified in~\cite{Kachru:1995wm}. We start by reviewing this heterotic theory which we refer to as $\mathtt{H}_1$ theory. The singular divisor $\Delta_1$ is intersected normally by the type II$_0$ divisor $\Delta_0$. The main goal of this section is to identify the heterotic dual that becomes weakly coupled near $\Delta_0$; we refer to it as $\mathtt{H}_0$ theory.

\subsubsection{Type II\texorpdfstring{$_1$}{1} Locus}\label{ssec:type_II_1_locus_het_dual}
The heterotic dual along the type II$_1$ locus has already been extensively studied in the literature, starting from~\cite{Kachru:1995wm, Kachru:1995fv}, and it represents one of the first instances of Type II--heterotic duality in four dimensions. The $\mathtt{H}_1$ theory was identified to be the heterotic string (either HO or HE) compactified on K3$\times T^2$ at the $T=U$ locus in the torus moduli space, for which we get an enhancement of the 8d perturbative gauge group to $G=G_{(E/O)}\times SU(2)\times U(1)^3$. The bundle data that  matches the dimensions of the vector and hypermultiplet moduli spaces are
\begin{equation}\label{eq:bundle_data_type_II_1}
    V^{(\mathtt{H}_1)}=V_G^{(\mathtt{H}_1)} \oplus V_{SU}^{(\mathtt{H}_1)} , \ c_2(V_G^{(\mathtt{H}_1)}) = 20 ,\ c_2(V_{SU}^{(\mathtt{H}_1)}) = 4 , 
\end{equation}
where $V_{ SU}^{(\mathtt{H}_1)}$ is an $SU(2)$ bundle on K3 and $V_G^{(\mathtt{H}_1)}$ is an $E_8 \times E_8$ or $SO(32)$ bundle. Indeed, from~\eqref{eq:indexthm} one can check that
\begin{equation}
    \dim_{\bbH} \cM_{V^{(\mathtt{H}_1)}} = 129 \ ,
\end{equation}
which reproduces the correct number of hypermultiplets, and, at the same time, yields $G\rightarrow U(1)^3$, matching $n_V=2$. This choice, similarly to the $X_{3,3}$ case, doesn't fix any K3 modulus since $c_1(V^{(\mathtt{H}_1)})=0$ and~\eqref{eq:muvalue} is trivially satisfied, and thus admits a decompactification limit to eight dimensions in the hypermultiplet moduli space. The duality has been tested far beyond a simple match of the massless modes. A match of the monodromies of the two sides has been performed in~\cite{Kachru:1995fv,Klemm:1995tj}, while one-loop corrections have been shown to agree in~\cite{Antoniadis:1995zn,Kaplunovsky:1995tm}.\footnote{For a related match of the monodromies in another two-parameter model with a different number of hypermultiplets see e.g.~\cite{Antoniadis:1995cy}. See also~\cite{Harvey:1995fq} for an extended discussion of loop corrections and string dualities in this context.} For this reason, we do not test the duality between Type IIB on $X_{12}$ and the $\mathtt{H}_1$ any further here, but move to the $\mathtt{H}_0$ theory.

\subsubsection{Type II\texorpdfstring{$_0$}{0} Locus}

Let us now consider the type II$_0$ degeneration locus $\Delta_0$ and construct the corresponding dual heterotic theory $\mathtt{H}_0$. Following the analysis in~\cite{Hassfeld:2025uoy,Monnee:2025msf}, along $\Delta_0$ the Calabi--Yau threefold undergoes a type II$_0$ degeneration of Tyurin type, implying that the transcendental lattice of the K3 surface has signature ${\rm sgn}\,\Lambda_{\rm trans}=(2,0)$. The LMHS is then characterized by $\dim{\rm Gr}_{l=2,4}(\Delta_0)=2$ and $\dim{\rm Gr}_{l=3}(\Delta_0)=2$, as can also be verified via the explicit form of the periods presented in~\eqref{eq:periods_two_parameter} and their monodromies. 

As was already identified in~\cite{Monnee:2025msf}, the associated dual heterotic theory should be given by the heterotic string compactified on K3$\times T^2_{T=U=i}$, which leads to an additional gauge enhancement of the form $U(1)\rightarrow SU(2)$ for the second left moving $U(1)_L$ factor. However, according to the analysis in~\cite{Hassfeld:2025uoy}, the latter $SU(2)$ gauge theory has to be Higgsed by a suitable gauge background, leading to a $U(1)^2$ perturbative heterotic gauge group and a breaking pattern of the form
\begin{equation}
    G=G_{(E/O)}\times SU(2)^2_L\times U(1)^2_R\xrightarrow[]{V = V_G\oplus V_{SU(2)^2}}U(1)^2_R\, .
\end{equation}
The missing $U(1)s$ associated with ${\rm Gr}_3(\Delta_0)$ are instead identified with a spacetime-filling NS5-brane wrapping $T^2$. We now investigate how the proposal of~\cite{Monnee:2025msf} can be realized in practice.

\paragraph{(1) Matching the massless spectrum.} Having identified the special locus in the torus moduli space for the $\mathtt{H}_0$ theory, we have to determine a suitable gauge background such that the resulting 4d $\mathcal{N}=2$ low-energy limit reproduces the supergravity limit of Type IIB on $X_{12}$. The gauge background for the $\mathtt{H}_0$ theory will turn out to be very similar to the one for the heterotic dual to Type IIB on $X_{4,4}$ analysed in Section \ref{sec:x44}. The main difference between the two lies in the presence of an NS5-brane on the heterotic side. 

We again make the ansatz for the gauge data of the $\mathtt{H}_0$ theory to be given by 
\begin{equation}
    E^{(\mathtt{H}_0)}= E_U \oplus E_G \,,
\end{equation}
with 
\begin{equation}
    v(E_G) = (r_G, 0, s_G) \,,\qquad v(E_U) = (2, c_U, s_U) \,.
\end{equation}
Since the $\mathtt{H}_0$ can be smoothly connected to the $\mathtt{H}_1$ theory in the $X_{12}$ complex structure moduli space, this must also be true from the heterotic side. For this reason, a good starting point for the bundle data of $\mathtt{H}_0$ is the gauge bundle  $V_G^{(\mathtt{H}_1)}$ in~\eqref{eq:bundle_data_type_II_1}. Notice that in this case, and opposed to the $X_{3,3}$ and $X_{6,6}$ cases, we are free to embed $E_G$ in $SO(32)$ or $E_8\times E_8$, which is also consistent with the fact that the heterotic torus is fixed to self-T-dual radius $T=U=i$. 

To break the additional $U(1)$ it is not sufficient to simply consider the direct sum of the $SU(2)$ bundle of the $\mathtt{H}_1$ theory together with a line bundle. Instead, we have to consider an irreducible gauge background $E_U$ with $U(2)$ structure group embedded in the $SU(2)^2$ factor of the gauge group. Nonetheless, differently from the previous discussion, in order to satisfy the Bianchi identity and match the dimension of the hypermultiplet moduli space we have to take into account the presence of the NS5-brane, which contributes one vector multiplet and one hypermultiplet. The former arises from the dimensional reduction of the tensor on the NS5-brane (in the $E_8\times E_8$ case) or the vector multiplet on the SO(32) NS5-brane. The hypermultiplet, instead, comes from the reduction of the hypermultiplet of the 6d $\mathcal{N}=(1,0)$ worldvolume theory of the NS5-brane in either HE or HO theory. The Type IIB low energy supergravity approximation is then reproduced by the heterotic configuration with
\begin{equation}\label{MukaivectorX12}
    v(E_G) = (18, 0, -2) \,,\qquad v(E_U) = (2, 1, -1) \,.
\end{equation}
Notice that, since $c_U$ is odd, the above gauge background cannot be decomposed into a $SU(2)$ bundle and a line bundle. Indeed, up to the NS5-brane contribution, this choice satisfies the Bianchi identity\footnote{Since we are considering a spacetime-filling NS5-brane wrapping $T^2$, from the point of view of the internal space K3 this can be regarded as localized at one point $p\in H_0(K3,\bbZ)$. Its contribution to the tadpole condition~\eqref{eq:tadpole} will then be given in terms of the associated Poincaré dual element $[P.D.(p)]\in H^4(K3,\bbZ)$.} 
\begin{equation}
    -(s_G - r_G) - (s_U-r_U) = 23\,. 
\end{equation}
and, at the same time, the resulting hypermultiplet moduli space has dimension
\begin{equation}
        \dim_\bbH \CM_{V^{(\mathtt{H}_0)}} =\overbrace{1}^{NS5}\, + \,\overbrace{2\times 52}^{E_G}\, +\, \overbrace{5}^{E_U }\, + \,\overbrace{19}^{K3} = 129 \ ,
\end{equation}
which precisely matches the counting from the Type IIB side in~\eqref{eq:numb_modulii_2_parameter}. Here, we again used that since $c_U^2>0$ the overall K3 volume modulus cannot be a marginal deformation on the string worldsheet and hence does not yield a hypermultiplet in spacetime. Again, we do not directly infer worldsheet supersymmetry and absence of worldsheet anomalies from the heterotic data; this is instead ensured by the Type IIB realization of the EFT string. We hence continue by showing that the given heterotic configuration reproduces the properties expected from the worldsheet theory on the EFT string. 

Unlike in the $X_{4,4}$ and $X_{6,6}$ cases, the Mukai vector $v(E_U)$ has stable rather than just semi-stable representatives. To see this, we notice that at the Gepner point any integer quantized Mukai vector $v=(r,c,s)$ has central charge
\begin{equation}
    Z(v) = -s-2c +i(r+2c) \in \mathbb{Z}+ i\mathbb{Z}\,. 
\end{equation}
The Mukai vector $v(E_U)$ is hence primitive in the central charge plane, meaning that there cannot be two objects $E_1$ and $E_2$ such that $v(E_U)=v(E_1)+v(E_2)$ and for which the phases align with $v(E_U)$ at the Gepner point. Since furthermore 
\begin{equation}
    v^2(E_U) = 8 >-2\,,
\end{equation}
the general results on K3-stability imply that there exist stable objects with this Mukai vector. The proposed heterotic configuration is hence stable, satisfies the Bianchi identity and reproduces the massless spectrum of Type IIB on $X_{12}$. 

\paragraph{(2) Matching the perturbative BPS spectrum.} The exact mass of BPS D3-brane states on the Type IIB side can be read-off from the periods in~\eqref{eq:periods_two_parameter}. Near the $\Delta_0$ locus, their leading order expansion reads
\begin{equation}\label{eq:asymp_period_two_param}
    \Pi = \begin{pmatrix}
        X^0\\X^1\\X^2 \\F_0 \\F_1 \\F_2
    \end{pmatrix} \sim \begin{pmatrix}
       1 \\
        i \\
        -2{\beta} \sqrt{x_0} \\
        \frac{1 }{2\pi i}\log(x_0^2 )\\
        \frac{\log(x_0^2 )}{2\pi} \\
        \frac{\beta}{\pi i} \log(x_1) \sqrt{x_0} + \frac{4\beta}{i\pi} \sqrt{x_0} 
    \end{pmatrix}\, .
\end{equation}
The $(X^0,X^1)$ periods correspond to the perturbative heterotic BPS spectrum studied in Section~\ref{ssec:hetDualII0}. Following the discussion therein, a non-trivial test of the proposed duality is to check whether the lattice of perturbative heterotic BPS states agrees with the integer lattice of special Lagrangian 3-cycles on $X_{12}$ that shrink to zero size along $\Delta_0$. To match the two, however, it is important to account for the fact that in the Type IIB basis given in~\eqref{eq:asymp_period_two_param} the symplectic pairing is $\hat \Sigma =-2\Sigma$, as mentioned in Section~\ref{sec:p11226}.\footnote{This factor of two is in fact the origin of the no-factorization result for the gauge group found in~\cite{Hattab:2026lho}.} Indeed, recall that the heterotic momentum and winding lattice in~\eqref{eq:momwind_hetunits} is identified with the integer homology lattice of electric 3-cycles, which in terms of the symplectic basis at hand reads\footnote{Indeed, by decomposing a 3-cycle $\Gamma=e_IA^I + m^JB_J$, with $e_I,\, b^I \in \bbZ$ and  $B_J\cdot A^I=-A^I \cdot B_J = 2\delta^{I}_J$, we obtain
\begin{equation}
q_I = B_I \cdot \Gamma= 2 e_I \quad {\rm and} \quad p^I = A^I \cdot \Gamma =-2m^I\, .
\end{equation}
The identification of the momentum and winding lattice with the integer homology lattice of electric 3-cycles, i.e. $(n_1,w_2) \cong (e_0,e_1)$, then leads~\eqref{eq:el_identification}.}
\begin{equation}\label{eq:el_identification}
    2(n_1,w_2)\cong (q_0,q_1)\, . 
\end{equation}
This ensures the matching of the electric periods on the two sides of the duality.

Since the heterotic model is realized at $T=U=i$, the locus $\Delta_0$ has orbifold order $n=2$. The asymptotic central charges for the heterotic BPS states then agree with the D3-brane central charges computed from~\eqref{eq:asymp_period_two_param} upon identifying the K-point modulus with $t\cong \frac{1}{2\pi i}\log x_0$. Thus, the perturbative string spectrum of the heterotic configuration indeed matches the light D-brane spectrum of Type IIB on $X_{12}$ in the vicinity of $\Delta_0$. 

\paragraph{(3) Non-perturbative tests of the duality.} 
Unlike the one-parameter models, in addition to the non-perturbative states coming from the magnetic dual lattice to~\eqref{eq:momwind_hetunits}, the heterotic dual of $X_{12}$ also contains a non-gravitational sector associated with the NS5-brane. The latter has to reproduce the Type IIB sector corresponding to the periods $(X^2,F_2)$. As discussed in~\cite{Monnee:2025msf}, the heterotic theory realizes this sector in the following way. Consider for definiteness the HE theory. The six-dimensional worldvolume theory of NS5-branes in the HE theory has an anti-self-dual two tensor under which E-strings are charged. In the Ho\v{r}ava--Witten realization of the HE theory, the NS5-brane can be viewed as an M5-brane localized at a point along the interval, with the E-strings corresponding to M2-branes stretching between the M5-brane and the 9-branes localized at the end of the interval. Considering the NS5-brane wrapping the $T^2$, we obtain two gauge fields under which the E-strings wrapping a 1-cycle of the $T^2$ factor are charged. As discussed in~\cite{Monnee:2025msf}, these E-string states are dual to the D3-brane states in $\mathrm{Gr}_3(\Delta_0)$. Notice that, in the full theory, the E-string states are not necessarily electro-magnetic dual to each other. On the Type IIB side, this is reflected by the fact that in the chosen basis the pairing  $(X^2,F_2)$ is not canonical, such that $(X^2,F_2)$ do not describe electro-magnetic dual periods. Identifying the heterotic dilaton $S\cong\frac{1}{2\pi i}\log x_0$, the latter transform under $S\to S+1$ as $(X^2,F_2) \to -(X^2,F_2) $. As remarked in~\cite{Monnee:2025msf}, the coordinate $x_1$ parametrizes instead the location of the NS5-brane on the Ho\v{r}ava-Witten interval in the M-theory picture. Thus, the duality between Type IIB on $X_{12}$ and the heterotic configuration proposed above also passes this non-perturbative test. 

We can say even more and reproduce the full monodromy around $\Delta_0$ from the heterotic side following the strategy outlined in Section~\ref{ssec:nonperttest}. Let us start by considering the unipotent monodromy. Following the discussion around~\eqref{eq:inverse_Gram_matrix_monodromy} with $n=2$, one expects that in the $(X^0,X^1,F_0,F_1)$ sector it acts as 
\begin{equation}
    \tilde T_{u} = \begin{pmatrix}
        {\rm Id}_{2} & 0 \\ 
        {\rm Id}_{2} & {\rm Id}_{2} 
    \end{pmatrix}\, .
\end{equation}
However, unlike in the $X_{4,4}$ case, one must also account for the identification~\eqref{eq:el_identification}, according to which the magnetic charges are identified as $p^I = q_{\rm n.p}^I$ in \eqref{eq:non-pert-charges}, in order to match their Dirac pairing on the Type IIB side.\footnote{Indeed, by requiring the Dirac pairing in the Type IIB basis with symplectic form $\hat \Sigma$ to be consistent with the heterotic dual counterpart, we have to impose ${\bf q}^T_{\rm het}\,\Sigma\, {\bf q}_{\rm het}={\bf q}^T_{\rm IIB}\,\hat\Sigma\, {\bf q}^T_{\rm IIB}$. By requiring the relation between the electric charges to be as in \eqref{eq:el_identification}, this automatically fixes the one between the magnetic dual ones. By the same reasoning, this also reflects in the relation between the periods on the two sides.} At the level of the period vector,~\eqref{eq:el_identification} thus leads to 
\begin{equation}
    X^I_{\rm IIB } = X^{I}_{\rm het}\, , \quad F_I^{\rm IIB} = 2 F_{I}^{\rm het}\, .
\end{equation}
All in all, noticing that the non-gravitational E-string states factorize as in~\eqref{eq:monodromy_factorization}, we get
\begin{equation}
    \tilde T_u^2= \begin{pmatrix}\label{eq:X12_unipot_het}
 1 & 0 & 0 & 0&0&0 \\
 0 & 1 & 0 & 0&0&0 \\
 0 & 0 & 1 & 0&0&0  \\
 2 & 0 & 0 & 1 & 0& 0  \\
 0&2& 0& 0& 1 &0\\
 0&0&0&0&0&1
    \end{pmatrix}\, ,
\end{equation}
matching the Type IIB unipotent monodromy $T_{\Delta_0,u}$ in~\eqref{eq:2-param-monodromy}. Notice that the identification $T_{\Delta_0,u}=\tilde T_u^2$ is compatible with the fact that the Type IIB modulus $x_0$ is already identified with the heterotic dilaton as $S=\frac{1}{2\pi i}\log x_0$, which, following the discussion in Section~\ref{ssec:nonperttest}, is sensitive to the finite cover with respect to the orbifold monodromies of the torus factor. The fact that there is no action on the non-gravitational sector is due to its polynomial dependence on $x_0$, which maps to an exponential dependence on the dilaton factor as $x_0\sim e^{2\pi iS}$.

The semi-simple part of the monodromy deserves more attention. From the discussion in Section~\ref{ssec:nonperttest}, one identifies $T_{\rm n.g}=-\mathrm{Id}_2$ acting on $(X^2,F_2)$ as the unique order-2, non-trivial transformation that can affect this sector. This follows from the factorization of the non-perturbative monodromies in~\eqref{eq:monodromy_factorization}, together with the requirement that they preserve the symplectic pairing $\hat \Sigma$. Explicitly, imposing a non-trivial monodromy on the non-gravitational sector hence leads to
\begin{equation}
    \tilde T_{\rm n.g} = \begin{pmatrix}
  1 & 0 & 0 & 0&0&0 \\
 0 & 1 & 0 & 0&0&0 \\
 0 & 0 & -1 & 0&0&0  \\
 0 & 0 & 0 & 1 & 0& 0  \\
 0&0& 0& 0& 1 &0\\
 0&0&0&0&0&-1
    \end{pmatrix}\, ,
\end{equation}
which matches $T_{\Delta_0,\bbZ_2}$ in~\eqref{eq:2-param-monodromy}. 

In addition to the $\bbZ_2$ monodromy associated with the non-gravitational sector, one expects a further order-two contribution arising from the stabilizer action on the torus moduli. Notice that the Type IIB monodromies computed from encircling $x_0$ are insensitive to these symmetries, as they act only multiplicatively on the period and require an analysis of the LMHS automorphisms together with its boundary data, similarly to the treatment for one-parameter models in~\cite{Bastian:2021eom}.  Motivated by the duality with the heterotic description, we however expect an additional semi-simple factor $\tilde T_{ss}$ to emerge from a more refined analysis of the Type IIB LMHS. In particular, the perturbative semi-simple part of the monodromy is derived by following the steps outlined for $X_{4,4}$, where the matrix $T_{ss}$ was computed in~\eqref{eq:x44_het_monod}. Adapted to the present context, this reads
\begin{equation}
    \tilde T_{ss} = \begin{pmatrix}\label{eq:X12_ss_het}
 0 & 1 & 0 & 0&0&0 \\
 -1 & 0 & 0 & 0&0&0 \\
 0 & 0 & 1 & 0&0&0  \\
 p & q & 0 & 0 & 1& 0  \\
 r&s& 0& -1& 0 &0\\
 0&0&0&0&0&1
    \end{pmatrix}\, , \quad p,q,r,s \in \bbQ\, .
\end{equation}
Imposing that $\tilde T_{\Delta_0,u} \tilde T_{ss} \in Sp(4,\bbZ)$ and $\tilde T_{ss}^2 = -\mathrm{Id}$ leads to
\begin{equation}\label{eq:boundary_data_X12}
    r=q\, ,\quad s=-p\, ,\quad p,q \in \bbZ\, .
\end{equation}
If one performs an integral change of basis, the values of these parameters can be further shifted into a conventional, restricted range. Specifically, let us choose
\begin{equation}
    \tilde X = \begin{pmatrix}
        \mathrm{Id}_3 & 0 \\ X & \mathrm{Id}_3 
    \end{pmatrix}\, , \qquad X=\begin{pmatrix}
        x_1 & x_2 & 0\\ x_2 & x_3 &0 \\ 0 &0& 0 
    \end{pmatrix}\, ,
\end{equation}
which implements the shifts $p \to 2x_2, q \to q+x_1 -x_3$. Then, $(p,q)$ can be brought to the domain $(p,q) \in [0,2) \times [0,1)$. Notice that the matrix above commutes with $\tilde T_u$, thus leaving it unchanged. Since $p,q$ are constrained to be integers, there are only two inequivalent choices:
\begin{equation}
    (p,q)=(0,0)\, \quad \mathrm{or} \quad (p,q)=(1,0)\, .
\end{equation}
In summary, the two possible choices of semi-simple monodromy factor are 
\begin{equation}
    \tilde T_{ss}^{(1)}=\begin{pmatrix}
 0 & 1 & 0 & 0 & 0 & 0\\
 -1 & 0 & 0 & 0 & 0 & 0\\
 0 & 0 & 1 & 0 & 0 & 0\\
 0 & 0 & 0 & 0 & 1 & 0\\
 0 & 0 & 0 & -1 & 0 & 0\\
 0 & 0 & 0 & 0 & 0 & 1
\end{pmatrix},\qquad
\tilde T_{ss}^{(2)}=\begin{pmatrix}
 0 & 1 & 0 & 0 & 0 & 0\\
 -1 & 0 & 0 & 0 & 0 & 0\\
 0 & 0 & 1 & 0 & 0 & 0\\
 1 & 0 & 0 & 0 & 1 & 0\\
 0 & -1 & 0 & -1 & 0 & 0\\
 0 & 0 & 0 & 0 & 0 & 1
\end{pmatrix}\, .
\end{equation}
From the heterotic perspective, these two choices cannot be distinguished at the level of the duality match performed in this paper. Notice that the two choices of semisimple monodromy matrices are again a consequence of the non-canonical symplectic pairing appearing on the Type IIB side; indeed, it is clear that the difference between~\eqref{eq:boundary_data_X12} and~\eqref{eq:boundary_data_X44} can be traced back to the period basis~\eqref{eq:asymp_period_two_param}.

\paragraph{(4) Heterotic worldsheet consistency from Type IIB.} Finally, similarly to the $X_{4,4}$ and $X_{6,6}$ cases, we have not shown that the Mukai vector~\eqref{MukaivectorX66} leads to a supersymmetric and anomaly-free worldsheet for the heterotic string. However, given that this configuration satisfies all the above tests to qualify it as the worldsheet theory of the Type IIB EFT string, we can use the same arguments as in the $X_{4,4}$ and $X_{6,6}$ cases. Namely, we exploit the Type IIB realization of the EFT string to argue that~\eqref{MukaivectorX66} indeed leads to an anomaly-free Gepner model worldsheet theory with $\cN=(0,4)$ supersymmetry.  

\subsection{Smooth Nucleation of Fivebranes}
The results of the previous section confirm that the worldsheet theories on the EFT strings associated with the loci $\Delta_0$ and $\Delta_1$ in $\cM_{\rm c.s.}(X_{12})$ both correspond to heterotic strings, albeit with very distinct worldsheet theories. The Type IIB formulation of the theory clearly demonstrates that these two heterotic string theories are part of the same moduli space and in fact mutually weakly coupled. By mutual weakly coupled we mean that the weak-coupling loci for both heterotic strings have a normal crossing intersection in the vector multiplet moduli space of the 4d $\cN=2$ effective theory of gravity and, furthermore, their primary type does not enhance at such intersection. Hence, not only the two divisors, but also their intersection is compatible with an heterotic string becoming weakly coupled as we move towards the singularity. In this sense we can view the duality between the $\mathtt{H}_0$ and $\mathtt{H}_1$ theory as a weak-weak coupling duality. This duality has to be contrasted with heterotic strong-weak self-duality. This latter kind of duality occurs already in higher dimensions. For example, the heterotic string on K3 with symmetric $(12,12)$ instanton embedding has a strong-weak self-duality~\cite{Polchinski:1995df}. In this case, the heterotic string $\mathtt{S}_1$, with dilaton $e^{2\phi_1}$, is identical with a heterotic string $\mathtt{S}_2$,  with dilaton $e^{2\phi_2}=e^{-2\phi_1}$, such that the strong-coupling limit for $\mathtt{S}_1$ is identified with the weak-coupling limit for $\mathtt{S}_2$. 

The duality between the $\mathtt{H}_0$ and $\mathtt{H}_1$ theory is different. In terms of the local coordinates on $\cM_{\rm c.s.}(X_{12})$, the four-dimensional heterotic dilaton of the respective theories are given by (cf.~\cite{Monnee:2025msf})
\begin{equation}
    S_0 = \frac{1}{2\pi i} \log x_0 \,,\qquad S_1 = \frac{1}{2\pi i} \log x_1\,, 
\end{equation}
such that $S_0, S_1\to i\infty$ is possible simultaneously. As a consequence, and as anticipated in~\cite{Monnee:2025msf}, there is a process that allows to transition between the $\mathtt{H}_0$ and $\mathtt{H}_1$ theories while maintaining a weakly coupled description. 

\begin{figure}[!t]
    \centering
    \includegraphics[width=0.85\linewidth]{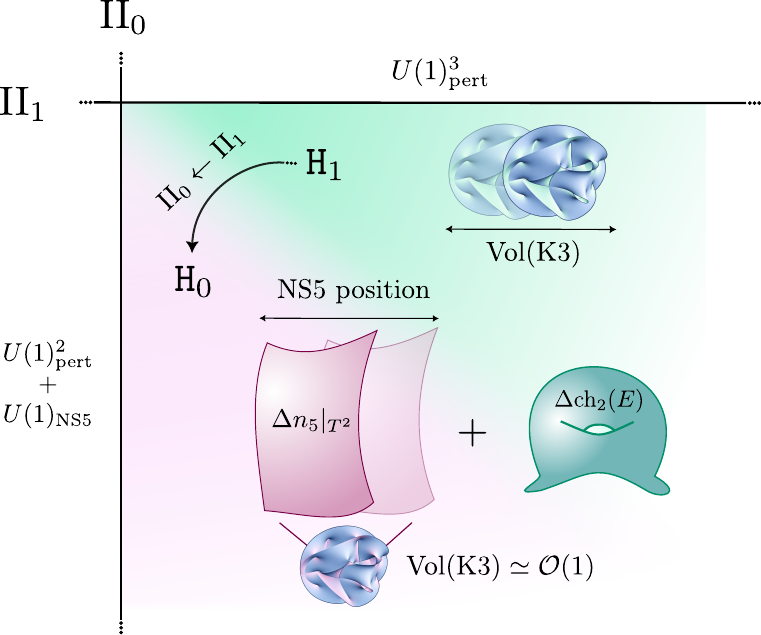}
    \caption{Heterotic phase transition relating the two type II loci intersecting normally in the complex structure moduli space of a Calabi--Yau threefold. The primary singularity-type associated with the intersection does not change, allowing for the transition to happen at weak coupling.  The heterotic model nucleates both NS5-branes, $\Delta n_5|_{T^2}$, and gauge data, $\Delta \ch_2(E)$, as to cancel the anomaly and keep the number of hypermultiplets and vector multiplets fixed. In particular, the K3 volume modulus is traded for the position of the NS5-brane. The process is smooth from the EFT perspective. On each divisor, we reported the heterotic origin of the low-energy $U(1)^3$ gauge group, where $\rm pert$ denotes a perturbative origin, while $NS5$ is reserved for $U(1)$s associated to the spacetime-filling NS5-brane. }
    \label{fig:het_transition}
\end{figure}

The results of this section give us further insights into this transition, depicted in Figure \ref{fig:het_transition}. In particular, we identified the worldsheet theories for the EFT strings associated with $\Delta_0$ and $\Delta_1$. The change in heterotic data going from $\mathtt{H}_1$ to $\mathtt{H}_0$ can be summarized as 
\begin{equation}\begin{aligned}
    \Delta \,\ch_2(E) &= \ch_2 (E^{(\mathtt{H}_0)}) - \ch_2(V^{(\mathtt{H}_1)}) =-[T^2] \in H^2({\rm K3\times T^2},\mathbb{Z})\,,\\  \Delta n_5|_{T^2}&=1\,,
\end{aligned}\end{equation}
where the second row measures the difference in the number of NS5-branes wrapping the torus in $\mathtt{H}_1$ and $\mathtt{H}_0$. The transition between $\mathtt{H}_1$ and $\mathtt{H}_0$ thus shares many features with small instanton transitions in compactifications of the heterotic $E_8\times E_8$ string on K3 leading to 6d $\mathcal{N}=(1,0)$ theories~\cite{Witten:1995gx}. 

However, there are also striking differences with the latter. First, in six dimensions, gravitational anomaly cancellation severely constrains the possible numbers $(n_V,n_T,n_H)$ of massless vector, tensor and hypermultiplets as 
\begin{equation}\label{eq:6danomaly}
    (n_H-n_V) = 273 - 29 n_T\,. 
\end{equation}
For small instanton transitions in six dimensions one has to first move to special loci in the hypermultiplet/bundle moduli space corresponding to the field-theoretic Higgs branch. Let us denote the codimension of this locus in the perturbative hypermultiplet moduli space by ${\rm codim}_{\mathbb{H}}= -\Delta n_H^{\rm pert}>0$. At this locus, a number $\Delta n_V$ of vector multiplets becomes massless, enhancing the perturbative heterotic gauge group. If in the small instanton transition $n$ NS5-branes are nucleated, these give rise to $n$ additional tensor and $n$ additional hypermultiplet. The net change in the number of hypermultiplets is then $\Delta n_H = \Delta n_H^{\rm pert}+n$. For the small instanton transition involving the nucleation of $n$ NS5-branes to be possible, the constraint 
\begin{equation}
    (\Delta n_V - \Delta n_H^{\rm pert}-n) = 29n \,, \quad n\in  \mathbb{Z}_{>0}\,
\end{equation}
has to be satisfied. The small instanton transition in six dimensions is hence an extremal transition from the hypermultiplet to the tensor multiplet branch. 

The transition relating the $\mathtt{H}_0$ and $\mathtt{H}_1$ theories in our setup is different, since the number of hypermultiplets does not change and we do not move between different moduli space branches; indeed, $\Delta_0$ and $\Delta_1$ are simply two divisors intersecting normally in the same moduli space. The reason for which this transition is possible in four dimensions is that there are no tensor multiplets and the NS5-brane contributes instead just another vector multiplet. In four dimensions, there is thus no anomaly cancellation condition similar to~\eqref{eq:6danomaly} that requires every change in the number of NS5-branes to be accompanied by a change in the number of hypermultiplets. The transition from the $\mathtt{H}_0$ theory to the $\mathtt{H}_1$ theory is hence a genuinely four-dimensional effect, which is supported by the fact that the $\mathtt{H}_0$ theory does not have a six-dimensional origin and that the change in the gauge bundle that allows for nucleating an NS5-brane is for the enhanced gauge group of the torus isometries. 

Let us take a closer look to the transition. Here we first focus on the transition $\mathtt{H}_1\to \mathtt{H}_0$ in the strict weak coupling limit, i.e., we discuss the heterotic dual of moving along the divisor $\Delta_1$ inside $\cM_{\rm c.s.}(X_{12})$ towards the intersection with $\Delta_0$, and then along $\Delta_0$. On the heterotic side, this means that we start at a generic value for $T=U$ and move towards the self-dual radius $T=U=i$. This is a gauge enhancement point for the torus isometries, where $SU(2)\times U(1)\to SU(2)^2$. At this gauge enhancement point we can now trade the $SU(2)$ bundle of the $\mathtt{H}_1$ theory for the $U(2)$ gauge background of the $\mathtt{H}_0$ theory with 
\begin{equation}
    -\ch_2(V_{SU}^{(\mathtt{H}_1)})= 4 \;\;\rightarrow\;\; -\ch_2(E_U^{(\mathtt{H}_0)}) = 3\,. 
\end{equation}
We thus lower the contribution of the gauge bundle to the Bianchi identity and at the same time decrease the rank of the surviving gauge group. At first sight this might appear counter-intuitive, but it is a consequence of $c_U\neq 0$ and $c_U^2>0$ in the Mukai vector of $E_U^{(\mathtt{H}_0)}$. 

Recall that the non-zero $c_U$ confines the heterotic model to a special point in the K3 moduli space, i.e., in the hypermultiplet moduli space of the $\mathtt{H}_1$ theory. This aspect of the transition requires some special attention. The loci $\Delta_0$ and $\Delta_1$ are divisors of the vector multiplet moduli space of the 4d $\mathcal{N}=2$ theory, such that the transition from $\mathtt{H}_1\to \mathtt{H}_0$ theory is expected to occur as we move in the vector multiplet moduli space. Due to the factorization of the vector and the hypermultiplet moduli space one then expects that this transition is independent of the hypermultiplet sector. However, from the perspective of the $\mathtt{H}_1$ theory, the $\mathtt{H}_0$ theory seems to be realized at a special locus in the hypermultiplet moduli space. But this does not mean that the $\mathtt{H}_0$ theory is realized at a ${\rm codim}_\mathbb{R}=4$ locus in the hypermultiplet moduli space, since the role of the overall volume modulus of the K3 surface in the $\mathtt{H}_1$ theory is taken over by the position modulus of the NS5-brane on the K3 in the $\mathtt{H}_0$ theory. The hypermultiplet moduli space of the $\mathtt{H}_1$ theory can hence be mapped one-to-one to the hypermultiplet moduli space of the $\mathtt{H}_0$ theory. In particular, a K\"ahler hypermultiplet has been traded for an hypermultiplet related to position of the NS5-brane.

The fact that the appearance of the $\mathtt{H}_0$ theory, which has a different bundle and non-gravitational sector, happens at a special point in the moduli space of the $\mathtt{H}_1$ theory is reminiscent of the duality between two heterotic theories in six dimensions. For heterotic compactifications on K3, it turns out that the theory with symmetric instanton embedding $(12,12)$ is dual to the theory with gauge embedding $(10,14)$~\cite{Kachru:1995wm}. This can be also seen from the F-theory perspective, where a certain smooth elliptically fibered Calabi--Yau threefold $T^2 \hookrightarrow X_0 \to \mathbb{F}_0$, with base given by the Hirzebruch surface $\mathbb{F}_0$, can be equivalently described as an elliptic fibration $T^2 \hookrightarrow X_2 \to \mathbb{F}_2$ with base $\mathbb{F}_2$~\cite{Morrison:1996na}. From the perspective of F-theory on $X_0$, the geometry $X_2$ is realized at a special locus in the complex structure moduli space at which an additional $(-2)$-curve in the base is holomorphic. Thus, it seems that naively the theory obtained from $X_2$ is realized at $\text{codim}_{\mathbb{R}}=4$ in the hypermultiplet moduli space of the theory obtained from $X_0$. However, as explained in~\cite{Morrison:1996na}, there is a so-called non-polynomial deformation\footnote{See \cite{Anderson:2026gvz} for a recent account of these deformations from the perspective of GLSMs.} of $X_2$ which accounts for the missing hypermultiplet such that, as expected, the dimension of the hypermultiplet moduli space of the theory obtained from $X_2$ agrees with that of $X_0$ and one can simply transition from one theory to the other if the hypermultiplets are correctly identified. From the heterotic perspective, this smooth transition is surprising because the heterotic dual of the theory on $X_2$ differs from the dual of the theory on $X_0$ by discrete bundle data. Still, the dual F-theory description ensures that there is a smooth transition between the two theories at zero energies which from the heterotic perspective is non-perturbative; see also~\cite{MojoDojo1} for a recent discussion. 

As far as the hypermultiplet branch is concerned, the transition $\mathtt{H}_1 \to \mathtt{H}_0$ is similar to the one discussed above: the relation between two theories that differ by discrete bundle data is apparent at a special locus in the hypermultiplet moduli space but extends, upon suitably identifying the moduli, to the full hypermultiplet moduli space. This ensures that, as far as the hypermultiplet sector is concerned, a smooth transition between the two theories is realized. The conceptually new aspect of the transition $\mathtt{H}_1\to \mathtt{H}_0$ is the nucleation of the NS5-branes in the transition. We therefore recall that in the 6d transition reviewed above, the gauge group into which the two bundles are embedded is the same for both theories. In the four-dimensional transition that we analyse here, the gauge sector of the $\mathtt{H}_0$ theory is enhanced by an $SU(2)$ factor, compared to $\mathtt{H}_1$. For this reason, we also have to tune the vector multiplet moduli of the $\mathtt{H}_1$ theory to the special value $T=U=i$ to be able to transition to $\mathtt{H}_0$. Using the Ho{\v r}ava--Witten (HW) description of the heterotic $E_8\times E_8$ string, a 5-brane is nucleated inside one of the 9-branes at the end of the HW interval if we move along $\Delta_1$ towards the intersection $\Delta_1 \cap \Delta_0$. Continuing along $\Delta_0$ then corresponds to moving along the HW interval which, however, shrinks to zero size in the weak-coupling limit for the heterotic string. This shrinking is reflected in the fact that, prior to resolution, the Type IIB moduli space has a point of tangency between the conifold singularity and $\Delta_1$. The resolution corresponds to a double-scaling limit in which the HW interval shrinks to zero but the information about the location of the 5-brane is conserved. 

Recall that the part of the gauge data that has a non-trivial role in the transition is $E_U$, which is not embedded in the $E_8\times E_8$ gauge group realized on the 9-branes at the end of the HW interval. Therefore, there is no reason for the 5-brane to be nucleated inside one of the 9-branes. In principle, it can be nucleated at any point along the interval. To understand how this is possible, we can use the Type IIB realization of the $\mathtt{H}_1\to \mathtt{H}_0$ transition. We therefore consider Type IIB on $X_{12}$ in the vicinity of $\Delta_1$ but not exactly on $\Delta_1$ such that we are in the weak-coupling regime but not at the strict locus $S=i\infty$. We first notice that from a point in the vicinity of $\Delta_1$ it is possible to smoothly move to a point in the vicinity of $\Delta_0$. Thus, the Type IIB realization of the theory implies that, as far as the vector multiplet sector is concerned, the transition $\mathtt{H}_1\to \mathtt{H}_0$ between the heterotic theories can be smooth. Indeed, suppose we start at a point 
\begin{equation}
    p_0=(x_0=\bar{x}_0 , x_1=\epsilon) \,,\qquad 0<|\epsilon| \ll 1\,,
\end{equation}
in $\cM_{\rm c.s.}(X_{12})$. If $|\bar{x}_0|\gg 0$ we are in the regime where the theory is best described by the weakly coupled $\mathtt{H}_1$ theory. Let us now move to the point \begin{equation}
    p_1 = (x_0=\delta,x_1=\epsilon) \,,\qquad  0<|\delta|\ll |\epsilon|
\end{equation}
In this process we smoothly transition into the regime that is best described by the weakly coupled $\mathtt{H}_0$ theory. This means that, by going from $p_0$ to $p_1$, the remaining left-moving heterotic $U(1)$ must enhance to $SU(2)$ and then be broken by the gauge configuration $E_U$. In the $\mathtt{H}_1$ theory, the torus volume is parametrized by $x_0$ with $x_0=0$ corresponding to $T=U=i$. This identification is, however, only valid at $x_1=0$. For finite $x_1$, i.e., for finite heterotic string coupling, there must hence be corrections to the dictionary with $T=U=i$ corresponding to some finite value $x_0 \sim x_1$. At this value, the gauge data is exchanged for a 5-brane whose the position along the HW interval is determined by $x_1=\epsilon$. Let us stress again that the entire process is smooth as the Type IIB realization demonstrates and therefore there is not an actual phase transition between the $\mathtt{H}_1$ and $\mathtt{H}_0$ theory, but merely a crossover. From the heterotic perspective, the effects that smooth out this process have to be non-perturbative in nature since in perturbative string theory the nucleation of a 5-brane is a highly singular process. This is reminiscent of the results~\cite{Cvetic:2024wsj,Cvetic:2025nfx} where, in four dimensions, non-perturbative effects in heterotic string theory smooth out process that are singular in the analogous six-dimensional theories. 

We thus conclude that there is a smooth transition from the $\mathtt{H}_1$ theory to the $\mathtt{H}_0$ theory which shares some features with small instanton transitions in six dimensions, namely the nucleation of NS5-branes. However, the process discussed here also has crucial differences with the six-dimensional counterpart, since the heterotic gauge group that is participating in this process arises from the torus isometries and hence the HW 5-brane does not have to be nucleated at the end of the HW interval. The resulting process is smooth and shares also some features with the transition between the heterotic compactifications on K3 with instanton embedding $(12,12)$ and $(10,14)$, with the important difference that in the four-dimensional case an NS5-brane is involved. We can hence view the transition $\mathtt{H}_1 \to \mathtt{H}_0$ as a special property of four-dimensional theories that unifies the heterotic--heterotic duality and small instanton transitions. 

Let us close by commenting on related transitions in four-dimensional heterotic theories. First, there are the heterotic conifold transitions, as discussed in~\cite{Anderson:2022bpo}, where also bundle data are traded for NS5-branes in compactifications of the heterotic string on Calabi--Yau threefolds. A similar process occurs for both the (0,2) GLSM correspondences, discussed in~\cite{Distler:1995bc, Blumenhagen:1997vt},\footnote{These works are particularly relevant in the context of $(0,2)$ mirror symmetry~\cite{Blumenhagen:1996vu,Adams:2003zy,Melnikov:2012hk}.} and the backgrounds studied in~\cite{Anderson:2019axt}, where, from the heterotic perspective, both the topology of the target space and of the bundle change as the transition occurs. The important difference to the case discussed here is that for these processes the underlying Calabi--Yau threefold must develop a conifold singularity that is then resolved or deformed on the two sides of the transition. Instead, here the heterotic threefold ${\rm K3}\times T^2$ remains smooth, since in the $\mathtt{H}_0$ theory the torus is confined to the elliptic point $T=U=i$ and the K3 is confined to the Landau-Ginzburg point, both of which sit in smooth regions of moduli space. For this reason, the heterotic transition is more similar to the heterotic dual of the transition discussed in~\cite{MojoDojo1} in F-theory compactification on Calabi--Yau fourfolds. If these F-theory setups have a heterotic dual given by the heterotic string compactified on a Calabi--Yau threefold, the transition corresponds to the nucleation of an NS5-brane wrapping a curve inside the Calabi--Yau threefold. For related processes involving the F-theory dual of heterotic theories in the presence of NS5-branes, see~\cite{Braun:2018ovc}. As is the case in the $\mathtt{H}_1\to \mathtt{H}_0$ theory discussed here, the transitions analysed in~\cite{MojoDojo1} do not require a singular conifold-like geometry but are smooth. Similar, less-conventional, smooth transitions have also been considered between K\"ahler and non-K\"ahler geometries~\cite{Becker:2007ea}. It would be very interesting to study the relation between these kinds of heterotic processes further.

\section{Conclusions}\label{sec:conclusions}
\noindent String dualities are among the few tools available for probing non-perturbative aspects of gravitational physics. In this work, we explored Type IIB–heterotic duality in four-dimensional theories preserving eight supercharges near type II$_0$ loci in the complex structure moduli space of Calabi–Yau threefolds. These loci, which in one-parameter models are also referred to as K-points, have previously been shown to correspond to emergent string limits~\cite{Hassfeld:2025uoy}. However, while the results of~\cite{Hassfeld:2025uoy} demonstrate that in these limits some heterotic string exists, the detailed realization of this string, in particular its gauge sector, remain undetermined.  
Namely,~\cite{Hassfeld:2025uoy} identifies the EFT string solution that becomes tensionless near type II loci in Type IIB moduli spaces with a critical, weakly-coupled heterotic string. In this work, we determine the concrete heterotic realization of this string which, in turn, specifies the worldsheet theory on the Type IIB EFT string. In this way, we go beyond the existence proof of a weakly coupled duality frame associated with the type II limits and establish new string dualities relating the Type IIB and the heterotic string.

More precisely, in our analysis, we focus on four examples of Calabi--Yau threefold compactifications of Type IIB string theory for which we work out the heterotic dual. The Calabi--Yau threefolds correspond to the three so-called hypergeometric one-parameter models, $X_{3,3}$, $X_{4,4}$ and $X_{6,6}$, for which the moduli space contains a K-point, as well as the two-parameter model $X_{12}$. The corresponding geometries are reviewed in Section \ref{sec:het}. The identification between the massless worldsheet spectrum of the heterotic and EFT strings, as discussed in~\cite{Hassfeld:2025uoy}, shows the existence of light states compatible with an heterotic dual description. Here, we turn this existence proof into a new concrete string duality by explicitly constructing the dual heterotic theory, which we require to pass the three levels of duality tests described in Section \ref{sec:HetDualTypeII0} and summarized in Figure \ref{fig:level_duality}. Specifically, the worldsheet theories that we identify $i)$ have to reproduce the massless spectrum of the low-energy 4d $\cN=2$ theory of supergravity, $ii)$ their spectrum of perturbative BPS states has to match the light D3-brane states of Type IIB in the vicinity of the type II locus and $iii)$ must reproduce the duality/monodromy group determining the global structure of the vector multiplet moduli space. The heterotic backgrounds thus constructed are characterized by an internal interacting CFT with target space K3$\times T^2$, with appropriate gauge theory data chosen as to satisfy spacetime anomaly cancellation and leading to a well-defined supersymmetric theory. These constraints are described in Section \ref{sec:HetDualTypeII0}. Moreover, the heterotic torus is always fixed at a self-dual value of its moduli space, and hence these backgrounds do not possess a decompactification limit to ten-dimensional heterotic supergravity. Turning the argument around, it would be interesting to understand how the constraints on the heterotic side of the duality translate into geometric constraints on the existence of type II loci in the complex structure moduli space of Calabi--Yau threefolds with given Hodge numbers.

For Type IIB on $X_{3,3}$, the dual heterotic background admits an interpretation as a compactification of an eight-dimensional $SO(32)$ heterotic theory on K3, equipped with the vector bundle data described in Section \ref{sec:x33}. 
This construction hence shares several features with the classic example of Type II--heterotic duality near the type II$_1$ locus of $X_{12}$~\cite{Kachru:1995fv} reviewed in Section~\ref{ssec:type_II_1_locus_het_dual}. Unlike the locus studied there, however, the K-point does not admit a large-volume mirror description in Type IIA. 

For $X_{4,4}$, $X_{6,6}$ and $X_{12}$ the heterotic theories dual to the type II$_0$ loci are instead \emph{genuinely} four-dimensional. Indeed, the consistency of the gauge sector restricts the K3 theory to the interior of its moduli space, specifically to the Landau-Ginzburg/Gepner phase, where the required gauge data do not admit a realization as a smooth supersymmetric vector bundle on a geometric K3. Given the non-geometric nature of the model, a complete description therefore requires a construction of the interacting heterotic worldsheet CFT, taking into account all perturbative and non-perturbative  $\alpha'$-corrections. We describe how these corrections affect the consistency conditions of the gauge background in Section \ref{sec:4dHetVac}, which are required for supersymmetry and absence of anomalies on the worldsheet. 

In this work, however, we do not establish the consistency of the worldsheet theories directly on the heterotic side. Instead, we show that there is a unique choice for the topological data characterizing the heterotic gauge background that reproduces the required dimension of the hypermultiplet moduli space that, in addition, does not obviously contradict supersymmetry and anomaly cancellation. To do so, we first exploit the duality between the $SO(32)$ heterotic string and Type I one to argue for a suitable notion of non-perturbative stability away from the geometric regime. We then show that the worldsheet theory with this gauge theory data passes the (non-)perturbative duality tests outlined in Section~\ref{ssec:hetDualII0} and~\ref{ssec:nonperttest}. This demonstrates that, indeed, the worldsheet theory of the EFT string in the Type IIB realization is characterized by the proposed gauge data. Using that the Type IIB EFT string is a BPS solution of a UV complete string theory, we then conclude that this worldsheet theory is supersymmetric and anomaly free. We thus use the Type IIB realization of this theory to conclude that a heterotic string with the given gauge theory data flows to a consistent worldsheet theory. In this sense, the Type IIB realization of the EFT string ensures that the gauge data solve the $\alpha'$-corrected heterotic D-term condition and allows to consistently define fermions in spacetime. It would be interesting to use this insight from the Type IIB side to, e.g., infer the exact form of the (non-)perturbative $\alpha'$ corrections to the heterotic D-term. 

For the $X_{12}$ model presented in Section \ref{sec:twoparametermodelshet}, the type II$_0$ and II$_1$ loci both admit an heterotic dual theory, albeit with different topological data. Moreover, the type II$_0$ dual background is characterized by a spacetime-filling NS5-brane, which is absent in the description near the type II$_1$ locus. Since from the Type IIB perspective the two loci intersect in the complex structure moduli space, the transition from one degeneration locus to the other is therefore smooth and remains in the weak-coupling regime. In this way, we identify a type of smooth transition that involves the nucleation of both perturbative (gauge background data) and non-perturbative (NS5-brane) ingredients. The crossover shares similarities with known phase transitions involving heterotic theories, such as the nucleation of five-branes in the small instanton transition~\cite{Witten:1995gx}. Differently from the latter, however, the type $\mathrm{II}_0 \to$ type $\mathrm{II}_1$ transition is smooth. In this regard, it is similar to both the F-theory transition $\mathbb{F}_0 \to \mathbb{F}_2$ in six dimensions~\cite{Morrison:1996na}, where the heterotic dual interpretation is a change of instanton numbers $(12,12) \to (10,14)$ for bundles over K3, and the (0,2) GLSM correspondences described in~\cite{Distler:1995bc}. In both cases, as well as in the one analysed here, the gauge bundle data change between the two descriptions. Finally, the transition analysed here also has some similarities to the heterotic conifold transition studied in~\cite{Anderson:2022bpo} and the one discussed in~\cite{MojoDojo1}. It would be interesting to address these points in the future and study the relation with these transitions more thoroughly.

\paragraph{Acknowledgements.}
We thank Mirjam Cveti\v{c}, Damian van de Heisteeg, Lukas Kauf-mann, Jeroen Monnee, Eran Palti, Cumrun Vafa, and Timo Weigand for useful discussions. CA would like to thank the II. Institute for Theoretical Physics at the University of Hamburg and the DESY Theory Group for hospitality, where this work 
was initiated. The work of CA is supported through the grants CEX2020-001007-S, PID2021-123017NB-I00 and PID2024-156043NB-I00 funded by MCIN/AEI/10.13039/501100011033 and ERDF. The work of GM and MW is supported by  Deutsche Forschungsgemeinschaft through the Emmy Noether program 557478919, by Deutsche Forschungsgemeinschaft under Germany’s Excellence Strategy EXC 2121 Quantum Universe 390833306, and by Deutsche Forschungsgemeinschaft through the Collaborative Research Center 1624 “Higher Structures, Moduli Spaces and Integrability.”

\appendix

\section{Bundles on K3 from 10d}\label{app:A}
The gauge data over K3 discussed in the main text break the 8d perturbative gauge group arising from the dimensional reduction of the heterotic theory on $T^2$. The torus moduli lie in vector multiplets that are removed by the choice of gauge bundle and consequent St\"uckelberg of the leftover abelian factors, bringing the gauge symmetry down to $U(1)^2$, as discussed in Section~\ref{sec:HetDualTypeII0}. This suggests that these construction should fix the torus moduli also from a ten-dimensional perspective. In this appendix, we describe the geometric realization of this mechanism and its extension to the enhanced current algebra of the torus worldsheet theory. 

We first recall that the eight-dimensional enhanced gauge groups $G$ discussed around \eqref{eq:gauge_group_enhanc_torus} arise from the current algebra of the heterotic torus sigma model. For definiteness, we suppress the 10d gauge group $G_E$ in the following. From the worldsheet perspective, constructing a gauge bundle associated to these symmetries amounts to fibering the current algebra associated to $G$ over the K3 sigma model, giving rise to a gauged WZW model, as described in~\cite{Melnikov:2012cv}. From the target space perspective instead, the counterpart of this mechanism is the fibering of the $T^2$ factor over K3, namely $T^2 \hookrightarrow M \to \ $K3. As the isometries of the torus are abelian, the latter construction is allowed only for an abelian current algebra. Nonetheless, since we will not make use of the full machinery of GLSMs in this work,\footnote{Which, moreover, should be modified to account for the possible presence of NS5-branes~\cite{Blaszczyk:2011ib}.} we take the simpler approach of discussing this geometric realization first and then extend it to the enhancement loci in the torus moduli space. The resulting backgrounds belong to the class of constructions commonly referred to as Fu--Yau manifolds~\cite{Becker:2006et}. 
We briefly describe it mainly following~\cite{Israel:2015aea, Melnikov:2014ywa}. Let us take a torus of complex structure $T$ and K\"ahler modulus $U$. The metric on the $T^2$ fibration is 
\begin{equation}
    {\rm d} s^2 = \frac{U_2}{T_2 }|{\rm d}x_1 + T {\rm d}x_2 + \alpha|^2 + e^{2A(y)}{\rm d}s_{\rm K3}^2 \ ,
\end{equation}
where $A$ depends only on the K3 coordinates, and $\alpha=\alpha_1+T\alpha_2$ is the complex combination of the two circle connections. The non-trivial fibration and the varying conformal factor generally lead to a non-closed Hermitian form. The resulting non-K\"ahler geometry, supported by $H$-flux, is described by the Hull–Strominger system~\cite{Becker:2008rc}.
Starting from this metric, we then construct the $U(1)$ field strengths from the connection $\alpha$ as 
\begin{equation}
F=\frac{1}{2\pi}{\rm d}\alpha=F_1+TF_2 \, , \qquad
F_i=\frac{1}{2\pi}{\rm d}\alpha_i \, , \quad
[F_i]\in H^2(K3,\mathbb Z) \, .
\end{equation}
Hence, $F_{1,2}$ are two real two-forms on K3, whereas $F$ is generically a complex cohomology class. For backgrounds that preserve eight supercharges, supersymmetry requires this to be only supported on K3 and to satisfy
\begin{equation}
    F\wedge J_{\rm K3} = 0\, , \quad F \in H^{1,1}({\rm K3},\bbZ)\, .
\end{equation}
Equivalently, the curvatures are anti-self dual with respect to the K3 metric. Notice that the first condition, integrated over K3, is precisely the classical DUY equation~\eqref{eq:muvalue} from the $8d$ perspective.

The ten-dimensional anomaly-cancellation condition~\eqref{eq:tadpole} can be rewritten as
\begin{equation}
    {\rm d} H = \frac{\alpha'}{4} \left[ \tr(R^2) - \tr(\mathcal F^2) \right] \, ,
\end{equation}
where $\cal F$ and $R$ are respectively the ten-dimensional gauge and spacetime curvatures. The field strengths $F_{1,2}$, which are geometrized in 10d, contribute to $R$, which enters the anomaly cancellation condition. Equivalently, their Chern--Simons term $\mathrm{CS}_{3}(\alpha)$ contributes to the $H$-flux, since:
\begin{equation}
   H = {\rm d}B_{\rm base} -\frac{\alpha' }{4}\mathrm{CS}_3(\alpha) + \mathrm{spacetime\ and \ gauge}\, .
\end{equation}
All in all, we thus have that~\cite{Israel:2015aea}
\begin{equation}
    \frac4{\alpha'}\,{\rm d}H =- \tr\left[\cF^2\right] +  \frac{U_2}{T_2}F\wedge \bar F + \tr(R^2) \ ,
\end{equation}
where $\CF$ is the $E_8\times E_8$ field strength, and 
\begin{equation}
    F\wedge \bar F = F_1\wedge \bar F_1 + 2T_1F_1\wedge F_2
+|T|^2F_2\wedge F_2 \, .
\end{equation}
The tadpole condition integrated over K3 reads instead 
\begin{equation}
-\ch_{2}(\mathcal{F})-\frac{U_2}{T_2}\int_{K3}F\wedge\bar F+N_5=24 \, ,
\end{equation}
with $\ch_{2}(\mathcal{F})$ the (integrated) instanton contribution from the 10d gauge group. 
One can see that anomaly cancellation implies that for a non-trivial bundle for $F$ the area of the $T^2$ has to be quantized in units of $\alpha'$, as we checked happening along type II$_0$ loci discussed in the main text. Compatibly, as discussed in~\cite{Melnikov:2014ywa}, only the gravitational right-moving $U(1)^2_R$ current algebra is preserved on the worldsheet. Aspects of this quantization have been further discussed in~\cite{Melnikov:2012cv, Melnikov:2014ywa}. 

Lastly, to extend this discussion to enhancement loci, one must include the additional momentum–winding states that become massless there and are not captured by the ten-dimensional supergravity description. At these loci, the Abelian connections may be embedded into the enhanced algebra, and more general non-Abelian bundles become available. Operationally, one replaces $F\wedge F \to \tr (F\wedge F)$ and takes $F$ to be the appropriate Lie algebra-valued field strength. From the worldsheet perspective, fibering this algebra over K3 gives the appropriate worldsheet generalization of the construction. The corresponding connection must satisfy the supersymmetry conditions on K3, and its Chern–Simons form contributes to $H$, with normalization determined by the affine level and the bundle embedding. The reader interested in the GLSM realization of this model can consult~\cite{Melnikov:2012cv} and~\cite{Israel:2015aea}.

\section{Heartshorne--Serre Correspondence}\label{app:B}
The Heartshorne--Serre correspondence generalizes the classical correspondence between effective divisors and line bundles to subschemes of higher codimension. On a smooth projective surface $X$, it relates the construction of rank two vector bundles with zero-dimensional subschemes, i.e. points, and line bundles. In this appendix, we apply the construction to $X=K3$ and derive the cohomological results used in the main text. The interested reader can consult~\cite{arrondo2006homemadehartshorneserrecorrespondence} for a deeper review of the construction.  

\paragraph{Construction. }Let $Z\subset X$ be a local complete intersection subscheme of codimension 2. For K3, this is a collection of points $Z=\{p_1,\dots,p_\ell\}$, which can be taken at generic positions on the surface. We call $\ell(Z)\equiv \dim H^0(Z)$ the \emph{length} of the scheme, that is, the number of points. If there exists a line bundle $\cal L$ such that 
\begin{equation}\label{eq:CB}
h^0(X,\mathcal{L}) = \ell \ , \quad \ell(Z) > \ell \ ,
\end{equation}
then the extension 
\begin{equation}
0 \rightarrow   \mathcal{O}_X \rightarrow V \rightarrow \mathcal{L}\otimes I_Z \rightarrow 0 \ 
\end{equation} 
is a short exact sequence and $V$ is locally free sheaf of rank two, i.e. a \emph{bona fide} vector bundle~\cite{arrondo2006homemadehartshorneserrecorrespondence}.\footnote{Another relevant reference in this context is~\cite{Liu2023MuStable}. See in particular Remark 2 and Proposition 0.2 therein.}  In the cases of interest, $c_1(\mathcal{L})>0$, $\mathcal{L}$ is ample and thus $H^0(X,\mathcal{L})=\frac12 c_1(\mathcal{L})^2 + 2 $ from the index theorem~\eqref{eq:indexthm}. In the above expression, $\mathcal{O}_X$ is the structure sheaf on $X$ and $I_Z$ the ideal sheaf defining the subscheme $Z$. Moreover, it follows from the construction that the bundle $V$ is characterized by the data
\begin{equation}
    c_1(V)=c_1(\mathcal{L})\, , \quad  c_2(V) = \ell(Z)\, . 
\end{equation}
Notice that the condition on $\ell(Z)$ above translates to 
\begin{equation}
c_2(V) > \frac{1}{2}c_1(V)^2 + 2 \ ,
\end{equation}
for the construction to be consistent. Condition~\eqref{eq:CB} is denoted as Calay-Bacharach condition and it is satisfied for all bundles constructed in the main text. Importantly, as remarked in~\cite{arrondo2006homemadehartshorneserrecorrespondence}, the above bundle can be shown to be $\mu$-slope stable, in the sense described in Section~\ref{sec:HetDualTypeII0}. The Heartshorne--Serre construction can thus be used to define every bundle used in the main text, as well as the gauge data of the non-geometric examples of Section \ref{sec:4dHetVac}. For the constructions that are characterized by a non-trivial $c_1(V)$, the Kodaira vanishing theorem automatically implies that due to $c_1(V) = n[H]$ $n\in \bbZ$ being ample, the dimension of the cohomology group $H^1(\det V)$ is $h^1(\det V) = 0$.

\paragraph{Cohomology of the bundle dual to $X_{6,6}$.} For the bundle constructed for $X_{6,6}$ in Section~\ref{ssec:x66}, we are also interested in computing $h^1(X,V\otimes \det V)$. To this end, we tensor the defining sequence of $V$ with $\det V$. Since $\det V$ is a line bundle, and hence locally free, tensoring preserves exactness. Denoting $E\equiv V\otimes \det V$, the relevant bundle can be defined through the sequence
\begin{equation}
0 \rightarrow \mathcal{L} \rightarrow E \rightarrow \mathcal{L}^2 \otimes I_Z \rightarrow 0 \ ,
\end{equation}
where we have used the fact that $\det V=\mathcal{L}$ by construction, together with the canonical isomorphism $\mathcal{F}\otimes \mathcal{O}_X \simeq \mathcal{F}$ for any $\cal F$. Furthermore, a short exact sequence always induces a long exact sequence in cohomology
\begin{equation}
0 \rightarrow H^0(\cL) \rightarrow H^0(E) \rightarrow H^0(\cL^2 \otimes I_Z) \rightarrow H^1(\cL) \rightarrow \dots
\end{equation}
For the ample line bundle $\cL$\footnote{Recall that, in all our constructions, the first Chern class of $\cL$ is proportional to the only curve class of the quartic K3 described in Section~\ref{ssec:nonGeomComp}.}, characterized by $H^1(\cL)=H^2(\cL)=0$, the sequence splits as
\begin{equation}
0 \rightarrow H^1(E) \rightarrow H^1(\cL^2 \otimes I_Z) \rightarrow 0 \ ,
\end{equation}
which induces an isomorphism of groups of the form
\begin{equation}\label{eq:isomorphism}
    H^1(E) \cong H^1(\cL^2 \otimes I_Z)\, .
\end{equation}
Having established the isomorphism, let us compute $H^1(\cL^2\otimes I_Z)$. To do that, let us use the sequence defining the ideal
\begin{equation}
0 \rightarrow I_Z \rightarrow \cO_X \stackrel{\rm ev}\rightarrow \cO_Z \rightarrow 0 \ .
\end{equation}
The map $\rm ev$ is called evaluation map, as it evaluates the trivial bundle on $X$ on the points defining $Z$. Exactness implies that $I_Z$ is precisely the kernel of this evaluation map\footnote{i.e its kernel consists precisely of functions vanishing on $Z$.}. Now, tensoring the sequence above with $\cL^2$ and taking the long sequence in cohomology, we obtain the following split:
\begin{equation}
0 \rightarrow H^0(\cL^2 \otimes I_Z) \rightarrow H^0(\cL^2) \rightarrow H^0(\cL^2|_Z) \rightarrow H^1(\cL^2 \otimes I_Z) \rightarrow 0 \ , 
\end{equation}
where we used that $H^1(\cL^2)=0$ and that $\cO_Z\otimes \cL^2 = \cL^2|_Z$, which defines the restriction of $\cL^2$ to the subscheme $Z$. Moreover, $H^0(\cL^2|_Z) \cong \mathbb{C}^{\mathrm{length}(Z)}$ since it is once again a collection of points. Notice that the map 
\begin{equation}
    \mathrm{ev:}\ H^0(\cL^2) \longrightarrow H^0(\mathcal L^2|_Z) \simeq\mathbb{C}^{\mathrm{length}(Z)}
\end{equation} 
is the lift of the evaluation map. Exactness means that
\begin{equation}
H^1(I_Z\otimes \cL^2) \cong \mathrm{coker}(\mathrm{ev}) \ \implies  \ h^1(I_Z \otimes \cL^2) = \mathrm{length}(Z) - \dim (\mathrm{im}(\mathrm{ev})) \ .
\end{equation}
Seen as a map between linear spaces given by the sections in $H^0$, the lift of the evaluation map has maximal possible rank when the choice of points is generic. Indeed, by choosing a basis of sections $\{ s_i\}_{i=1}^N$ of $H^0(\cL^2)$ and a basis of $\mathbb{C}^{\mathrm{length}(Z)}$, the $m\times N$ matrix defining the evaluation map is given by \begin{equation}
    M_{ji}=s_i(p_j)\, .
\end{equation} 
The kernel of this map defines precisely the linear combinations of global sections that vanish at all points on $Z$, such that $c^T M= 0$ for a set of coefficients $c_i$. Each point of $Z$ imposes one linear condition on a global section, namely that it vanish at that point. For a generic configuration, these conditions are independent and the space of sections satisfying all of them simultaneously is empty. The behaviour of this map thus depends on the number of sections versus the number of points. If we have more points than sections, there are generically no sections that pass through all of them, and the map is injective with maximal rank. If this is not the case, the generic position of the points implies that the map is instead surjective and each point imposes independent conditions. All in all, we have
\begin{equation}
h^1(I_Z \otimes \cL^2) = \mathrm{length}(Z) - \mathrm{min}(\mathrm{length}(Z), h^0(\cL^2)) \ ,
\end{equation}
In terms of the Chern classes, recalling the isomorphism~\eqref{eq:isomorphism}, it reads
\begin{equation}
h^1(E) = c_2(V) - \mathrm{min}(c_2(V), 2c_1(V)^2 + 2) \ .
\end{equation}
For the $X_{6,6}$ bundle considered in Section~\ref{ssec:x66} and defined by the data~\eqref{eq:bundle_choice_x66}, it yields $h^1(E)=0$ as required.

\bibliography{references}
\bibliographystyle{JHEP}

\end{document}